\documentclass[12pt]{article}
\pdfoutput=1

\usepackage{cite}
\usepackage{setspace}
\usepackage{xspace}
\usepackage{color}
\usepackage{amsmath, amsfonts, amssymb, amsthm, upgreek, amstext, bm}
\usepackage[font=small,labelfont=bf]{caption}
\usepackage{slashed}
\usepackage{subfig}
\usepackage{multirow}
\usepackage{array} 
\usepackage{bbold}
\usepackage{physics}
\usepackage{booktabs}
\usepackage{fontawesome}
\usepackage[english]{babel}
\usepackage[utf8]{inputenc}
\usepackage[colorlinks=true,linkcolor=rossos,anchorcolor=black,citecolor=blue,filecolor=cyan,menucolor=red,runcolor=filecolor,urlcolor=blue,bookmarks=true,bookmarksnumbered=true]{hyperref}
\usepackage[dvipsnames]{xcolor}
\usepackage{doi}

\usepackage{ifpdf}
\ifpdf
\usepackage[pdftex]{graphicx}
\usepackage{epstopdf}
\else
\usepackage[dvips]{graphicx}
\fi

\definecolor{rosso}{cmyk}{0,1,1,0.4}
\definecolor{rossos}{cmyk}{0,1,1,0.55}
\definecolor{rossoc}{cmyk}{0,1,1,0.2}
\definecolor{blu}{cmyk}{1,1,0,0.3}
\definecolor{blus}{cmyk}{1,1,0,0.6}
\definecolor{bluc}{cmyk}{1,1,0,0.1}
\definecolor{verde}{cmyk}{0.92,0,0.59,0.25}
\definecolor{verdec}{cmyk}{0.92,0,0.59,0.15}
\definecolor{verdes}{cmyk}{0.92,0,0.59,0.4}

\newcommand{\be}{\begin{equation}}
\newcommand{\ee}{\end{equation}}
\newcommand{\bea}{\begin{eqnarray}}
\newcommand{\eea}{\end{eqnarray}}
\newcommand{\bry}{\begin{array}}
\newcommand{\ery}{\end{array}}
\newcommand{\bag}{\begin{align}}
\newcommand{\eag}{\end{align}}
\newcommand{\dst}{\displaystyle}

\usepackage{marginnote}

\def\f{\frac}
\def\l{\label}
\def\({\left(}
\def\){\right)}

\title{
\vspace{-1.5cm}
{\normalsize{\hfill CERN-TH-2019-187}}\\
\vspace{1.5cm}
\vspace{0.0 cm}
{\bf The DNNLikelihood: enhancing likelihood distribution with Deep Learning}}

\author{
Andrea Coccaro$^{a}$, Maurizio Pierini$^{b}$, Luca Silvestrini$^{b,c}$, and Riccardo Torre$^{a,b}$\\ \\
{\small\emph{$^a$ INFN, Sezione di Genova, Via Dodecaneso 33, I-16146 Genova, Italy}}\\
{\small\emph{$^b$ CERN, 1211 Geneva 23, Switzerland}}\\
{\small\emph{$^c$ INFN, Sezione di Roma, P.le A. Moro, 2, I-00185 Roma, Italy}}\\
}
\date{}

\begin{document}
\maketitle

%
%

\begin{abstract}
\noindent We introduce the DNNLikelihood, a novel framework to easily
encode, through Deep Neural Networks (DNN), the full experimental
information contained in complicated likelihood functions (LFs). We
show how to efficiently parametrise the LF, treated as a multivariate
function of parameters and nuisance parameters with high
dimensionality, as an interpolating function in the form of a DNN
predictor. We do not use any Gaussian approximation or dimensionality
reduction, such as marginalisation or profiling over nuisance
parameters, so that the full experimental information is retained. The
procedure applies to both binned and unbinned LFs, and allows for an
efficient distribution to multiple software platforms, e.g.\ through
the framework-independent ONNX model format. The distributed
DNNLikelihood can be used for different use cases, such as re-sampling
through Markov Chain Monte Carlo techniques, possibly with custom
priors, combination with other LFs, when the correlations among
parameters are known, and re-interpretation within different
statistical approaches, i.e.\ Bayesian vs frequentist. We discuss the
accuracy of our proposal and its relations with other approximation
techniques and likelihood distribution frameworks. As an example, we
apply our procedure to a pseudo-experiment corresponding to a
realistic LHC search for new physics already considered in the
literature. 

\end{abstract}

\clearpage

\setcounter{equation}{0}
\setcounter{footnote}{0}

\newpage


\section{Introduction} \label{intro}
The Likelihood Function (LF) is the fundamental ingredient of any statistical inference.
It encodes the full information on experimental measurements and
allows for their interpretation both from a frequentist (e.g.\ Maximum
Likelihood Estimation (MLE)) and a Bayesian (e.g.\ Maximum a Posteriori
(MAP)) perspectives.\footnote{To be precise, the LF does not contain
  the full information in the frequentist approach, since the latter
  does not satisfy the Likelihood Principle (for a detailed comparison
  of frequentist and Bayesian inference see, for instance,
  refs. \cite{Stuart1996,OHagan2004}). In particular, frequentists
  should specify assumptions about the experimental setup and about
  experiments that are not actually performed which are not relevant
  in the Bayesian approach. Since these assumptions are however
  usually well spelled in fundamental physics and astrophysics (at
  least when classical inference is carefully applied), we ignore this
  issue and assume that the LF encodes the full experimental
  information.}

On top of providing a description of the combined conditional
probability distribution of data given a model (or vice versa, when
the prior is known, of a model given data), and therefore of the
relevant statistical uncertainties, the LF may also encode, through
the so-called {\it nuisance parameters}, the full knowledge of systematic uncertainties and additional constraints (for instance coming from the measurement of fundamental input parameters by other experiments) affecting a given measurement or observation, as for instance discussed in ref.~\cite{ATLAS:2011tau}.

Current experimental and phenomenological results in fundamental physics and astrophysics typically involve complicated fits with several parameters of interest and hundreds of nuisance parameters.
Unfortunately, it is generically considered a hard task to provide all the information encoded in the LF in a practical and reusable way.
Therefore, experimental analyses usually deliver only a small fraction of the full information contained in the
LF, typically in the form of confidence intervals obtained by
profiling the LF on the nuisance parameters (frequentist approach), or
in terms of probability intervals obtained by marginalising over
nuisance parameters (Bayesian approach), depending on the statistical
method used in the analysis. This way of presenting results is very
practical, since it can be encoded graphically into simple plots
and/or simple tables of expectation values and correlation matrices
among observables, effectively making use of the Gaussian
approximation. However, such `partial' information can hardly be used
to reinterpret a result within a different physics scenario, to combine
it with other results, or to project its sensitivity to the future.
These tasks, especially outside experimental collaborations, are usually
done in a na\"{i}ve fashion, trying to reconstruct an approximate
likelihood for the quantities of interest, employing a Gaussian
approximation, assuming full correlation/uncorrelation among
parameters, and with little or no control on the effect of systematic
uncertainties. Such control on systematic uncertainties could be
particularly useful to project the sensitivity of current analyses to
future experiments,
an exercise particularly relevant in the context of future collider
studies
\cite{Abada:2019lih,Behnke:2013xla,Aicheler:2012bya}.
One could for instance ask how a certain
experimental result would change if a given systematic uncertainty
(theoretical or experimental) is reduced by some amount. This kind of question is usually not addressable using only public results for the aforementioned reasons. This and other limitations could of course be overcome if the full LF were available as a function of the observables and of the elementary nuisance
parameters, allowing for:
\begin{enumerate}
\item the combination of the LF with other LFs involving (a subset of) the same observables and/or nuisance parameters;
\item the reinterpretation of the analysis under different theoretical assumptions (up to issues with unfolding);
\item the reuse of the LF in a different statistical framework;
\item the study of the dependence of the result on the prior knowledge of the observables and/or nuisance parameters.
\end{enumerate}
A big effort has been put in recent years into improving the distribution of information on the experimental LFs, usually in the form of binned histograms in mutually exclusive categories, or, even better, giving information on the covariance matrix between them. An example is given by the Higgs Simplified Template Cross Sections~\cite{Berger:2019wnu}. Giving only information on central values and uncertainties, this approach makes intrinsic use of the Gaussian approximation to the LF without preserving the original information on the nuisance parameters of each given analysis.
A further step has been taken in refs. \cite{Fichet:2016gvx,Buckley:2018vdr,CMS:SLnote} where a simplified parameterisation in terms of a set of ``effective'' nuisance parameters was proposed, with the aim of catching the main features of the true distribution of the observables, up to the third moment.\footnote{This approach is similar to the one already proposed in ref.~\cite{Cranmer:2013hia}.}
This is a very
practical and effective solution, sufficiently accurate for many use cases. On the other hand, its underlying approximations come short whenever
the dependence of the LF on the original nuisance parameters is needed, and 
with highly non-Gaussian (e.g.\ multi-modal) LFs.

Recently, the  ATLAS collaboration has taken a major step forward, releasing
the full experimental LF of an analysis \cite{Aad:2019pfy} on
HEPData~\cite{hepdata} through the HistFactory
framework~\cite{Cranmer:1456844}, with the format presented in
ref.~\cite{ATL-PHYS-PUB-2019-029}. Before this, the release of the
full experimental likelihood was advocated several times as a
fundamental step forward for the HEP community (see e.g.\ the panel
discussion in ref.~\cite{James:2000et}), but so far it was not followed up with a concrete commitment. The lack of a concrete effort in this direction was usually attributed to technical difficulties and indeed, this fact was our main motivation when we initiated the work presented in this paper.

In this work, we propose to present the full LF, as used by experimental collaborations to produce the results of their analyses, in the form of a suitably trained Deep Neural Network (DNN), which is able to reproduce the original LF as a function of physical and nuisance parameters with the accuracy required to allow for the four aforementioned tasks.

The DNNlikelihood approach offers at least two remarkable practical
advantages. First, it does not make underlying assumptions on the
structure of the LF (e.g.\ binned vs unbinned), extending to use cases
that might be problematic for currently available alternative
solutions. For instance, there are extremely relevant analyses that
are carried out using an unbinned LF, notably some Higgs study in the
four-leptons golden decay mode~\cite{Sirunyan:2017exp} and the
majority of the analyses carried out at B-physics experiments. Second,
the use of a DNN does not impose on the user any specific software
choice. Neural Networks are extremely portable across multiple
software environments (e.g.\ C++, \textsc{Python}, Matlab, R, or
Mathematica) through the ONNX format~\cite{ONNX}. This aspect could be
important whenever different experiments make different choices in
terms of how to distribute the likelihood.

In this respect, we believe that the use of the DNNLikehood could be
relevant even in a future scenario in which every major experiment has
followed the remarkable example set by ATLAS. For instance, it could
be useful to overcome some technical difficulty with specific classes
of analyses. Or, it could be seen as a further step to take in order
to import a distributed likelihood into a different software
environment. In addition, the DNNLikehood could be used in other
contexts, e.g.\ to distribute the outcome of phenomenological analyses
involving multi-dimensional fits such as the Unitarity Triangle
Analysis
\cite{Ciuchini:2000de,Hocker:2001xe,Charles:2004jd,Bona:2005vz,Bona:2007vi},
the fit of electroweak precision data and Higgs signal strengths
\cite{Ciuchini:2013pca,Baak:2014ora,deBlas:2016ojx,Falkowski:2017pss,Ellis:2018gqa},
etc.

There are two main challenges associated to our proposed strategy:
on one hand, in order to design a supervised
learning technique,  an accurate sampling of the LF is needed for the training of
the DNNLikelihood. On the other hand a (complicated) interpolation
problem should be solved with an accuracy that ensures a real
preservation of all the required information on the original
probability distribution.

The first problem, i.e.\ the LF sampling, is generally easy to solve
when the LF is a relatively simple function which can be fastly
evaluated in each point in the parameter space. In this case, Markov
Chain Monte Carlo (MCMC) techniques \cite{Clark2016} are usually
sufficient to get dense enough samplings of the function, even for
very high dimensionality, in a reasonable time. However the problem
may quickly become intractable with these techniques when the LF is
more complicated, and takes much longer time to be evaluated.
This is typically the case when the sampling workflow requires
  the simulation of a data sample, including computationally costly
 corrections (e.g.\ radiative corrections) and/or a
simulation of the detector response, e.g.\ through Geant4
\cite{Agostinelli:2002hh}. In these cases, evaluating the LF for a
point of the parameter space may require $\mathcal{O}$(minutes) to go through the full chain of generation,
  simulation, reconstruction, event selection, and likelihood
  evaluation, 
making the LF sampling with standard MCMC techniques
impractical.
To overcome this difficulty,
several ideas have recently been
proposed, inspired by Bayesian optimisation and Gaussian processes,
known as Active Learning (see, for instance,
refs.~\cite{Kandasamy2017,Caron:2019xkx} and references therein). These
techniques, though less robust than MCMC ones, allow for a very
``query efficient'' sampling, i.e.\ a sampling that requires the
smallest possible number of evaluations of the full LF. Active
Learning applies machine learning techniques to design the proposal function of
the sampling points and can be shown to be much more query efficient
than standard MCMC techniques. Another possibility would be to employ
deep learning and MCMC techniques together in a way similar to Active
Learning, but inheriting some of the nice properties of MCMC. We defer
a discussion of this new idea to a forthcoming publication
\cite{toappear_sampling}, while in this work we focus on the second of
the aforementioned tasks: we assume that an accurate sampling of the LF
is available and design a technique to encode it and distribute it
through DNNs.

A \textsc{Jupyter} notebook and the \textsc{Python} source files which
allow to reproduce all results presented in this paper are available
on GitHub \href{https://github.com/riccardotorre/DNNLikelihood}{\faGithub}. A dedicated
\textsc{Python} package allowing to sample LFs and to build, optimize,
train, and store the corresponding DNNLikelihoods is in
preparation. This will allow not only allow to construct and
distribute DNNLikelihoods, but also to use them for inference both in
the Bayesian and frequentist frameworks.

This paper is organized as follows. In Section \ref{sec:regr} we discuss the issue of interpolating the LF from a Bayesian and frequentist perspective and set up the procedure for producing suitable training datasets.
In Section \ref{sec:toy} we describe a benchmark example, consisting
in the realistic LHC-like New Physics (NP) search proposed in
ref.~\cite{Buckley:2018vdr}. In Section \ref{sec:DNNLik} we show our
proposal at work for this benchmark example, whose LF depends on one
physical parameter, the signal strength $\mu$, and $94$ nuisance parameters.
Finally, in Section \ref{sec:conclusion} we conclude and discuss some interesting ideas for future studies.

\section{Interpolation of the Likelihood Function}\l{sec:regr}
The problem of fitting high-dimensional multivariate functions is a classical interpolation problem, and it is nowadays widely known that DNNs provide the best solution to it.
Nevertheless, the choice of the loss function to minimise and of the
metrics to quantify the performance, i.e.\ of the
``distance'' between the fitted and the true function, depends
crucially on the nature of the function and its properties. The LF is
a special function, since it represents a probability distribution. As
such, it corresponds to the integration measure over the probability
of a given set of random variables.
The interesting regions of the LF are twofold.
From a frequentist perspective the knowledge of the local maxima, and
of the global maximum, is needed.
This requires a good knowledge of the LF in regions of the parameter space with high probability (large likelihood), and, especially for high dimensionality, very low probability mass (very small prior volume).
These regions are therefore very hard to populate via sampling techniques \cite{Feroz:2011bj} and give tiny contributions to the LF integral, the latter being increasingly dominated by the ``tails'' of the multidimensional distribution as the number of dimensions grows.
From a Bayesian perspective the expectation values of observables or parameters, which can be computed through integrals over the probability measure, are instead of interest.
In this case one needs to accurately know regions of very small probabilities, which however correspond to large prior volumes, and could give large contributions
to the integrals.

Let us argue what is a good ``distance'' to minimise to achieve both
of the aforementioned goals, i.e.\ to know the function equally well in
the tails and close to the local maxima. Starting from the view of the
LF as a probability measure (Bayesian perspective), the quantity that
one is interested in minimising is the difference between the expectation
values of observables computed using the true probability distribution
and the fitted one.

For instance, in a Bayesian analysis one may be interested in using
the probability density $\mathcal{P} = \mathcal{L}\times \Pi$, where
$\mathcal{L}$ denotes the likelihood and $\Pi$ the prior, to estimate
expectation values as
\be
\l{eq:expf}
\bm{E}_{\mathcal{P}(\bm{x})}[f(\bm{x})] =\int f(\bm{x})\dd
\mathcal{P}(\bm{x}) = \int f(\bm{x}) \mathcal{P}(\bm{x})\dd \bm{x}\,,
\ee
where the probability measure is 
$\dd \mathcal{P}(\bm{x}) = \mathcal{P}(\bm{x})\dd \bm{x}$, and we
collectively denoted by the $n$-dimensional vector $\bm{x}$ the
parameters on which $f$ and $\mathcal{P}$ depend, treating on the same
footing the nuisance parameters and the parameters of interest. Let us
assume now that the solution to our interpolation problem provides a
predicted pdf $\mathcal{P}_{\mathrm{P}}(\bm{x})$, leading to an
estimated expectation value
\be
\l{eq:expfp}
\bm{E}_{\mathcal{P}_{\mathrm{P}}(\bm{x})}[f(\bm{x})] = \int
f(\bm{x}) \mathcal{P}_{\mathrm{P}}(\bm{x})\dd
\bm{x}\,. 
\ee
This can be rewritten, by defining the ratio
$r(\bm{x}) \equiv \mathcal{P}_{\mathrm{P}}(\bm{x})/\mathcal{P}(\bm{x})
= \mathcal{L}_{\mathrm{P}}(\bm{x})/\mathcal{L}(\bm{x})$, as
\be
\bm{E}_{\mathcal{P}_{\mathrm{P}}(\bm{x})}[f(\bm{x})] = \int
f(\bm{x}) r(\bm{x})\mathcal{P}(\bm{x}) \dd \bm{x} \,,
\ee so that the
absolute error in the evaluation of the expectation value is given by
\be
\l{eq:expfe}
\left|\bm{E}_{\mathcal{P}(\bm{x})}[f(\bm{x})]-\bm{E}_{\mathcal{P}_{\mathrm{P}}(\bm{x})}[f(\bm{x})]
\right| = \left|\int f(\bm{x})\qty(1-r(\bm{x}))\mathcal{P}\bm{(}x)\dd
  \bm{x}\right|\,.
\ee
For a finite sample of points $\bm{x}_{i}$, with $i=1,\ldots,N$, the
integrals are replaced by sums and eq.~\eqref{eq:expfe} becomes
\be
\l{eq:expfe2}
\left|\bm{E}_{\mathcal{P}(\bm{x})}[f(\bm{x})]-\bm{E}_{\mathcal{P}_{\mathrm{P}}(\bm{x})}[f(\bm{x})]\right|  
= \left|\frac{1}{N}\sum_{\bm{x}_{i}|_{\mathcal{U}(\bm{x})}} f(\bm{x}_{i})\qty(1-r(\bm{x}_{i}))\mathcal{F}(\bm{x}_{i})\right| \,.
\ee
Here, the probability density function $\mathcal{P}(\bm{x})$ has been
replaced with $\mathcal{F}(\bm{x}_{i})$, the frequencies with which
each of the $\bm{x}_{i}$ occurs, normalised such that
$\sum_{\bm{x}_{i}|_{\mathcal{U}(\bm{x})}}\mathcal{F}(\bm{x}_{i})=N$,
the notation $\bm{x}_{i}|_{\mathcal{U}(\bm{x})}$ indicates that the
$\bm{x}_{i}$ are drawn from a uniform distribution, and the $1/N$
factor ensures the proper normalisation of probabilities. This sum is
very inefficient to calculate when the probability distribution
$\mathcal{P}(\bm{x})$ varies rapidly in the parameter space, i.e.\ deviates strongly from a uniform distribution, since most of the $\bm{x}_{i}$ points drawn from the uniform distribution will correspond to very small probabilities, giving negligible contributions to the sum.  An example is given by multivariate normal distributions, where, increasing the dimensionality, tails become more and more relevant (see Appendix \ref{app:multvarnorm}). A more efficient way of computing the sum is given by directly sampling the $\bm{x}_{i}$ points from the probability distribution $\mathcal{P}(\bm{x})$, so that eq.~\eqref{eq:expfe2} can be rewritten as
\be
\l{eq:expfe3}
\left|\bm{E}_{\mathcal{P}(\bm{x})}[f(\bm{x})]-\bm{E}_{\mathcal{P}_{\mathrm{P}}(\bm{x})}[f(\bm{x})]\right|  
= \left|\frac{1}{N}\sum_{\bm{x}_{i}|_{\mathcal{P}(\bm{x})}} f(\bm{x}_{i})\qty(1-r(\bm{x}_{i}))\right|\,.
\ee
This expression clarifies the aforementioned importance of being able
to sample points from the probability distribution
$\mathcal{P}$ to efficiently discretize the integrals and compute
expectation values. The minimum of this function for any $f(\bm{x})$
is in $r(\bm{x}_{i})=1$, which, in turn, implies
$\mathcal{L}(\bm{x}_{i})=\mathcal{L}_{\mathrm{P}}(\bm{x}_{i})$. This
suggests that an estimate of the performance of the interpolated
likelihood is given by the Mean Percentage Error (MPE)
\be
\text{MPE}_{\mathcal{L}}\,\,=\,\,
\dst\frac{1}{N}\sum_{\bm{x}_{i}|_{\mathcal{P}(\bm{x})}}
\left(
  1-\f{\mathcal{L}_{\mathrm{P}}(\bm{x}_{i})}{\mathcal{L}(\bm{x}_{i})}
\right)=
\frac{1}{N}\sum_{\bm{x}_{i}|_{\mathcal{P}(\bm{x})}}
\left(
  1-r(\bm{x}_{i})
\right)\,.
\ee

Technically, formulating the interpolation problem on the LF itself
introduces the difficulty of having to fit the function over several
orders of magnitude, which leads to numerical instabilities. For this reason it is much
more convenient to formulate the problem using the natural logarithm
of the LF, the so-called log-likelihood $\log\mathcal{L}$.
Let us see how the error on the
log-likelihood propagates to the actual likelihood. Consider the Mean
Error (ME)
on the log-likelihood
\be \text{ME}_{\log\mathcal{L}}\,\,=
\,\,\dst\frac{1}{N}\sum_{\bm{x}_{i}|_{\mathcal{P}(\bm{x})}}\left(\log\mathcal{L}(\bm{x}_{i})-\log\mathcal{L}_{\mathrm{P}}(\bm{x}_{i})\right)
=\frac{1}{N}\sum_{\bm{x}_{i}|_{\mathcal{P}(\bm{x})}}\log
r(\bm{x}_{i})\,.
\label{eq:MELL}
\ee
The last logarithm can be expanded for $r(\bm{x}_{i}) \sim 1$ to give
\be
\text{ME}_{\log\mathcal{L}}\,\,\approx
\,\,\frac{1}{N}\sum_{\bm{x}_{i}|_{\mathcal{P}(\bm{x})}}
\left(
  1-r(\bm{x}_{i})
\right)=\text{MPE}_{\mathcal{L}}\,.
\ee

It is interesting to notice that $\text{ME}_{\log\mathcal{L}}$ defined
in eq.~(\ref{eq:MELL}) corresponds to the Kullback–Leibler divergence \cite{kullback1951},
or relative entropy, between $\mathcal{P}$ and
$\mathcal{P}_{\mathrm{P}}$:
\begin{equation}
  \label{eq:KL}
  D_{\mathrm{KL}} = \int \log(\frac{\mathcal{P}(\bm{x})}{\mathcal{P}_{P}(\bm{x})})
\mathcal{P}(\bm{x})\dd{x} = \frac{1}{N}\sum_{\bm{x}_{i}|_{\mathcal{P}(\bm{x})}}\log
r(\bm{x}_{i}) = \text{ME}_{\log\mathcal{L}}\,.
\end{equation}

While eq.~(\ref{eq:KL}) confirms that small values of
$D_{\mathrm{KL}}=\text{ME}_{\log\mathcal{L}}\sim\text{MPE}_{\mathcal{L}}$
correspond to a good performance of the interpolation,
$D_{\mathrm{KL}}$, as well as ME and MPE, do not satisfy the triangular inequality and therefore cannot be directly
optimised for the purpose of training and evaluating a
DNN. Eq.~(\ref{eq:MELL}) suggests however that the Mean Absolute Error
(MAE) or the Mean Square Error (MSE) on $\log\mathcal{L}$ should be
suitable losses for the DNN training: we explicitly checked
that this is indeed the case, with MSE performing slightly better for
well-known reasons. 

Finally, in the frequentist approach, the LF can be treated just as any function in a regression (or interpolation) problem, and, as we will see, the MSE provides a good choice for the loss function.

\subsection{Evaluation metrics}
\l{sec:evaluation}

We have argued above that the MAE or MSE on $\log \mathcal{L}(\bm{x}_{i})$ are the most suitable loss functions to train our DNN for interpolating the LF on the sample $\bm{x}_{i}$.
We are then left with the question of measuring the performance of our interpolation from the statistical point of view. 
In addition to $D_{\mathrm{KL}}$, several quantities can be computed to quantify the performance of the predictor. First of all, we perform a
Kolmogorov-Smirnov (K-S) two-sample test \cite{tKOL33a,smirnov1948} on all the marginalised one-dimensional distributions obtained using $\mathcal{P}$ and $\mathcal{P}_{P}$.
In the hypothesis that both distributions are drawn from the same pdf, the $p$-value should be distributed uniformly in the interval $[0,1]$.
Therefore, the median of the distribution of $p$-values of the one-dimensional K-S tests is a representative single number which allows to evaluate the performance of the model.
We also compute the error on the width of Highest Posterior Density Intervals (HPDI) $\mathrm{PI}^{i}$ for the
marginalised one-dimensional distribution of the $i$-th parameter, $E_{\mathrm{PI}}^{i} =\left\vert \mathrm{PI}^{i} - \mathrm{PI}^{i}_{P} \right\vert$, as well as the relative error on the median of each marginalised distribution. 
From a frequentist point of view, we are interested in reproducing as precisely as possible the test statistics used in classical inference.
In this case we evaluate the model looking at the mean error on the test statistics $t_{\mu}$, that is the likelihood ratio profiled over nuisance parameters.

To simplify the presentation of the results, we choose the best models according to the median K-S $p$-value, when considering bayesian inference, and the mean error on the $t_{\mu}$ test-statistics, when considering frequentist inference.
These quantities are compared for all the different models on an identical test set statistically independent from both the training and validation sets used for the hyperparameter optimisation.

\subsection{Learning from imbalanced data}\l{sec:imbdata}
The loss functions we discussed above are all averages over all
samples and, as such, will lead to a better learning in regions that
are well represented in the training set and to a less good learning
in regions that are under-represented. On the other hand, an unbiased
sampling of the LF will populate much more regions corresponding to a
large probability mass than regions of large LF. Especially in large
dimensionality, it is prohibitive, in terms of the needed number of
samples, to make a proper unbiased sampling of the LF, i.e.\ converging to the underlying probability distribution, while still covering the large LF region with enough statistics. In this respect, learning a multi-dimensional LF raises the issue of learning from highly imbalanced data. This issue is extensively studied in the ML literature for classification problems, but has gathered much less attention from the regression point of view \cite{Krawczyk2016,Branco2017,Ren2018}. 

There are two main approaches in the case of regression, both
resulting in assigning different weights to different examples. In the
first approach, the training set is modified by oversampling and/or
undersampling different regions (possibly together with noise) to
counteract low/high population of examples, while in the second
approach the loss function is modified to weigh more/less
regions with less/more examples. In the case where the theoretical
underlying distribution of the target variable is (at least
approximately) known, as in our case, either of these two procedures
can be applied by assigning weights that are proportional to the
inverse frequency of each example in the population. This approach,
applied for instance by adding weights to a linear loss function,
would really weigh each example equally, which may not be exactly what
we need. Moreover, in the case of large dimensionality, the
interesting region close to the maximum would be completely absent
from the sampling, making any reweighting irrelevant. In this paper we
therefore apply an approach belonging to the first class mentioned
above, consisting in sampling the LF in the regions of interest and in
constructing a training sample that effectively weighs the most
interesting regions.
As we clarify in Section \ref{sec:toy}, this procedure consists in building three samples: an unbiased sample, a biased sample and a mixed one. Training data will be extracted from the latter sample. Let us briefly describe the three:
\begin{itemize}
\item {\bf Unbiased sample}: a sample that has converged as accurately as possible to the true probability distribution. Notice that this sample is the only one which allows posterior inference in a Bayesian perspective, but would generally fail in making frequentist inference \cite{Feroz:2011bj}.
\item {\bf Biased sample}: a sample concentrated around the region of
  maximum likelihood. It is obtained by biasing the sampler in the
  region of large LF, only allowing for small moves around the
  maximum. Tuning this sample, targeted to a frequentist MLE
  perspective, raises the issue of coverage, that we discuss in
  Section \ref{sec:toy}. One has to keep in mind that the region of
  the LF that needs to be well known, i.e.\ around the maximum, is
  related to the coverage of the frequentist analysis being
  carried out.
\item {\bf Mixed sample}: this sample is built by enriching the unbiased sample with the biased one, in the region of large values of the LF. This is a tuning procedure, since, depending on the number of available samples and the statistics needed for training the DNN, this sample needs to be constructed for reproducing at best the results of the given analysis of interest both in a Bayesian and frequentist inference framework.
\end{itemize}
Some considerations are in order. The unbiased sample is enough if one
wants to produce a DNNLikelihood to be used only for Bayesian
inference. As we show later, this does not require a complicated
tuning of hyperparameters (at least in the example we consider) and
reaches very good performances, evaluated with the metrics that we
discussed above, already with a relatively small statistics in the
training sample (considering the high dimensionality). The situation
complicates a bit when one wants to be able to also make frequentist
inference using the same DNNLikelihood. In this case the mixed sample
(and therefore the biased one) is needed, and more tuning of the
network as well as more samples in the training set are required. In
order to optimise the predictions made by the DNNLikelihood close to
the maximum (or in other words to get an accurate estimate of the
$t_{\mu}$ test statistics), allowing for a reliable frequentist
inference, we average over a few models with the same architecture,
but trained with different randomly generated training sets, therefore
employing ensemble learning. This averaging can be done both at the
level of the predicted quantities, or at the level of the DNNs. As an
example we show the results obtained by averaging the values of the
$t_{\mu}$ test-statistics obtained with a few models trained with different
training sets. We have obtained very similar results also by stacking
these models together, training a small neural network to optimise
their combined prediction, and using this neural network to predict
$t_{\mu}$.

The final issue we have to address when training with the mixed
sample, which is biased by construction, is to ensure that the
DNNLikelihood can still produce accurate enough Bayesian posterior
estimates. This is actually guaranteed by the fact that a regression
(or interpolation) problem, contrary to a classification one, is
insensitive to the distribution in the target variable, since the
output is not conditioned on such probability distribution. This, as
can be clearly seen from the results presented in Section \ref{sec:toy}, is a crucial ingredient for our procedure to be useful, and leads to the main result of our approach: a DNNLikelihood trained with the mixed sample can be used to perform a new MCMC that converges to the underlying distribution, forgetting the biased nature of the training set.

In the next Section we give a thorough example of the procedure discussed here in the case of a prototype LHC-like search for NP corresponding to a $95$-dimensional LF.

\section{A realistic LHC-like NP search}\l{sec:toy}
In this section we
introduce
the prototype LHC-like NP search presented in
ref.~\cite{Buckley:2018vdr}, which we take as a representative example to
illustrate how to train the DNNLikelihood. We refer the reader to
ref.~\cite{Buckley:2018vdr} for a detailed discussion of this setup and
repeat here only the information that is strictly necessary to follow our analysis.

The toy experiment consists in a typical ``shape analysis'' in a given distribution aimed at extracting information on a possible NP signal from the Standard Model (SM) background. The measurement is divided in three different event categories, containing 30 bins each. The signal is characterized by a single ``signal-strength'' parameter $\mu$ and the uncertainty on the signal is neglected.\footnote{This approximation is done in ref.\cite{Buckley:2018vdr} to simplify the discussion, but it is not a necessary ingredient, neither there nor here.}
All uncertainties affecting the background are parametrised in terms of nuisance parameters, which may be divided into three categories:
\begin{enumerate}
\item fully uncorrelated uncertainties in each bin: they correspond to a nuisance parameter for each bin $\delta_{\text{MC},i}$, with uncorrelated priors, parametrising the uncertainty due to the limited Monte Carlo statistics, or statistics in a control region, used to estimate the number of background events in each bin.
\item fully correlated uncertainties in each bin: they correspond to a single nuisance parameter for each source of uncertainty affecting in a correlated way all bins in the distribution. In this toy experiment, such sources of uncertainty are the modeling of the Initial State Radiation and the Jet Energy Scale, parametrised respectively by the nuisance parameters $\delta_{\mathrm{ISR}}$ and $\delta_{\mathrm{JES}}$.
\item uncertainties on the overall normalisation (correlated among
  event categories): they correspond to the previous two nuisance parameters $\delta_{\mathrm{ISR}}$ and $\delta_{\mathrm{JES}}$, that, on top of affecting the shape, also affect the overall normalisation in the different categories, plus two typical experimental uncertainties, that only affect the normalisation, given by a veto efficiency and a scale-factor appearing in the simulation, parametrised respectively by $\delta_{\mathrm{LV}}$ and $\delta_{\mathrm{RC}}$.
\end{enumerate}
In summary, the LF depends on one physical parameter $\mu$ and $94$ nuisance parameters, that we collectively indicate with the vector $\bm{\delta}$, whose components are defined by $\delta_{i}=\delta_{\text{MC},i}$ for $i=1,\ldots,90$, $\delta_{91}=\delta_{\mathrm{ISR}}$, $\delta_{92}=\delta_{\mathrm{JES}}$, $\delta_{93}=\delta_{\mathrm{LV}}$, $\delta_{94}=\delta_{\mathrm{RC}}$.

The full model likelihood can be written as\footnote{There is a
  difference in the interpretation of this formula in the frequentist
  and Bayesian approaches: in a frequentist approach, the nuisance
  parameter distributions $\pi(\bm{\delta})$ do not constitute a
  prior, but should instead be considered as the likelihood of the
  nuisance parameters arising from other (auxiliary) measurements
  \cite{Tanabashi:2018oca}. In this perspective, since the product of
  two likelihoods is still a likelihood, the right hand side of
  eq.~\eqref{eq:lik} is the full likelihood. On the contrary, in a
  Bayesian perspective, the full likelihood is given by the product of
  probabilities in the right hand side of eq.~\eqref{eq:lik}, while
  the distributions $\pi(\bm{\delta})$ parametrise the {\it prior}
  knowledge of the nuisance parameters. Therefore, in this case,
  according to Bayes' theorem
  the right hand side of the equation should not be interpreted as the likelihood $\mathcal{P}(\text{data}|\text{pars})$, but as the full posterior probability $\mathcal{P}(\text{pars}|\text{data})$, up to a normalisation given by the Bayesian evidence $\mathcal{P}(\text{data})$. Despite this difference, in order to carry on a unified approach without complicating formul{\ae} too much, we abuse the notation and denote with $\mathcal{L}\(\mu,\bm{\delta}\)$ the frequentist likelihood and the Bayesian posterior distribution, since these are the two central objects from which frequentist and Bayesian inference are carried out, respectively.}
\be\l{eq:lik}
\mathcal{L}\(\mu,\bm{\delta}\)=\prod_{I=1}^{P}\mathrm{Pr}\(n_{I}^{\text{obs}}|n_{I}(\mu,\bm\delta)\)\pi\(\bm{\delta}\)\,,
\ee
where the product runs over all bins $I$.
The number of expected events in each bin
is given by $n_{I}(\mu,\bm\delta)=n_{s,I}(\mu)+n_{b,I}(\bm\delta)$, and the probability distributions are given by Poisson distributions in each bin
\be\l{eq:expobs}
\mathrm{Pr}\(n_{I}^{\text{obs}}|n_{I}\)=\f{(n_{I})^{n_{I}^{\text{obs}}}e^{-n_{I}}}{n_{I}^{\text{obs}}!}\,.
\ee
In this toy LF, the number of background events in each bin $n_{b,I}(\bm\delta)$ is known analytically as a function of the nuisance parameters, through various numerical parameters that interpolate the effect of systematic uncertainties. The parametrisation of  $n_{b,I}(\bm\delta)$ is such that the nuisance parameters $\bm\delta$ are normally distributed with vanishing vector mean and identity covariance matrix
\be\l{eq:prior}
\pi(\bm \delta)=\f{e^{-\f{1}{2}|\bm\delta|^{2}}}{(2\pi)^{\f{\dim(\bm\delta)}{2}}}\,.
\ee
Moreover, due to the interpolations involved in the parametrisation of the nuisance parameters, in order to ensure positive probabilities, the $\bm\delta$s are only allowed to take values in the range $[-5,5]$.

In our approach, we are interested in setting up a supervised
learning problem to learn the LF as a function of the
parameters. Independently of the statistical perspective, i.e.\ whether the parameters are treated as random variables or just variables, we need to choose some values to evaluate the LF. For the nuisance parameters the function $\pi(\bm \delta)$ already tells us how to choose these points, since it implicitly treats the nuisance parameters as random variables distributed according to this probability distribution. For the model parameters, in this case only $\mu$, we have to decide how to generate points, independently of the stochastic nature of the parameter itself. In the case of this toy example, since we expect $\mu$ to be relatively ``small'' and most probably positive, we generate $\mu$ values according to a uniform probability distribution in the interval $[-1,5]$. This could be considered as the prior on the stochastic variable $\mu$ in a Bayesian perspective, while it is just a scan in the parameter space of $\mu$ in the frequentist one.\footnote{Each different choice of $\mu$ corresponds, in the frequentist approach, to a different theoretical hypothesis. This raises the issue of generating pseudo-experiments for each different value of $\mu$, that we 
discuss further in Section \ref{sec:freq_inference} and Appendix \ref{app:coverage}.} Notice that we allow for small negative values of $\mu$.
Whenever the NP contribution comes from the on-shell production of some new physics, this assumption is not consistent. However, the ``signal" may come, in an Effective Field Theory (EFT) perspective, from the interference of the SM background with higher dimensional operators. This interference could be negative depending on the sign of the corresponding Wilson coefficient, and motivates our choice to allow for negative values of $\mu$ in our scan.
\begin{figure*}[t]
\begin{center}
\includegraphics[width=0.495\textwidth]{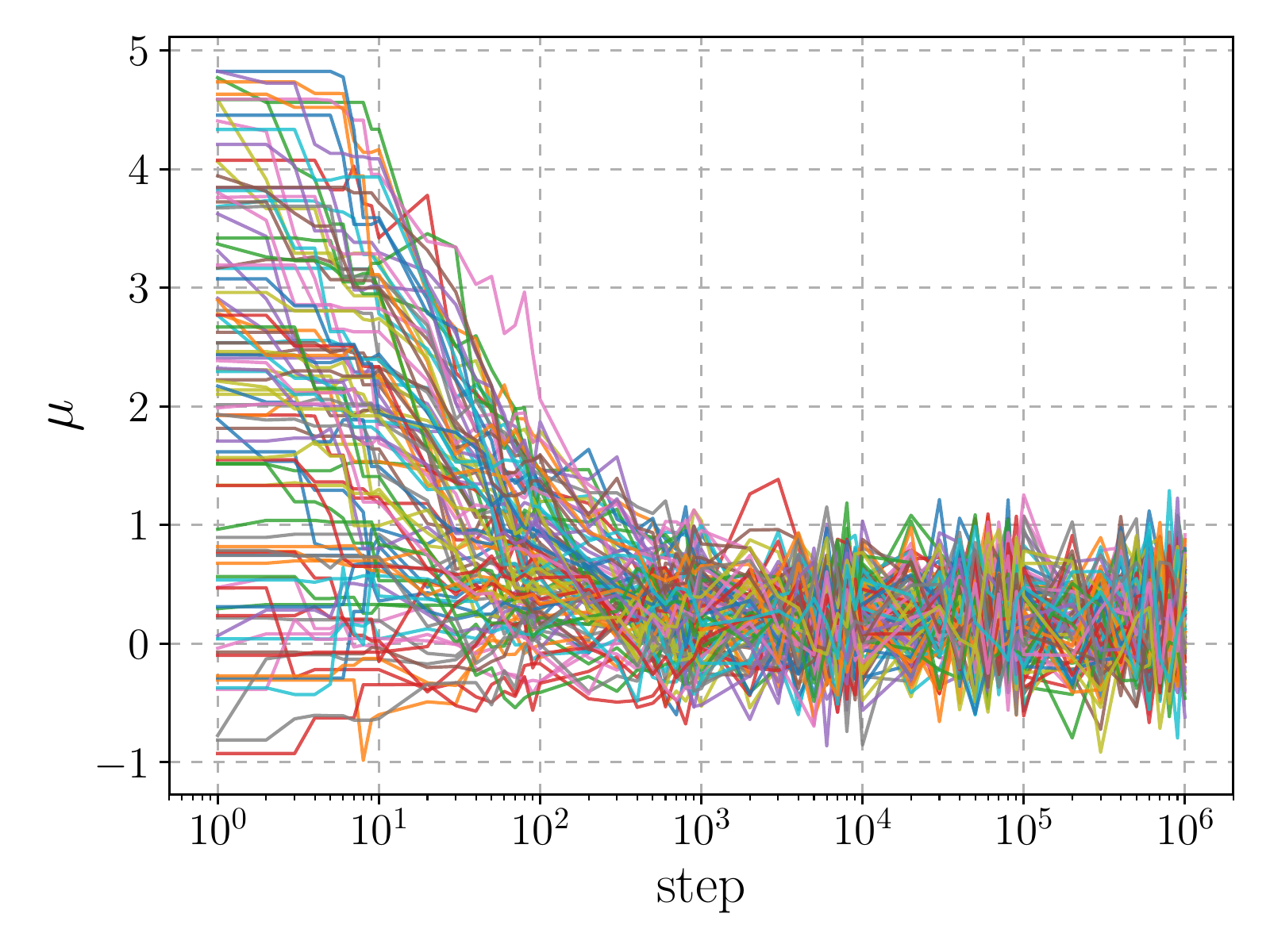}
\includegraphics[width=0.495\textwidth]{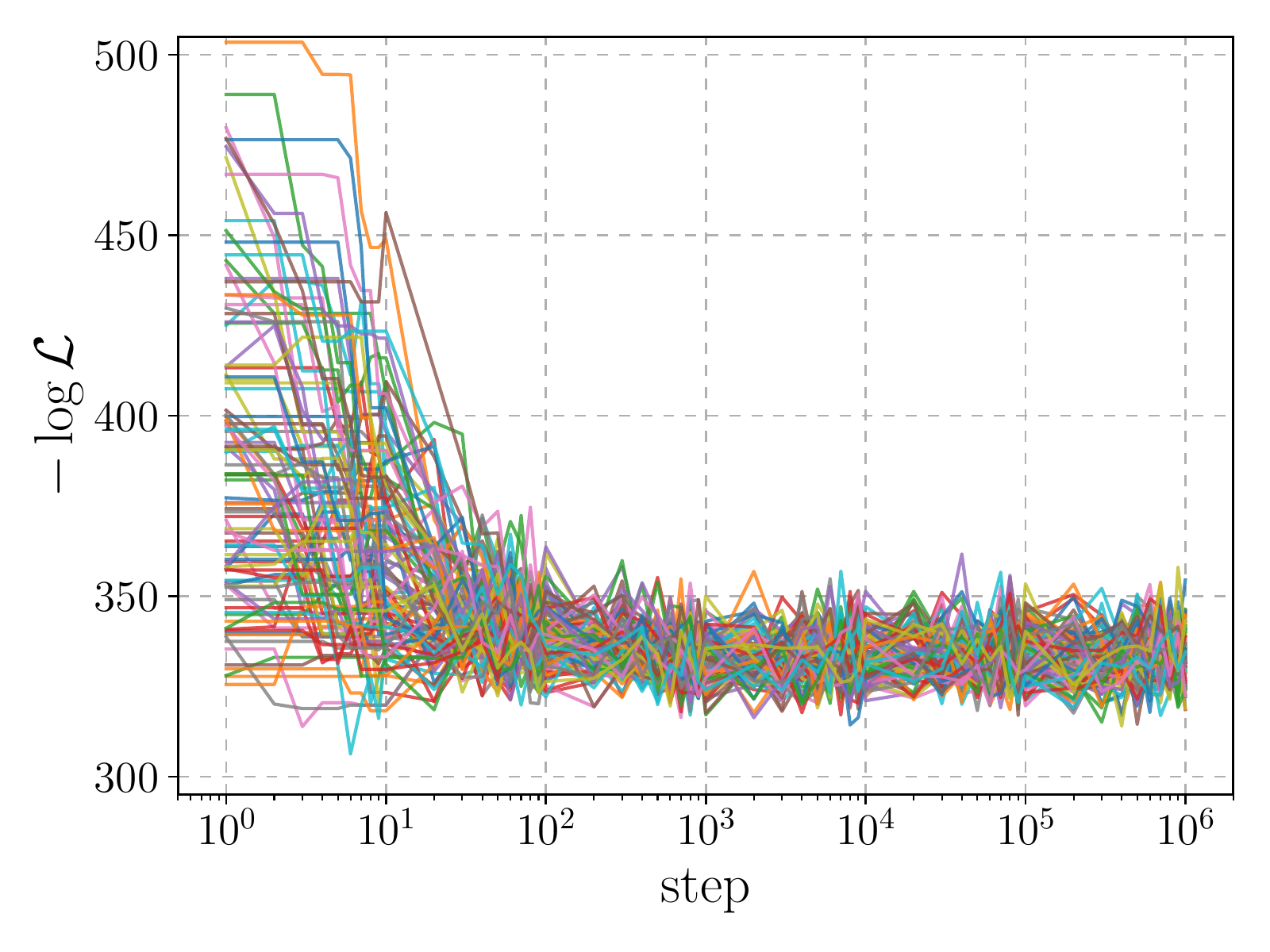}
\caption{Evolution of the chains in an {\sc emcee3} sampling of the LF in eq.~\eqref{eq:lik} with $10^{3}$ walkers and $10^{6}$ steps using the \textsc{StretchMove} algorithm with $a=1.3$. The plots show the explored values of the parameter $\mu$ (left) and of minus log-likelihood $-\log\mathcal{L}$ (right) versus the number of steps for a random subset of $10^{2}$ of the $10^{3}$ chains. For visualization purposes, values in the plots are computed only for numbers of steps included in the set $\{a\cdot 10^{b}\}$ with $a\in [1,9]$ and $b\in [0,6]$.}
\label{fig:mcmc_chains_sm}\end{center}
\end{figure*}

\subsection{Sampling the full likelihood}\l{sec:sampling}
To obtain the three samples discussed in Section \ref{sec:imbdata} from the full model LF in eq.~\eqref{eq:lik} we used the {\sc emcee3} Python package \cite{ForemanMackey:2012ig}, which implements the Affine Invariant (AI) MCMC Ensemble Sampler \cite{2010CAMCS...5...65G}. We proceeded as follows:
\begin{enumerate}
\item {\bf Unbiased sample $S_{1}$}\\
  In the first sampling, the values of the proposals have been updated using the default \textsc{StretchMove} algorithm implemented in {\sc emcee3}, which updates all values of the parameters (95 in our case) at a time. The default value of the only free parameter of this algorithm $a=2$ delivered a slightly too low acceptance rate $\epsilon\approx 0.12$. We have therefore set $a=1.3$, which delivers a better acceptance rate of about $0.36$. Walkers\footnote{Walkers are the analog of chains for ensemble sampling methods \cite{2010CAMCS...5...65G}. In the following, we
interchangeably use the words ``chains" and ``walkers" to refer to the same object.} have been initialised randomly according to the prior distribution of the parameters. The algorithm efficiently achieves convergence to the true target distribution, but, given the large dimensionality, hardly explores large values of the LF. 

In Figure \ref{fig:mcmc_chains_sm} we show the evolution of the walkers for the parameter $\mu$ (left) together with the corresponding values of $-\log\mathcal{L}$ (right) for an illustrative set of $100$ walkers. From these figures a reasonable convergence seems to arise already after roughly $10^{3}$ steps, which gives an empirical estimate of the autocorrelation of samples within each walker. 

\begin{figure*}[t]
\begin{center}
\includegraphics[width=0.325\textwidth]{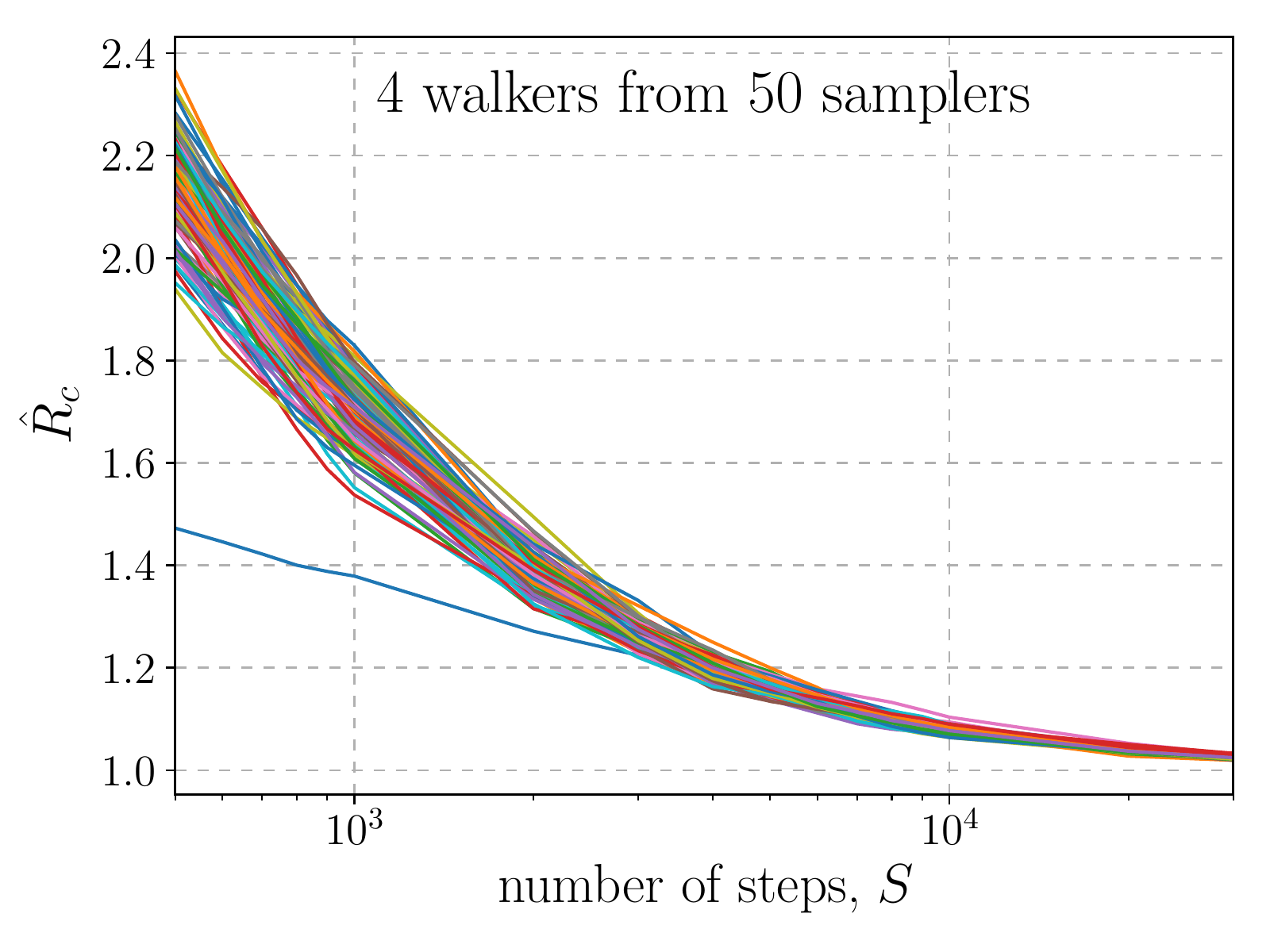}
\includegraphics[width=0.325\textwidth]{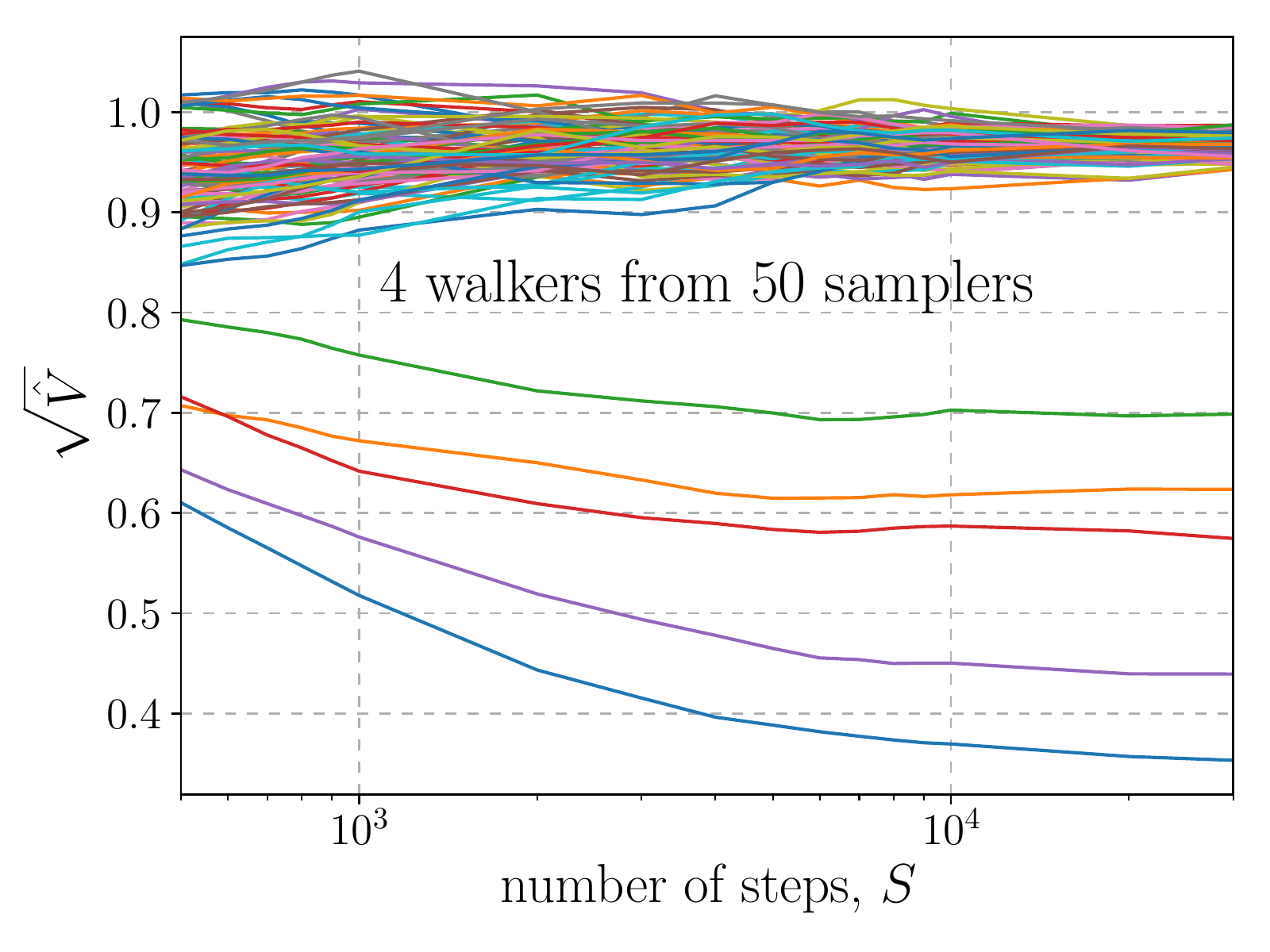}
\includegraphics[width=0.325\textwidth]{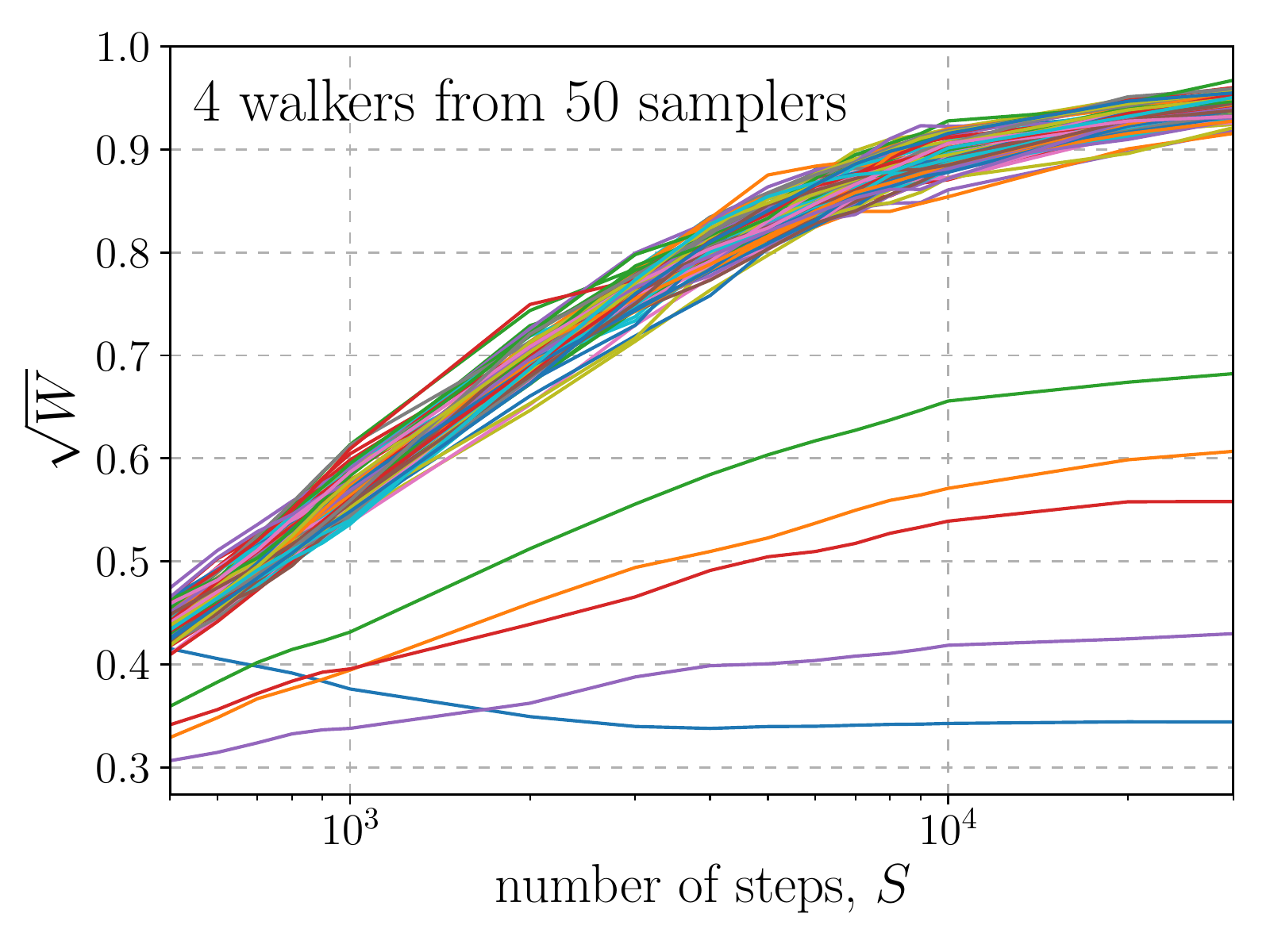}\\
\includegraphics[width=0.325\textwidth]{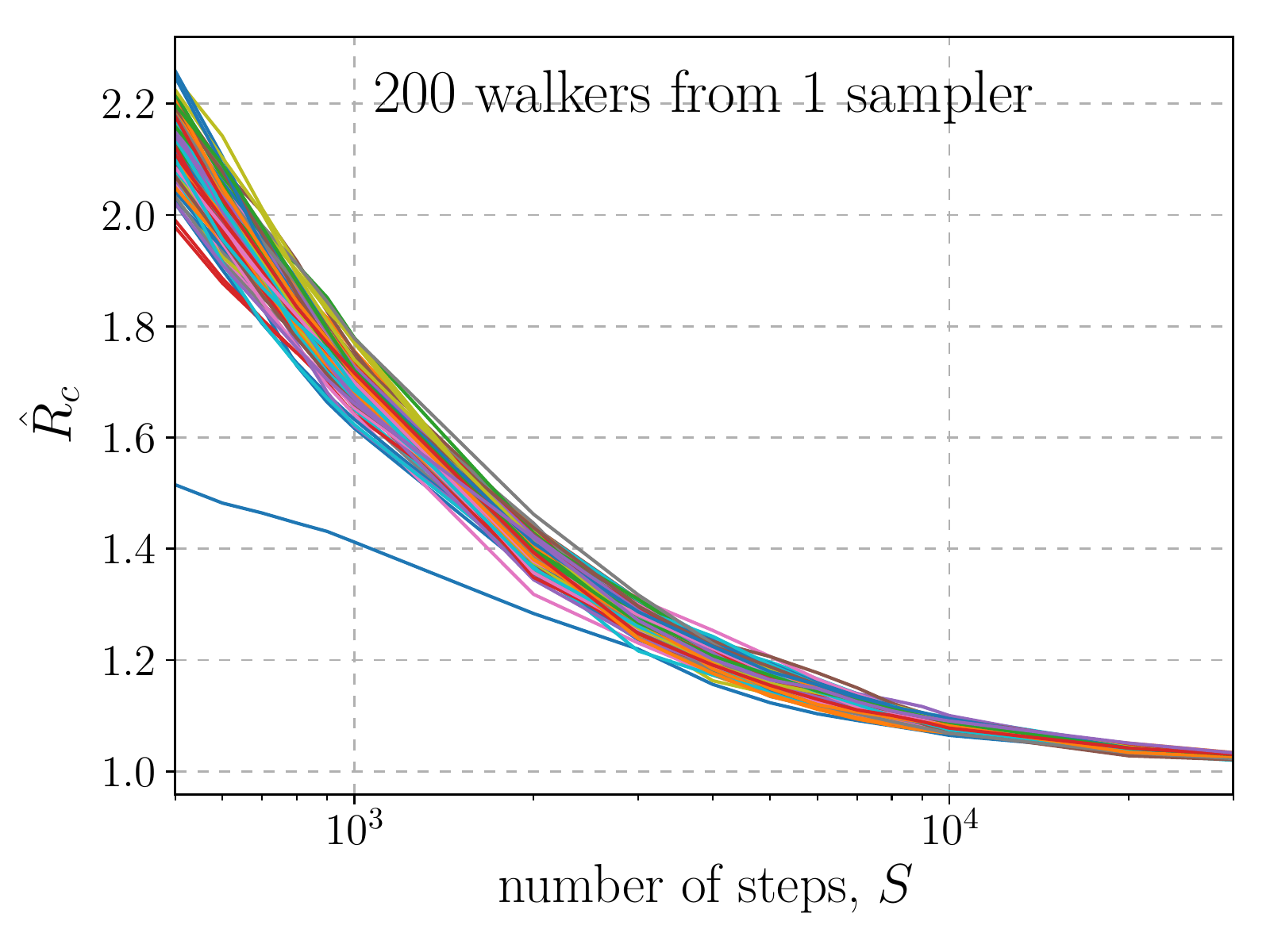}
\includegraphics[width=0.325\textwidth]{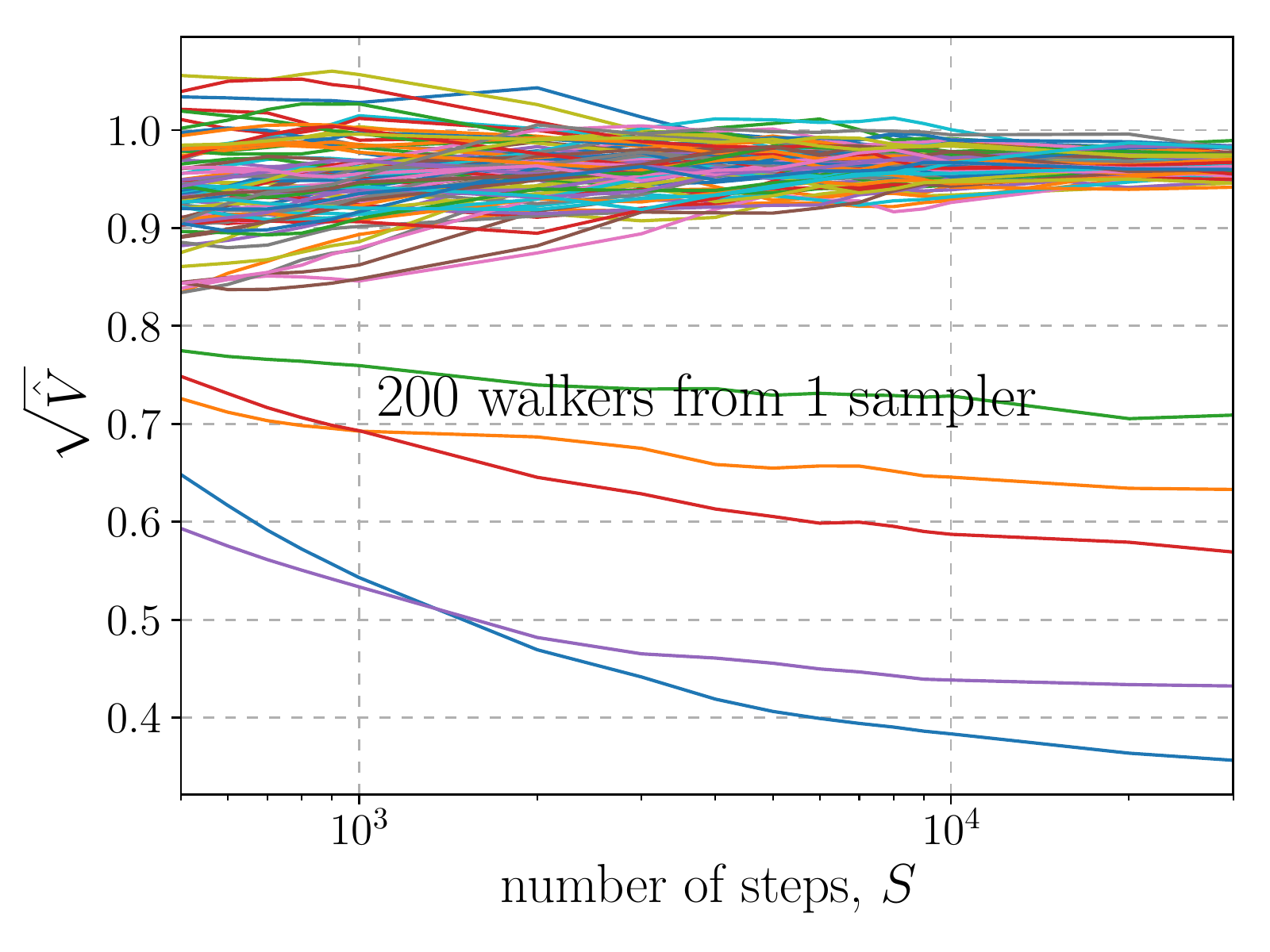}
\includegraphics[width=0.325\textwidth]{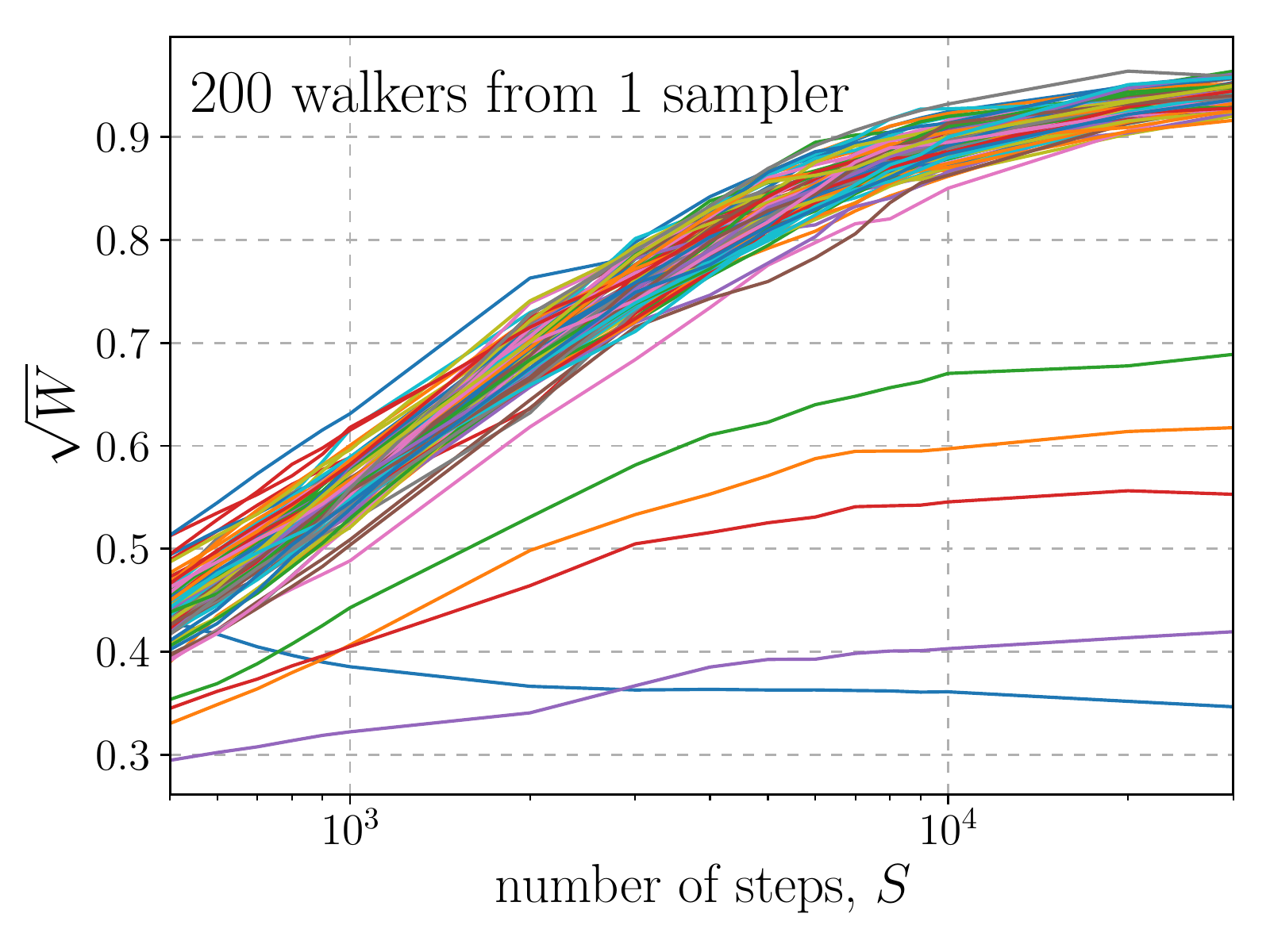}\\
\caption{{\bf Upper panel:} Gelman, Rubin and Brooks $\hat{R}_{c}$, $\sqrt{\hat{V}}$, and $\sqrt{W}$ parameters computed from an ensemble of $200$ walkers made by joining together $4$ samples extracted (randomly) from $50$ samplers of $200$ walkers each. {\bf Lower panel:} Same plots made using $200$ walkers from a single sampler.}
\label{fig:gelman_rubin}\end{center}
\end{figure*}
Notice that, in the case of ensemble sampling algorithms, the usual Gelman, Rubin and Brooks statistics,
usually denoted as $\hat{R}_{c}$ \cite{Gelman:1992zz,Brooks_1998}, is not
expected to be a robust tool to asses convergence, due to the
correlation between different walkers. However, in order to reduce
this correlation, one could consider a number of independent samplers,
extract a few walkers from each run, and compute $\hat{R}_{c}$ for
this set \cite{Huijser2015}. Considering the aforementioned empirical
estimate of the number of steps for convergence, i.e.\ roughly few
$10^{3}$, we have run 50 independent samplers for a larger number of
steps ($3\cdot 10^{4}$), extracted randomly 4 chains from each, joined
them together, and computed $\hat{R}_{c}$ for this set. This is shown
in the upper-left plot of Figure \ref{fig:gelman_rubin}. With a
requirement of $\hat{R}_{c}<1.2$ \cite{Brooks_1998} we see that chains
have already converged after around $5\cdot 10^{3}$ steps, which is
roughly what we empirically estimated looking at the chains evolution
in Figure \ref{fig:mcmc_chains_sm}. An even more robust requirement
for convergence is given by $\hat{R}_{c}<1.1$, together with
stabilized evolution of both variances $\hat{V}$ and $W$
\cite{Brooks_1998}. In the center and right plots of Figure
\ref{fig:gelman_rubin} we show this evolution, from which we see that
convergence has robustly occurred after $2-3\cdot 10^{4}$ steps. In
order to check the statement that the $\hat{R}_{c}$ metric cannot be
directly applied to ensemble sampling techniques, we have performed
the same analysis using $200$ walkers from a single sampler. The
result is shown in the lower panels of Figure
\ref{fig:gelman_rubin}. As it can be seen comparing the upper and
lower plots, we would have got to the exact same conclusions about
convergence, showing that the correlation of walkers does not, at least in this case, affect diagnostics of convergence based on the $\hat{R}_{c}$ measure.

\begin{figure*}[t]
\begin{center}
\includegraphics[width=0.495\textwidth]{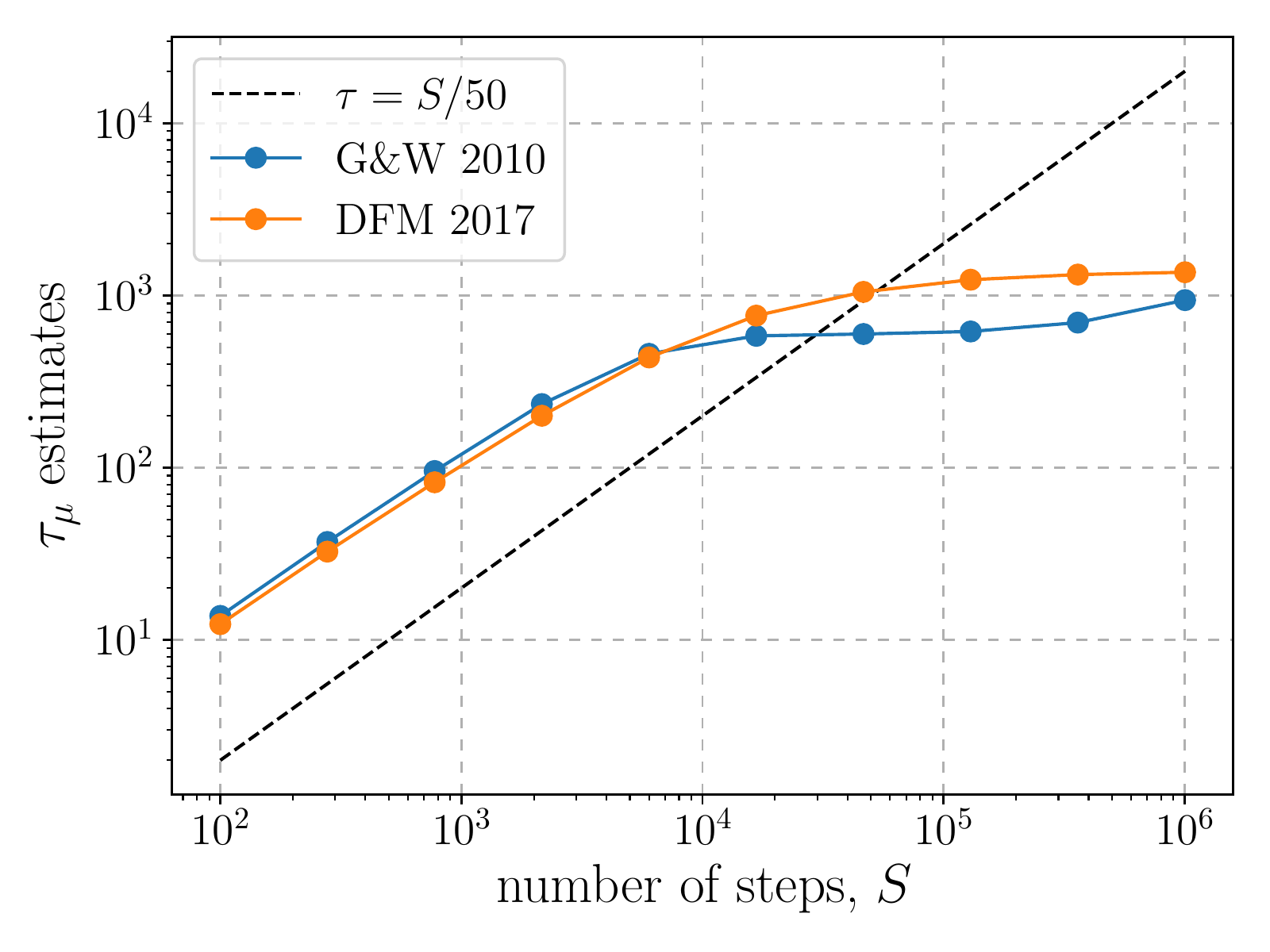}
\includegraphics[width=0.495\textwidth]{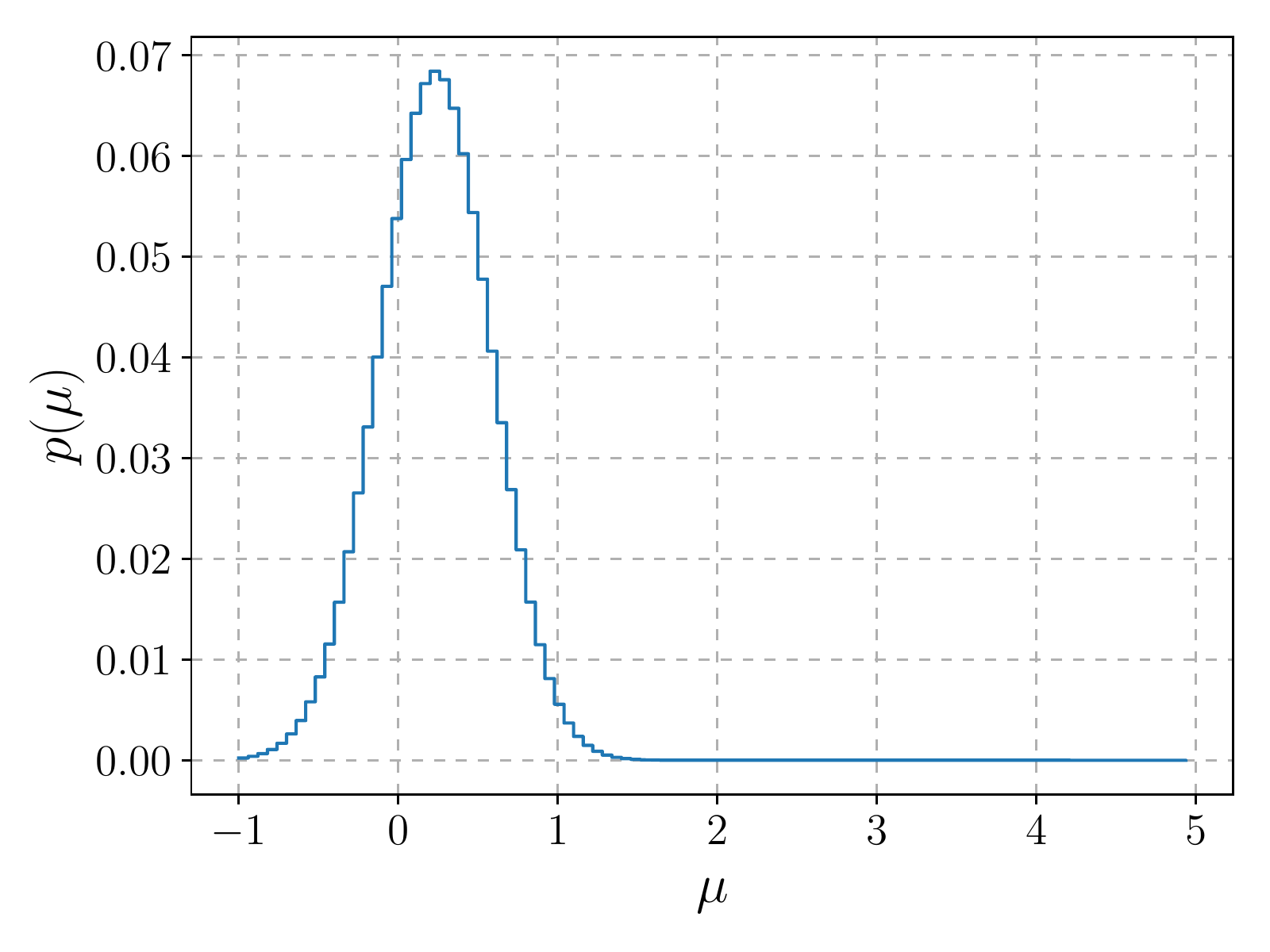}
\caption{Normalised histogram (right) and estimate of the autocorrelation time $\tau_{\mu}$ (estimated using both the original method proposed in ref.~\cite{2010CAMCS...5...65G} and the alternative one discussed by the {\sc emcee3} authors \cite{emcee3github,emcee3readthedocs}) as a function of the number of samples (left) for the parameter $\mu$.}
\label{fig:lik_sampling_emcee}\end{center}
\end{figure*}

An alternative and pretty general way to diagnose MCMC sampling is the
autocorrelation of chains, and in particular the Integrated
Autocorrelation Time (IAT). This quantity represents the average
number of steps between two independent samples in the chain. For
unimodal distributions, one can generally assume that after a few IAT
the chain forgot where it started and converged to generating samples
distributed according to the underlying target distribution. There are
more difficulties in the case of multimodal distributions, which are
however shared by most of the MCMC convergence diagnostics. We do not
enter here in such a discussion, and refer the reader to the overview
presented in ref.~\cite{Ford2015}. An exact calculation of the IAT for
large chains is computationally prohibitive, but there are several
algorithms to construct estimators of this quantity. The {\sc emcee3}
package comes with tools that implement some of these algorithms,
which we have used to study our sampling
\cite{emcee3github,emcee3readthedocs}. To obtain a reasonable estimate
of the IAT $\tau$, one needs enough samples, a reasonable empirical
estimate of which, that works well also in our case, is at least $50
\tau$. An illustration of this, for the parameter $\mu$, is given in
the left panel of Figure \ref{fig:lik_sampling_emcee}, where we show,
for a sampler with $10^{3}$ chains and $10^{6}$ steps, the IAT
estimated after different numbers of steps with two different
algorithms, ``G\&W 2010'' and ``DFM 2017'' (see
refs.~\cite{emcee3github,emcee3readthedocs} for details). It is clear
from the plot that the estimate becomes flat, and therefore converges
to the correct value of the IAT, roughly when the estimate curves
cross the empirical value of $50\tau$ (this is an order of magnitude
estimate, and obviously, the larger the number of steps, the better
the estimate of $\tau$). The best estimate that we get for this
sampling for the parameter $\mu$ is obtained with $10^{6}$ steps using
the ``DFM 2017'' method and gives $\overline{\tau}\approx 1366$,
confirming the order of magnitude estimate empirically extracted from
Figure \ref{fig:mcmc_chains_sm}. In the right panel of Figure
\ref{fig:lik_sampling_emcee} we show the resulting one-dimensional (1D) marginal posterior distribution of the parameter $\mu$ obtained from the corresponding run. 
Finally, we have checked that Figures \ref{fig:mcmc_chains_sm} and \ref{fig:lik_sampling_emcee} are quantitatively similar for all other parameters. 

As we mentioned above, the IAT gives an estimate of the number of steps between independent samples (it roughly corresponds to the period of oscillation, measured in number of steps, of the chain in the whole range of the parameter). Therefore, in order to have a true unbiased set of independent samples, one has to ``thin" the chain with a step size of roughly $\overline{\tau}$. This greatly decreases the statistics available from the MCMC run. Conceptually there is nothing wrong with having correlated samples, provided they are distributed according to the target distribution, however, even though this would increase the effective available statistics, it would generally affect the estimate of the uncertainties in the Bayesian inference \cite{Link2012,Owen2015}. We defer a careful study of the issue of thinning to a forthcoming publication \cite{toappear_thinning}, while here we limit ourselves to describe the procedure we followed to get a rich enough sample.

We have run {\sc emcee3} for $10^{6}+5\cdot 10^{3}$ steps with
$10^{3}$ walkers for $11$ times. From each run we have discarded a
pre-run of $5\cdot 10^{3}$ steps, which is a few times
$\overline{\tau}$, and thinned the chain with a step size of $10^{3}$,
i.e.\ roughly $\overline{\tau}$.\footnote{Even though the $\hat{R}_{c}$ analysis we performed suggests robust convergence after few $10^{4}$ steps, considering the length of the samplers we used ($10^{6}$ steps) and the large thinning value ($10^{3}$ steps), the difference between discarding a pre-run of $5\cdot 10^{3}$ versus a few $10^{4}$ steps is negligible. We have therefore set the burn-in number of steps to $5\cdot 10^{3}$ to slightly improve the efficiency of our MCMC generation.} Each run then delivered $10^{6}$ roughly independent samples. With parallelization, the sampler generates and stores about $22$ steps per second.\footnote{All samplings presented in the paper were produced with a SYS-7049A-T \sloppy{Supermicro\textsuperscript{\textregistered} workstation} configured as follows: Dual Intel\textsuperscript{\textregistered} Xeon\textsuperscript{\textregistered} Gold 6152 CPUs at $2.1$GHz (22 physical cores), 128 Gb of 2666 MHz Ram, Dual NVIDIA\textsuperscript{\textregistered} RTX 2080-Ti GPUs and 1.9Tb M.2 Samsung\textsuperscript{\textregistered} NVMe  PM963 Series SSD (MZ1LW1T9HMLS-00003). Notice that speed, in our case, was almost constant for a wide choice of the number of parallel processes in the range $\sim 30-88$, with CPU usage never above about $50\%$. We therefore conclude that generation speed was, in our case, limited by data transfer and not by CPU resources, making parallelization less than optimally efficient.\protect\label{foot:1}} The final sample obtained after all runs consists of $1.1\cdot 10^{7}$ samples. We stored $10^{6}$ of them as the test set to evaluate our DNN models, while the remaining $10^{7}$ are used to randomly draw the different training and validation sets used in the following.

\item {\bf Biased sample $S_{2}$}\\ 
The second sampling has been used to enrich the training and test sets
with points corresponding to large values of the LF, i.e.\ points close to the maximum for each fixed value of $\mu$. 
\begin{figure*}[t]
\begin{center}
\includegraphics[width=0.495\textwidth]{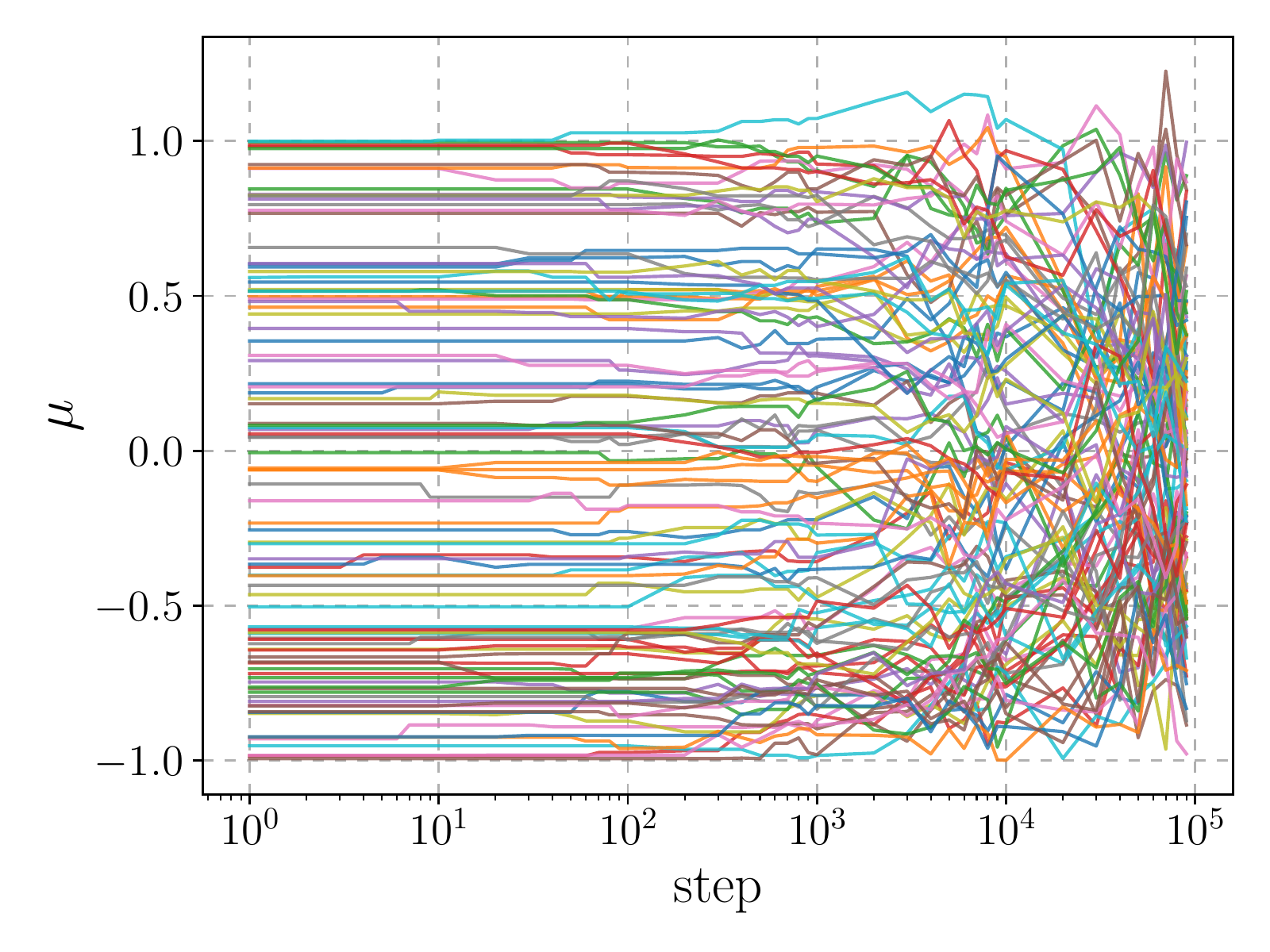}
\includegraphics[width=0.495\textwidth]{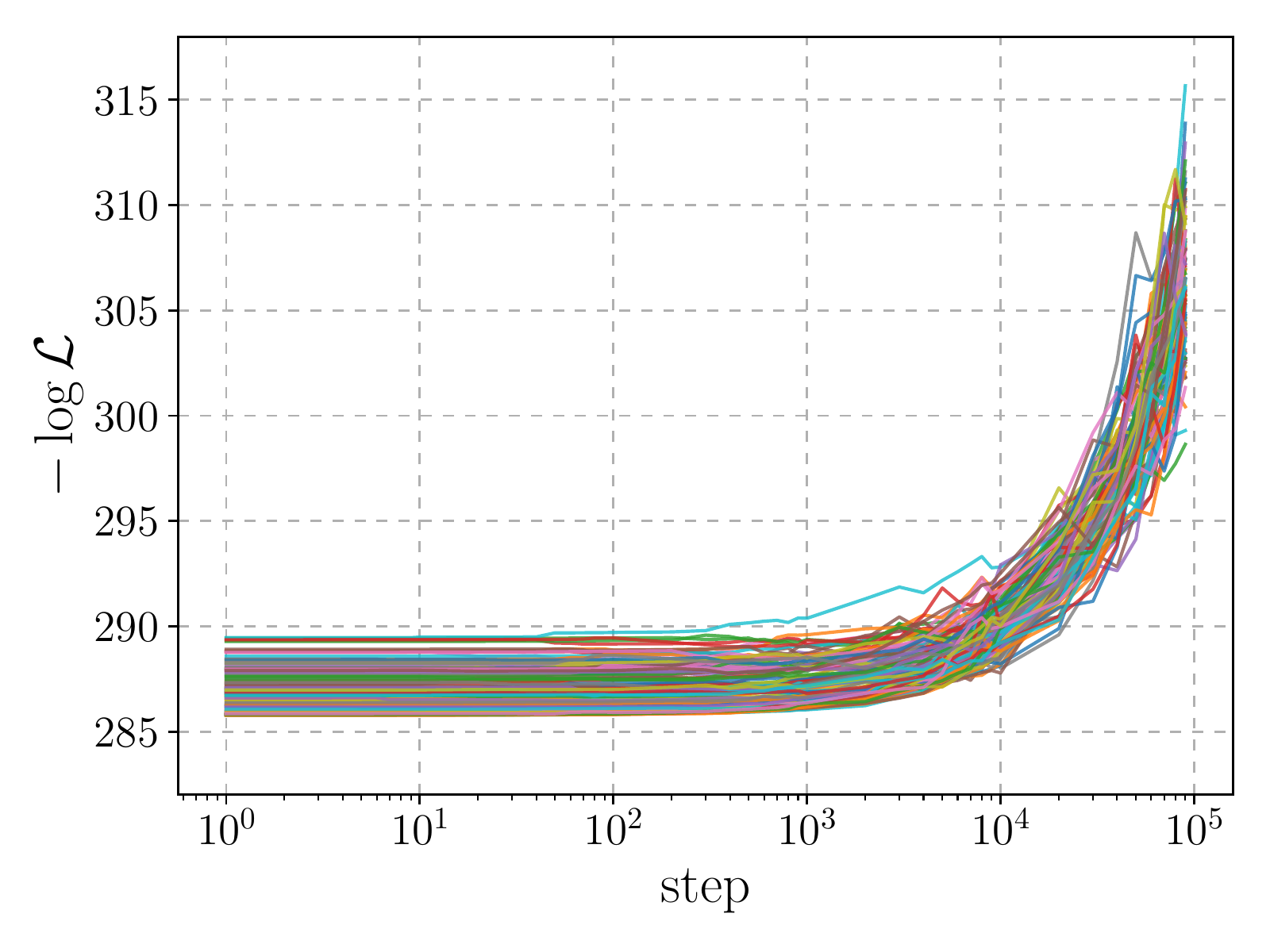}
\caption{Evolution of the chains in an {\sc emcee3} sampling of the LF in eq.~\eqref{eq:lik} with $200$ walkers and $10^{5}$ steps using the \textsc{GaussianMove} algorithm, that updated one parameter at a time, with a variance $5\cdot 10^{-4}$. The plots show the explored values of the parameter $\mu$ (left) and of minus log-likelihood $-\log\mathcal{L}$ (right) versus the number of steps for a random subset of $10^{2}$ of the $200$ chains. For visualization purposes, values in the plots are computed only for numbers of steps included in the set $\{a\cdot 10^{b}\}$ with $a\in [1,9]$ and $b\in [0,6]$.}
\label{fig:mcmc_chains_gm}\end{center}
\end{figure*}
In this case we initialised $200$ walkers in maxima of the LF
calculated for random values of $\mu$, extracted according to a
uniform probability distribution in the interval
$[-1,1]$.\footnote{This interval has been chosen smaller than the
  interval of $\mu$ considered in the unbiased sampling since values
  of $\mu$ outside this interval correspond to values of the LF much
  smaller than the global maximum, that are not relevant from the
  frequentist perspective. The range for the biased sampling can be
  chosen a posteriori by looking at the frequentist confidence
  intervals on $\mu$.} Moreover, the proposals have been updated using
a Gaussian random move with variance $5\cdot 10^{-4}$ (small moves) of
a single parameter at a time. In this way, the sampler starts
exploring the region of parameters corresponding to local maxima,
i.e.\ large values of the LF, and then slowly moves towards the tails. Once the LF gets further and further from the local maxima, the chains do not explore the region of local maxima anymore. Therefore, in this case we do not want to discard a pre-run, neither to check convergence, which implies that this sampling will have a strong bias (obviously, since we forced the sampler to explore only a particular region).

In Figure \ref{fig:mcmc_chains_gm} we show the evolution of the chains for the parameter $\mu$ (left panel) together with the corresponding values of $\log\mathcal{L}$ (right panel) for an illustrative (random) set of $100$ chains.
 Comparing Figure \ref{fig:mcmc_chains_gm} with Figure \ref{fig:mcmc_chains_sm}, we see that now the moves of each chain are much smaller and the sampler generates many points all around the local maxima at which the chains are initialised.

\begin{figure*}[t]
\begin{center}
\includegraphics[width=0.6\textwidth]{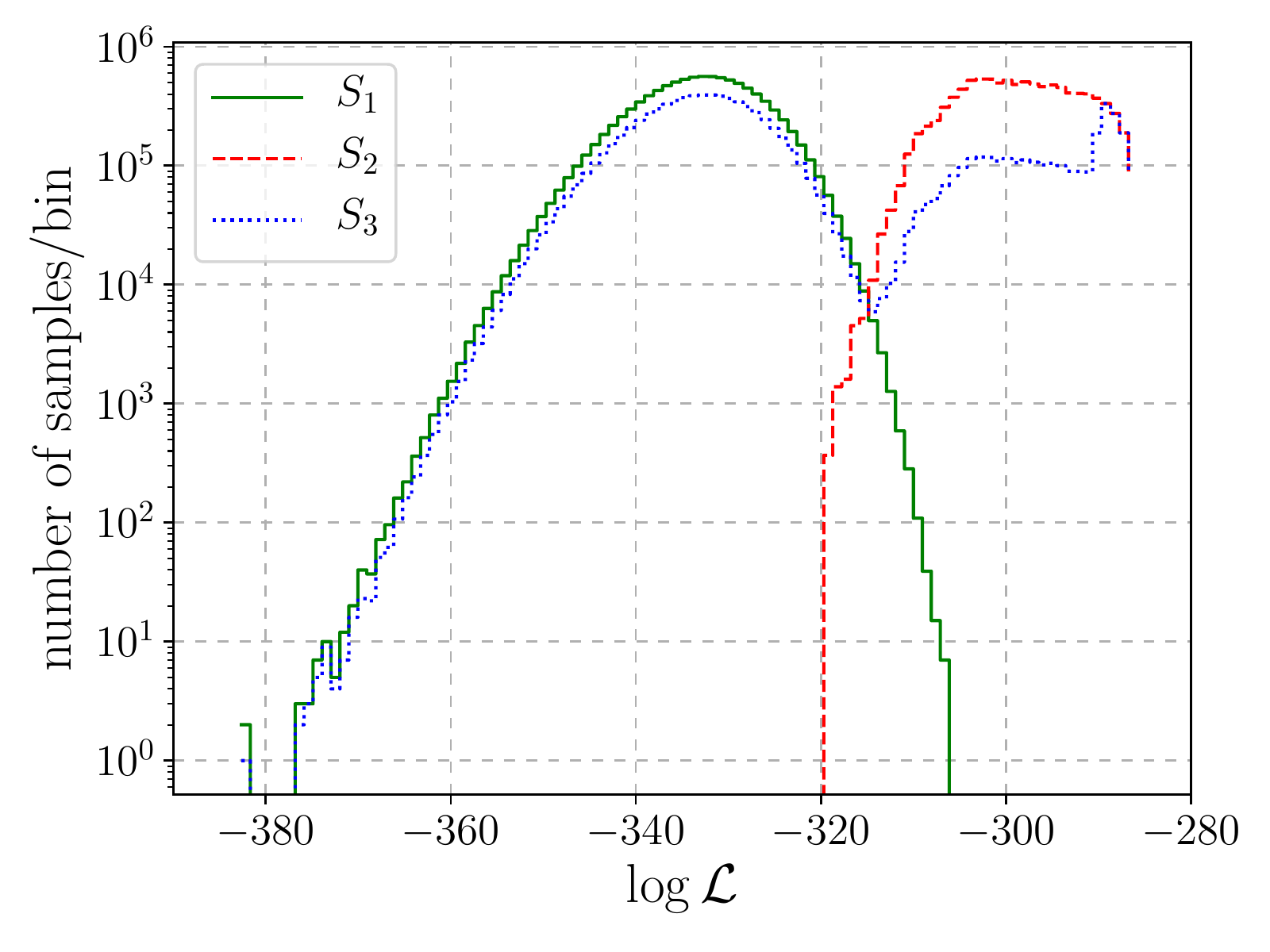}
\caption{Distribution of $\log\mathcal{L}$ values in $S_{1}$, $S_{2}$ and $S_{3}$. $S_{1}$ represents the unbiased sampling, $S_{2}$ is constructed close to the maximum of $\log\mathcal{L}$, and $S_{3}$ is obtained mixing the previous two as explained in the text.}
\label{fig:distr_toy_lik_sm_vs_gm_vs_mixed}
\end{center}
\end{figure*}

In order to ensure a rich enough sampling close to local maxima of the LF, we have made $10^{5}$ iterations for each walker. Since moves are much smaller than in the previous case (only one parameter is updated at a time), the efficiency in this case is very large, $\varepsilon \approx 1$.
We could therefore obtain a sampling of $1.1\cdot 10^{7}$ points by randomly picking points within the $10^{5}\cdot 200\cdot \varepsilon$ samples. As for $S_{1}$, two samples of $10^{6}$ and $10^{7}$ points have been stored separately: the first serves to build the test set, while the second is used to construct the training and validation sets. As mentioned before, this is a biased sample, and therefore should only be used to enrich the training sample to properly learn the LF close to the maximum (and to check results of the frequentist analysis), but it cannot be used to make any posterior inference. Due to the large efficiency, this sampling took less than one hour to be generated.
\item {\bf Mixed sample $S_{3}$}\\
The mixed sample $S_{3}$ is built from $S_{1}$ and $S_{2}$ in order to properly populate both the large probability mass region and the large log-likelihood region. Moreover, we do not want a strong discontinuity for intermediate values of the LF, which could become relevant, for instance, when combining with another analysis that prefers slightly different values of the parameters. For this reason, we have ensured that also intermediate values of the LF are represented, even though with a smaller effective weight, and that no more than a factor of 100 difference in density of examples is present in the whole region $-\log\mathcal{L}\in [285,350]$. Finally, in order to ensure a good enough statistics close to the maxima, we have enriched further the sample above $\log\mathcal{L}\approx -290$ (covering the region $\Delta\log\mathcal{L}\lesssim 5$). 

$S_{3}$ has been obtained taking all samples from $S_{2}$ with $\log\mathcal{L}> -290$ (around $10\%$ of all samples in $S_{2}$), $70\%$ of samples from $S_{1}$ (randomly distributed), and the remaining fraction, around $20\%$, from $S_{2}$ with $\log\mathcal{L}< -290$. With this procedure we obtained a total of $10^{7}$($10^{6}$) train(test) samples. We have checked that results do not depend strongly on the assumptions made to build $S_{3}$, provided enough examples are present in all the relevant regions in the training sample.

\end{enumerate}

The distribution of the LF values in the three samples are shown in Figure \ref{fig:distr_toy_lik_sm_vs_gm_vs_mixed} (for the $10^{7}$ points in the training/validation set).

We have used the three samples as follows: examples drawn from $S_{3}$ were used to train the full DNNLikelihood, while results have been checked against $S_{1}$ in the case of Bayesian posterior estimations and against $S_{2}$ (together with results obtained from a numerical maximisation of the analytical LF) in the case of frequentist inference. Moreover, we also present a ``Bayesian only" version of the DNNLikelihood, trained using only points from $S_{1}$.

\subsubsection{Bayesian inference}\l{sec:Bayes_inference}
In the Bayesian approach one is interested in marginal distributions, used to compute marginal posterior probabilities and credibility intervals.
For instance, in the case at hand, one may be interested in
two-dimensional (2D) marginal probability distributions in the parameter space $(\mu,\bm\delta)$, such as
\be
p(\mu,\delta_{i})=\int d\delta_{1}\cdots\int d\delta_{i-1}\int d\delta_{i+1}\cdots\int d\delta_{95} \mathcal{L}\(\mu,\bm{\delta}\)\pi\(\mu\)\,,
\ee
or in 1D HPDI corresponding to probabilities $1-\alpha$, such as
\be
1-\alpha = \int_{\mu_{\text{low}}}^{\mu_{\text{high}}}d\mu \int d\delta_{i}\, p(\mu,\delta_{i})\,.
\ee
All these integrals can be discretized and computed by just summing over quantities evaluated on a proper unbiased LF sampling.
\begin{figure*}[hp!]
\begin{center}
\includegraphics[width=0.93\textwidth]{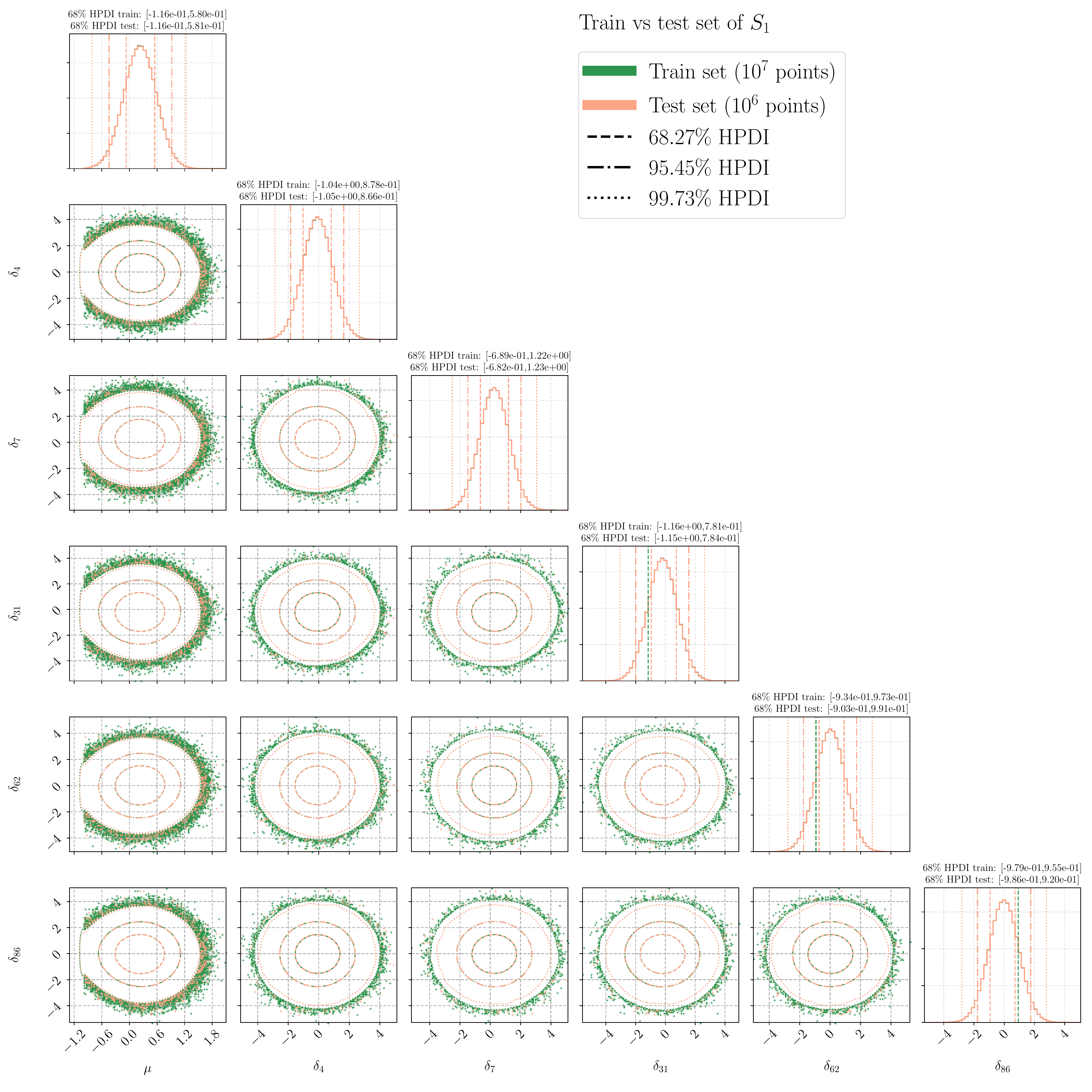}
\caption{1D and 2D posterior marginal probability distributions
  for a subset of parameters from the unbiased $S_{1}$. This gives a
  graphical representation of the sampling obtained through MCMC. The
  green (darker) and red (lighter) points and curves correspond to the
  training set ($10^{7}$ points) and test set ($10^{6}$ points) of
  $S_{1}$, respectively. Histograms are made with $50$ bins and
  normalised to unit integral. The dotted, dot-dashed, and dashed
  lines represent the $68.27\%,95.45\%,99.73\%$ 1D and 2D HPDI. The
  difference between green (darker) and red (lighter) lines gives an idea of the uncertainty on the HPDI due to finite sampling. Numbers for the $68.27\%$ HPDI for the parameters in the two samples are reported above the 1D plots.}
\label{fig:corner_toy_lik_params}\end{center}
\end{figure*}

This can be efficiently done with MCMC techniques, such as the one
described in Section \ref{sec:sampling}. For instance, using the
sample $S_{1}$ we can directly compute HPDIs for the parameters. Figure
\ref{fig:corner_toy_lik_params} shows the 1D and 2D posterior
marginal probability distributions of the subset of parameters
$(\mu, \delta_{4},\delta_{7},\delta_{31},\delta_{62},\delta_{86})$
obtained with the training set ($10^{7}$ points, green (darker)) and
test set ($10^{6}$ points, red (lighter)) of $S_{1}$.  Figure
\ref{fig:corner_toy_lik_params} also shows the 1D and 2D
$68.27\%,95.45\%,99.73\%$ HPDIs. All the HPDIs, including those shown
in Figure \ref{fig:corner_toy_lik_params}, have been computed by
binning the distribution with $50$ bins, estimating the interval, and
increasing the number of bins by $50$ until the interval splits due to
statistical fluctuations.

The results for $\mu$ with the assumptions $\mu>-1$ and $\mu>0$, estimated from the training set, which has the largest statistics, are given in Table \ref{tab:HPDI}. 
\begin{table}[t!]
\begin{center}
\begin{tabular}{c|c|c}
HPDI & $\mu>-1$ & $\mu>0$ \\
\hline
$68.27\%$  & $[-0.12, 0.58]$ & $0.48$ \\
$95.45\%$  & $[-0.47, 0.92]$ & $0.86$ \\
$99.73\%$  & $[-0.82, 1.26]$  & $1.22$
\end{tabular}
\end{center}\caption[]{HPDIs obtained using all $10^{7}$ samples from
  the training set of $S_{1}$. The result is shown both for $\mu>-1$
  and $\mu>0$ (only the upper bound is given in the latter case).}\l{tab:HPDI}
\end{table}
Figure \ref{fig:corner_toy_lik_params} shows how the 1D marginal probability distributions are extremely accurate up to the $99.73\%$, while the 2D ones, for the same interval, start showing differences, due to the finite sample size. 

Considering that the sample sizes used to train the DNN range from
$10^{5}$ to $5\cdot 10^{5}$, we do not consider probability intervals higher than $99.73\%$. Obviously, if one is interested in covering higher HPDIs, larger training sample sizes need to be considered (for instance, to cover a Gaussian $5\sigma$ interval, that corresponds to a probability $1-5.7\cdot 10^{-7}$, even only on the 1D marginal distributions, a sample with $\gg 10^{7}$ points would be necessary). We will not consider this case in the present paper.

\subsubsection{Frequentist inference}
\begin{figure*}[t]
\begin{center}
\includegraphics[width=0.325\textwidth]{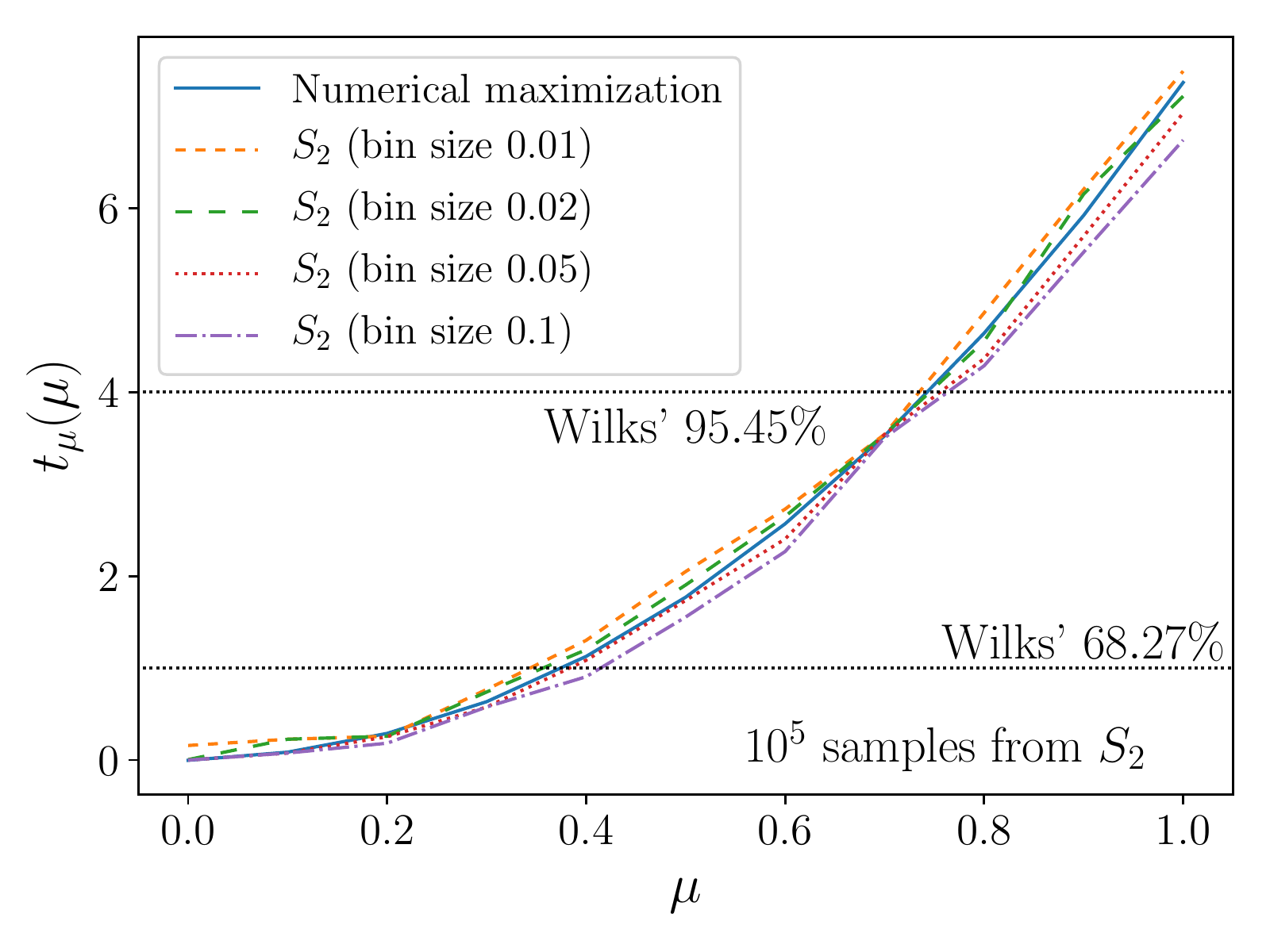}
\includegraphics[width=0.325\textwidth]{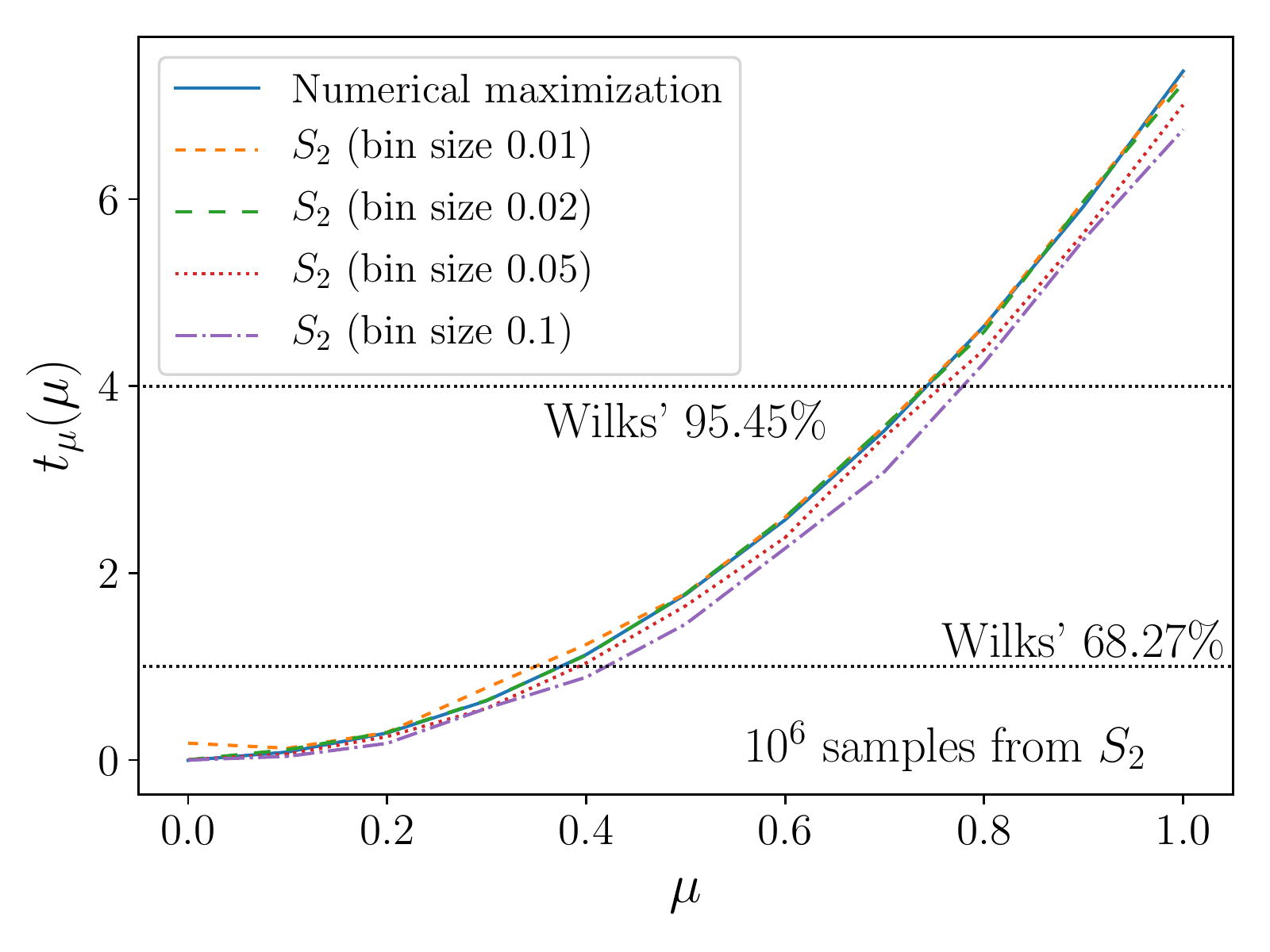}
\includegraphics[width=0.325\textwidth]{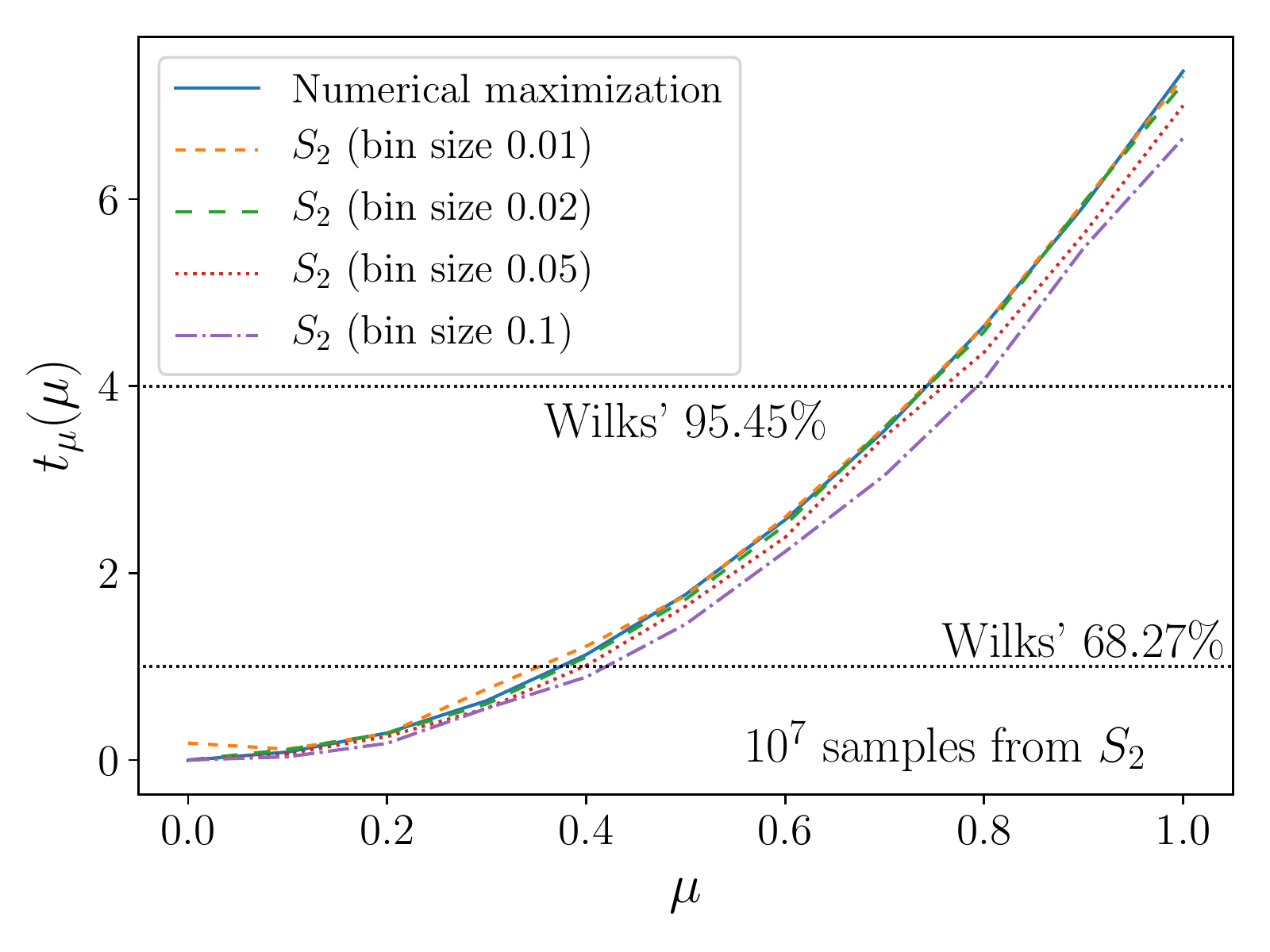}\\
\caption{Comparison of the $t_{\mu}$ test-statistics computed using numerical maximisation of eq.~\eqref{eq:tmu} and using a variable sample size from $S_{2}$. We show the result obtained searching from the maximum by usind different binning in $t_{\mu}$ with bin size $0.01,0.02,0.05,0.1$ around each value of $\mu$ (between $0$ and $1$ in steps of $0.1$).}
\label{fig:tmu_samp2}\end{center}
\end{figure*}

\l{sec:freq_inference}
In a frequentist inference one usually constructs a test statistics $\lambda(\mu,\bm\theta)$ based on the LF ratio
\be
\lambda(\mu,\bm\delta)=\frac{\mathcal{L}(\mu,\bm\delta)}{\mathcal{L}_{\text{max}}(\hat{\mu},\bm{\hat\delta})}\,.
\ee
Since one would like the test statistics to be independent of the nuisance parameters, it is common to use instead the profiled likelihood, obtained replacing the LF at each value of $\mu$ with its maximum value (over the nuisance parameters volume) for that value of $\mu$. One can then construct a test statistics $t_{\mu}$ based on the profiled (log)-likelihood ratio, given by
\be\l{eq:tmu}
t_{\mu}=-2\log\frac{\mathcal{L}_{\text{prof}}(\mu)}{\mathcal{L}_{\text{max}}}=-2\log\f{\sup_{\bm\delta}\mathcal{L}(\mu,\bm\delta)}{\sup_{\mu,\bm\delta}\mathcal{L}(\mu,\bm\delta)}=-2\(\sup_{\bm\delta}\log\mathcal{L}(\mu,\bm\delta)-\sup_{\mu,\bm\delta}\log\mathcal{L}(\mu,\bm\delta)\)\,.
\ee
Whenever suitable general conditions are satisfied, and in the limit of large data sample, by Wilks' theorem \cite{Wilks1938} the distribution of this test-statistics approaches a $\chi^{2}$ distribution that is independent of the nuisance parameters $\bm\delta$ and has a number of degrees of freedom equal to $\dim \mathcal{L}-\dim \mathcal{L}_{\text{prof}}$ \cite{Cowan:2010js}. In our case $t_{\mu}$ can be computed using numerical maximisation on the analytic LF, but it can also be computed from $S_{2}$ (and $S_{3}$, which is identical in the large likelihood region), which was constructed with the purpose of describing the LF as precisely as possible close to local maxima. In Figure \ref{fig:tmu_samp2} we show the result for $t_{\mu}$ using both approaches for different sample sizes drawn from $S_{2}$.
The three samples from $S_{2}$ used for the maximisation, with sizes
$10^{5}$, $10^{6}$, and $10^{7}$ (full training set of $S_{2}$),
contain in the region $\mu\in [0,1]$ around $5\cdot 10^{4}$, $5\cdot
10^{5}$, and $5\cdot 10^{6}$ points respectively, which results in increasing statistics in each bin and a more precise and stable prediction for $t_{\mu}$. As it can be seen $10^{5}$ points, about half of which contained in the range $\mu\in [0,1]$, are already sufficient, with a small bin size of $0.02$, to reproduce the $t_{\mu}$ curve with great accuracy. As expected, larger bin sizes result in too high local maxima estimates, leading to an underestimate of $t_{\mu}$. 

Under Wilks' theorem assumptions, $t_{\mu}$ should be distributed as a $\chi^{2}_{1}$ (1 d.o.f.) distribution, from which we can determine CL upper limits. The $68.27\%(95.45\%)$ CL upper limit (under the Wilks' hypotheses) is given by $t_{\mu}= 1(4)$, corresponding to $\mu< 0.37$($0.74$). These upper limits are compatible with the ones found in ref.~\cite{Buckley:2018vdr}, and are quite smaller than the corresponding upper limits of the HPDI obtained with the Bayesian analysis in Section \ref{sec:Bayes_inference} (see Table \ref{tab:HPDI}). This may suggest that the result obtained using the asymptotic approximation for $t_{\mu}$ is underestimating the upper limit (undercoverage). This is somehow expected for the search under consideration, given the large number of bins in which the observed number of events is below $5$ or even $3$ (see Figure 2 of ref.~\cite{Buckley:2018vdr}). Indeed, the true distribution of $t_{\mu}$ is expected to depart from a $\chi^{2}_{1}$ distribution when the hypotheses of Wilks' theorem are violated. The study of the distribution of $t_{\mu}$ is related to the problem of coverage of frequentist confidence intervals, and requires to perform pseudo-experiments. We present results on the distribution of $t_{\mu}$ obtained through pseudo-experiments in Appendix \ref{app:coverage}. The important conclusion is that using the distribution of $t_{\mu}$ generated with pseudo-experiments, CL upper limits become more conservative by up to almost a factor of two, depending on the choice of the approach used to treat nuisance parameters. This shows that the upper limits computed through asymptotic statistics undercover, in this case, the actual upper bounds on $\mu$.

\section{The DNNLikelihood}\label{sec:DNNLik}
The sampling of the full likelihood discussed above has been used to
train a DNN regressor constructed from multiple fully connected
layers, i.e.\ a multilayer perceptron (MLP). The regressor has been trained to predict values of the LF given a vector of inputs made by the physical and nuisance parameters. In order to introduce the main ingredients of our regression procedure and DNN training, we first show how models trained using only points from $S_{1}$ give reliable and robust results in the case of the Bayesian approach. Then we discuss the issue of training with samples from $S_{3}$ to allow for maximum likelihood based inference. Finally, once a satisfactory final model is obtained, we show again its performance for posterior Bayesian estimates.

\subsection{Model architecture and optimisation}\l{sed:DNNLik_architecture}
We used \textsc{Keras}~\cite{chollet2015keras} with \textsc{TensorFlow}~\cite{tensorflow} backend, through their \textsc{Python} implementation, to train a MLP and considered the following
hyperparameters to be optimised, the value of which defines what we call a {\it model} or a DNNLikelihood.
\begin{itemize}
\item {\bf Size of training sample}\\
In order to assess the performance of the DNNLikelihood given the
training set size we considered three different values: $10^{5}$,
$2\cdot 10^{5}$ and $5\cdot 10^{5}$. The training set
(together with an half sized evaluation set) has been randomly drawn
from $S_{1}$ for each model training, which ensures the absence of
correlation between the models due to the training data: thanks to the
large size of $S_{1}$ ($10^{7}$ samples) all the training sets can be
considered roughly independent. In order to allow for a consistent
comparison, all models trained with the same amount of training data
have been tested with a sample from the test set
of $S_{1}$, and with half the size of the training set. In general, and in particular in our interpolation problem, increasing the size of the training set allows to reduce the generalization error and
therefore to obtain the desired performance on the test set. 

\item {\bf Loss function}\\
In Section \ref{sec:regr} we have argued that both MAE and MSE are suitable loss functions to learn the log-likelihood function.
In our optimisation procedure we tried both, always finding (slightly) better results for the MSE.
We therefore choose the MSE as our loss function in all results presented here.

\item {\bf Number of hidden layers}\\
  From a preliminary optimisation we concluded that more than a single
  HL (deep network) always performs better than a single HL (shallow
  network). However, in the case under consideration, deeper networks
  do not seem to perform much better than $2$HL networks, even though
  they are typically much slower to train and to make
  predictions. Therefore, after this preliminary assessment, we
  focused on $2$HL architectures.
\item {\bf Number of nodes on hidden layers}\\
We considered architectures with the same number of nodes on the two hidden layers. The number of trainable parameters (weights) in the case of $n$ fully connected HLs with the same number of nodes $ d_{\text{HL}}$ is given by
\be
d_{\text{HL}}\(d_{\text{input}}+(n-1)d_{\text{HL}}+(n+1)\)+1\,,
\ee
where $d_{\text{input}}$ is the dimension of the input layer, i.e.\ the
number of independent variables, $95$ in our case. DNNs trained with
stochastic gradient methods tend to small generalization errors even
when the number of parameters is larger than the training sample size
\cite{Zhang2016}. Overfitting is not an issue in our interpolation
problem \cite{Belkin2018}. In our case we considered HLs not smaller than $500$ nodes,
which should ensure enough {\it bandwidth} throughout the network and
model {\it capacity}. In particular we compared results obtained with
$500$, $1000$, $2000$, and $5000$ nodes on each HL, corresponding to
$299001$, $1098001$, $4196001$, and $25490001$ trainable parameters.
\item {\bf Activation function on hidden layers}\\
We compared RELU~\cite{pmlr-v15-glorot11a}, ELU~\cite{2015arXiv151107289C},  and SELU~\cite{2017arXiv170602515K}
activation functions and the latter one showed to fit better our problem. In order to correctly implement the SELU activation in Keras
we initialised all weights using the \textsc{Keras} ``lecun\_normal'' initialiser
\cite{2017arXiv170602515K,LeCun:1998:EB:645754.668382}.
\item {\bf Batch size}\\
When using a stochastic gradient optimisation technique, of which
\textsc{Adam} is an example, the minibatch size is an
hyperparameter. For the training to be stochastic, the batch size
should be much smaller than the training set size, so that each
minibatch can be considered roughly independent. Large batch sizes
lead to more accurate weight updates and, due to the parallel
capabilities of GPUs, to faster training time. However, smaller batch
sizes usually contribute to regularize and avoid overfitting. After a
preliminary optimisation obtained changing the batch size from $256$
to $4096$, we concluded that the best performances were obtained by
keeping the number of batches roughly fixed to $200$ when changing the
training set size. In particular, choosing batch sizes among powers of
two, we have used $512$, $1024$ and $2048$ for $10^{5}$, $2\cdot10^{5}$ and $5\cdot 10^{5}$ training set sizes respectively. Notice that increasing the batch size when enlarging the training set, also allowed us to keep the initial learning rate fixed~\cite{Smith2017}. Similar results could be obtained by keeping a fixed batch size of $512$ and reducing the starting learning rate when enlarging the training set.
\item {\bf Optimiser}\\
We used the \textsc{Adam} optimiser with default parameters, and in particular
with learning rate $\epsilon=0.001$. We reduced the learning rate by a
factor $0.2$ every $40$ epochs without improvements on the validation
loss within an absolute amount (\textsc{min$\_$delta} in \textsc{Keras}) $1/N_{\text{points}}$, with $N_{\text{points}}$ the training set size. Indeed, being the \textsc{Keras} \textsc{min$\_$delta} parameter absolute and not relative to the value of the loss function, we needed to reduce it when getting smaller losses (better models). We have found that $1/N_{\text{points}}$ corresponded roughly to one to few permil of the best minimum validation loss obtained for all different traning set sizes. This value turned out to give the best results with reasonably low number of epochs (fast enough training). Finally, we performed early stopping \cite{Yao2007,raskutti2014early} using the same \textsc{min$\_$delta} parameter and no improvement in the validation loss for $50$ epochs. This ensured that training did not go on for too long without substantially improving the result. We also tested the newly proposed \textsc{AdaBound} optimiser~\cite{Luo2019AdaBound} without seeing, in our case, large differences.
\end{itemize}

Notice that the process of choosing and optimising a model depends on the LF under consideration (dimensions, number of modes, etc.) and this procedure should be repeated for different LFs. However, good initial points for the optimisation could be chosen using experience from previously constructed DNNLikelihoods.

As we discussed in Section \ref{sec:regr}, there are several metrics that we can use to evaluate our model. 
Based on the results obtained by re-sampling the DNNLikelihood with {\sc emcee3}, we see a strong correlation between the quality of the
re-sampled probability distribution (i.e.\ of the final Bayesian inference results) and the metric corresponding to the median of the K-S test on the 1D posterior marginal distributions.
We therefore present results focusing on this evaluation metric. When dealing with the Full DNNLikelihood trained with the biased sampling $S_{3}$ we also consider the performance on the mean relative error on the predicted $t_{\mu}$ test statistics when choosing the best models.

\subsection{The Bayesian DNNLikelihood}\l{sec:Bayes_DNNLik}
From a Bayesian perspective, the aim of the DNNLikelihood is to be
able, through a DNN interpolation of the full LF, to generate a
sampling analog to $S_{1}$, which allows to produce Bayesian posterior
density distributions as close as possible to the ones obtained using
the true LF, i.e.\ the $S_{1}$ sampling. Moreover, independently on how complicated to evaluate the original LF is, the DNNLikelihood is extremely fast to compute, allowing for very fast sampling.\footnote{In this case the original likelihood is extremely fast to evaluate either, since it is known in analytical form. This is usually not the case in actual experimental searches involving theory and detector simulations.} The {\sc emcee3} MCMC package allows, through vectorization of the input function for the log-probability, to profit of parallel GPU predictions, which made sampling of the DNNLikelihood roughly as fast as the original analytic LF.

We start by considering training using samples drawn from the unbiased $S_{1}$. The independent variables all vary in a reasonably small interval around zero and do not need any preprocessing. However, the $\log\mathcal{L}$ values in $S_{1}$ span a range between around $-380$ and $-285$. This is both pretty large and far from zero for the training to be optimal. For this reason we have pre-processed data scaling them to zero mean and unit variance. Obviously, when predicting values of $\log\mathcal{L}$ we applied the inverse function to the DNN output.

We rank models trained during our optimisation procedure by the median $p$-value of 1D K-S test on all coordinates between the test set and the prediction performed on the validation set. The best models are those with the highest median $p$-value.
In Table \ref{tab:results_DNNLik_Bayes} we show results for the best model we obtained for each training sample size. Results have been obtained by training $5$ identical models and taking the best one. We call these four best models $B_{1}-B_{3}$ ($B$ stands for Bayesian). All three models have two HLs with $5\cdot 10^{3}$ nodes each, and are therefore the largest we consider in terms of number of parameters. However, it should be clear that the gap with smaller models is extremely small in some cases with some of the models with less parameters in the ensemble of $5$ performing better than some others with more parameters. This also suggests that results are not too sensitive to model dimension, making the DNNLikelihood pretty robust.
\begin{table}[ht!]
\begin{center}
\begin{tabular}{l|c|c|c}
Name											& $B_{1}$			& $B_{2}$			& $B_{3}$	\\
\hline\hline
Sample size ($\times 10^{5}$) 				& $1$					& $2$					& $5$				\\
Epochs											& $178$				& $268$				& $363$		\\
Loss train (MSE) ($\times 10^{-3}$)		& $0.14$				& $0.088$				& $0.054$				\\
Loss val (MSE) ($\times 10^{-3}$)			& $10.11$				& $6.66$				& $3.9$			\\
Loss test (MSE) ($\times 10^{-3}$)			& $10.02$				& $6.64$				& $3.9$				\\
ME train ($\times 10^{-3}$)					& $0.47$				& $0.53$				& $0.28$		\\
ME val	($\times 10^{-3}$)					& $5.44$				& $2.58$				& $1.76$		\\
ME test ($\times 10^{-3}$)					& $4.91$				& $2.31$				& $1.72$		\\
Median $p$-value of 1D K-S test vs pred. on train				& $0.41$				& $0.46$				& $0.39$\\
Median $p$-value of 1D K-S test vs. pred. on val.				& $0.24$				& $0.33$				& $0.43$\\
Median $p$-value of 1D K-S val vs. pred. on test				& $0.24$				& $0.40$				& $0.34$\\		
Training time (s)								& $1007$				& $2341$				& $8446$\\
Prediction time ($\mu\text{s}/\text{point}$)& $11.5$				& $10.4$				& $14.5$			\\
\end{tabular}
\end{center}\caption[]{Results for the best models (Bayesian
  DNNLikelihood) for different training sample size. All models have
  been trained for $5$ times to check the stability of the result and
  the best performing one has been quoted. Prediction time is evaluated on a test set with half the size of the training set using the same batch size used in training, and evaluating on a Nvidia Tesla V100 GPU with 32GB of RAM. All best models have $d_{\text{HL}}=5\cdot 10^{3}$.}\l{tab:results_DNNLik_Bayes}
\end{table}

\begin{figure*}[t!]
\begin{center}
\includegraphics[width=0.325\textwidth]{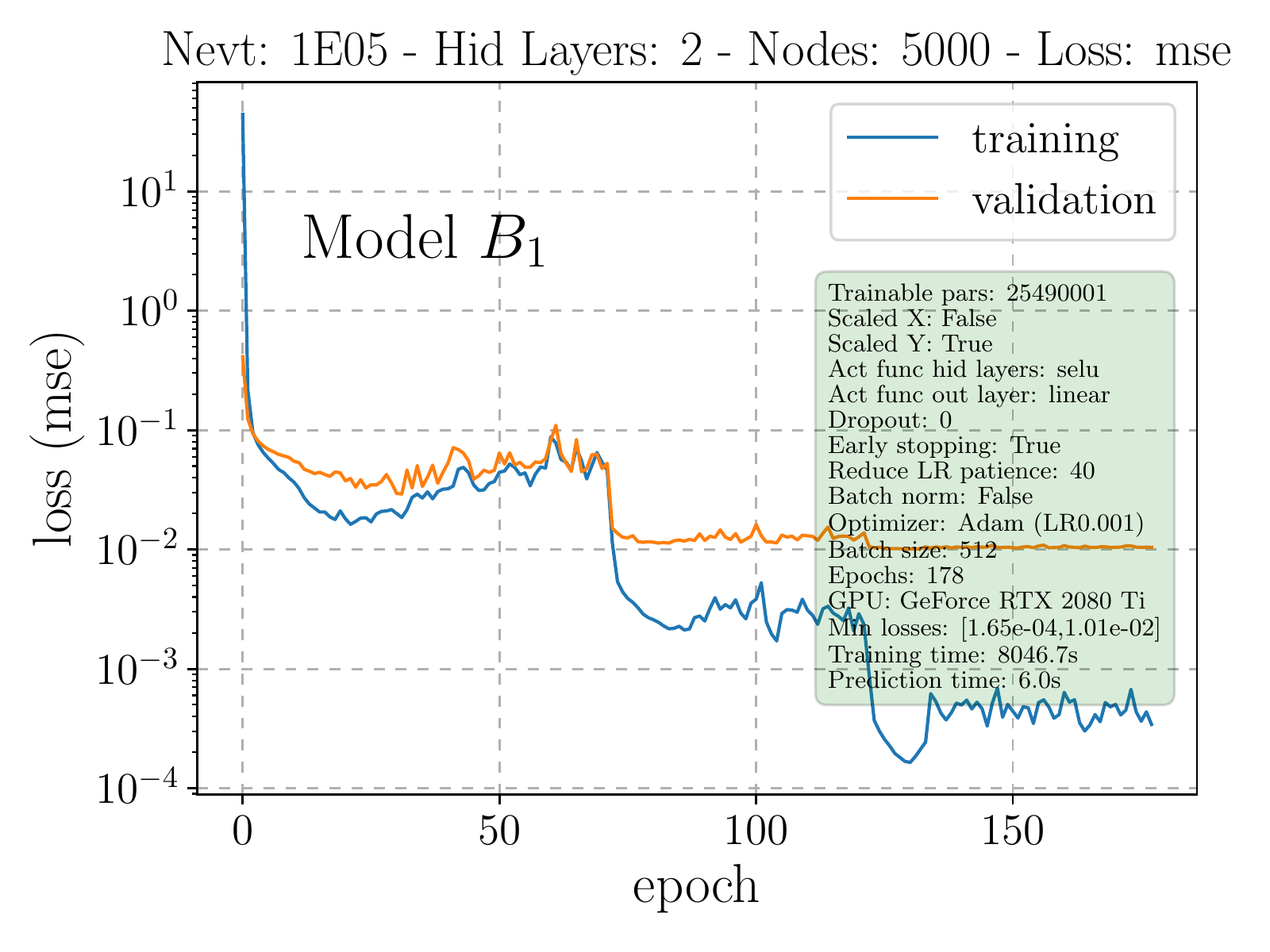}
\includegraphics[width=0.325\textwidth]{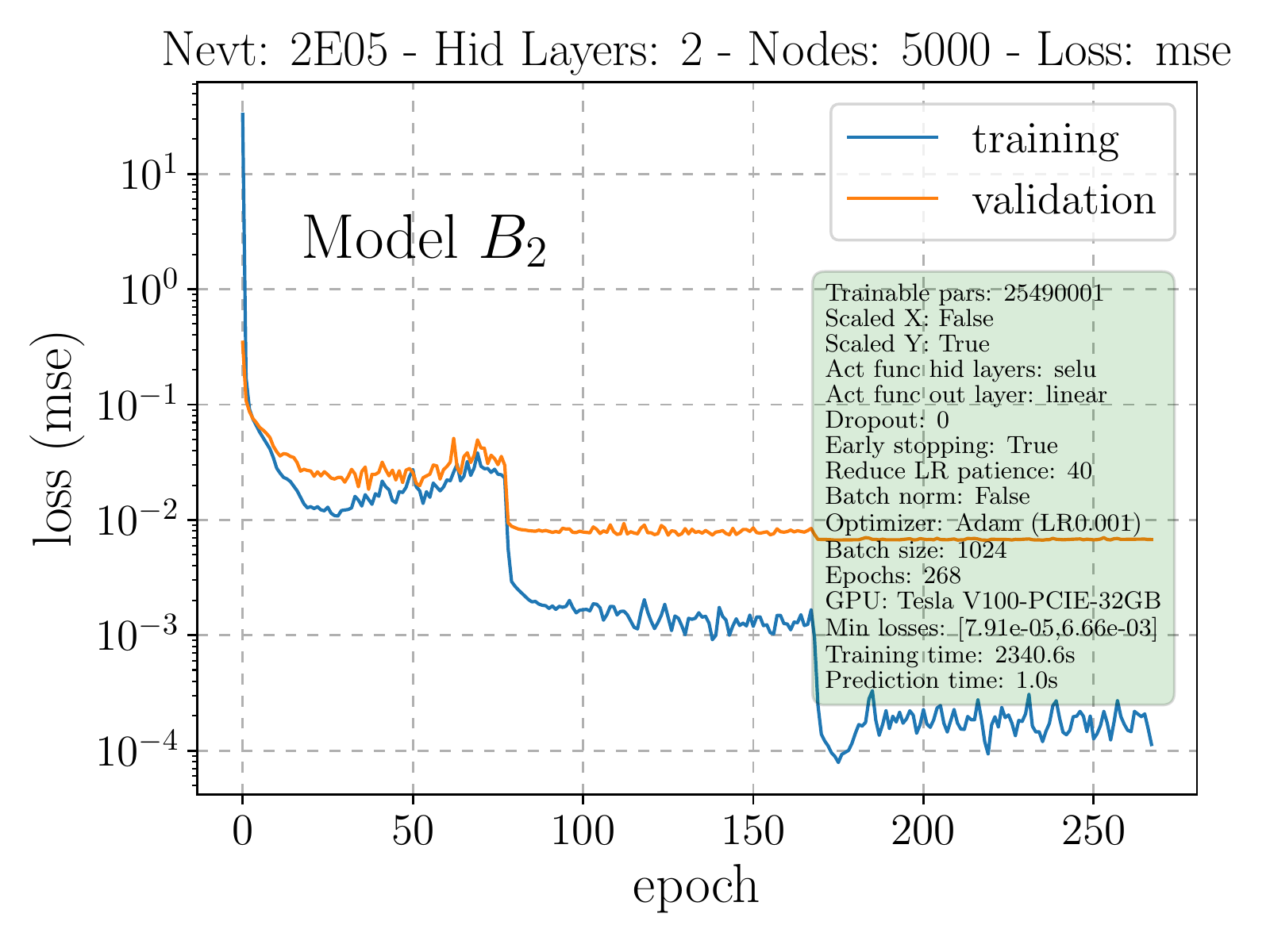}
\includegraphics[width=0.325\textwidth]{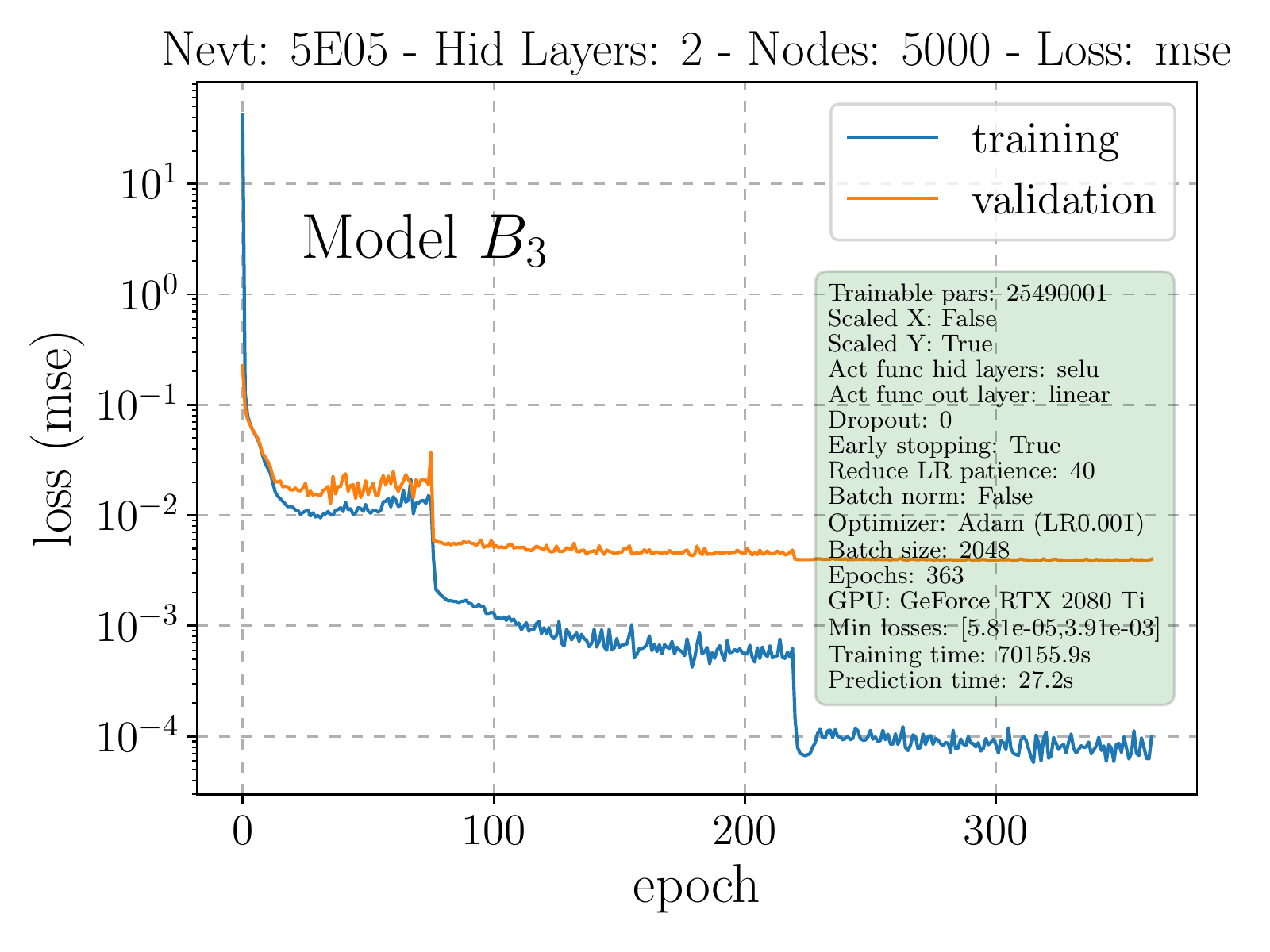}
\caption{Training and validation loss (MSE) vs number of training epochs for models $B_{1}-B_{3}$. The jumps correspond to points of reduction of the \textsc{Adam} optimiser learning rate.}
\label{fig:toy_training_results_Bayes}\end{center}
\end{figure*}

Figure \ref{fig:toy_training_results_Bayes} shows the learning curves obtained for the values of the hyperparameters shown in the legends. Early stopping is usually triggered after a few hundred epochs (ranging from around $200$ to $500$, with the best models around $200-300$) and values of the validation loss (MSE) that range in the interval $\approx [0.01, 0.003]$. Values of the validation ME, which, as explained in Section \ref{sec:regr} correspond to the K-L divergence for the LF, range in the $\approx [1, 5]\cdot 10^{-3}$, which, together with median of the $p$-value of the 1D K-S tests in the range $0.2-0.4$ deliver very accurate models. Training times are not prohibitive, and range from less than one hour to a few hours for the models we considered on a Nvidia Tesla V100 GPU with 32GB of RAM. Prediction times, using the same batch sizes used during training, are in the ballpark of $10-15\mu\text{s}/\text{point}$, allowing for very fast sampling and inference using the DNNLikelihood.
Finally, as shown in Table \ref{tab:results_DNNLik_Bayes}, all models
present very good generalization when going from the evaluation to the
test set, with the generalization error decreasing with the sample
size as expected.

In order to get a full quantitative assessment of the performances of
the Bayesian DNNLikelihood, we compared the results of a Bayesian
analysis performed using the test set of $S_{1}$ and each of the
models $B_{1}-B_{3}$. This was done in two ways. Since the model is
usually a very good fit of the LF, we reweighted each point in $S_{1}$
using the ratio between the original likelihood and the DNNLikelihood
(reweighting). This procedure is so fast that can be done for each
trained model during the optimisation procedure giving better insights
on the choice of hyperparameters. Once the best model has been chosen,
the result of reweighting has been checked by directly sampling the
DNNLikelihoods with {\sc emcee3}.\footnote{Sampling has been done on
  the same hardware configuration mentioned in Footnote
  \ref{foot:1}. However, in this case log-probabilities have been
  computed in parallel on GPUs (using the "vectorize" option of
  \textsc{emcee3}).} We present results obtained by sampling the
DNNLikelihoods in the form of 1D and 2D marginal posterior density
plots on a chosen set of parameters $(\mu,
\delta_{10},\delta_{40},\delta_{70},\delta_{95})$. 

\begin{figure*}[htbp!]
\begin{center}
\includegraphics[width=0.93\textwidth]{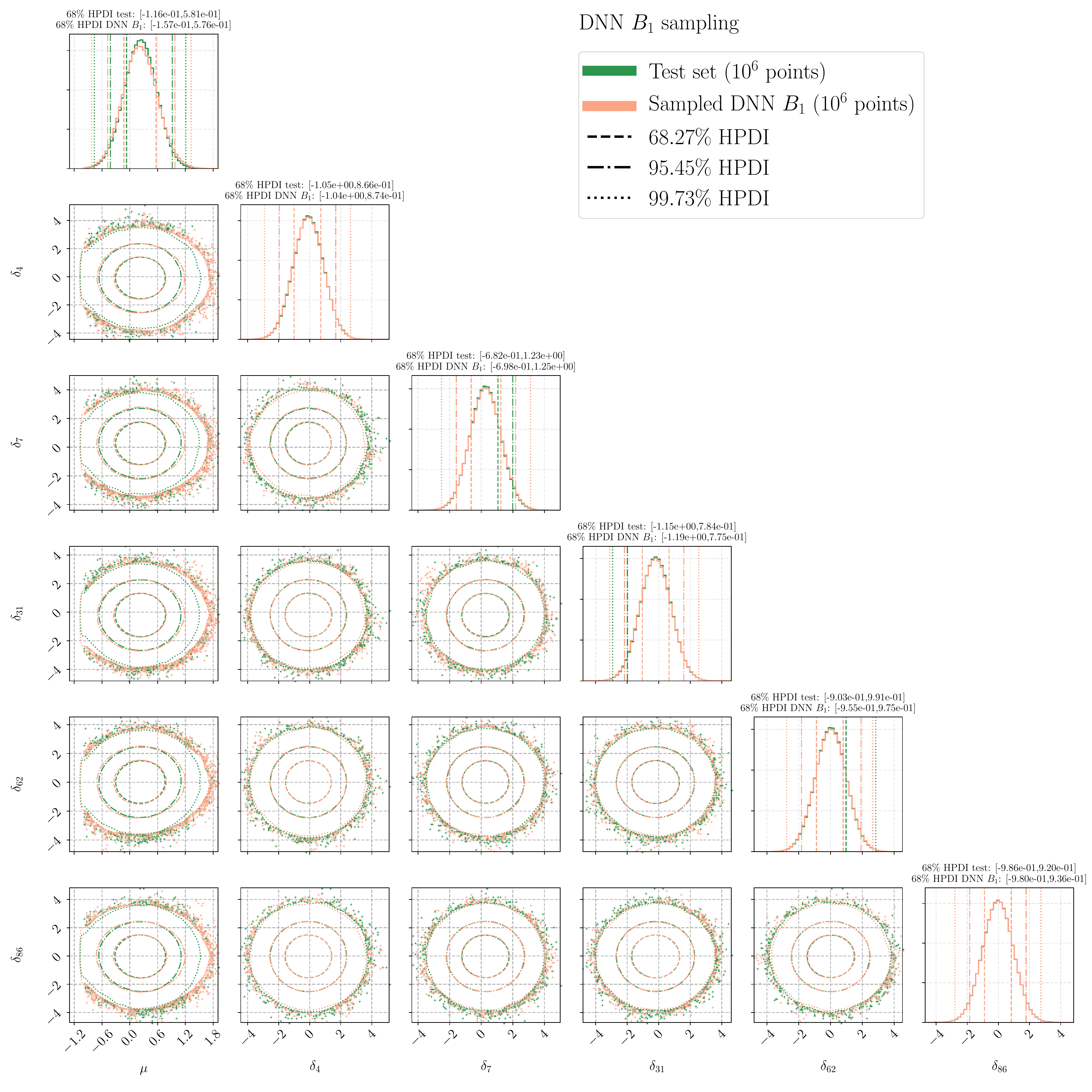}
\caption{1D and 2D posterior marginal probability distributions
  for a subset of parameters from the unbiased $S_{1}$. The green
  (darker) distributions represent the test set of $S_{1}$, while the
  red (lighter) distributions are obtained by sampling the
  DNNLikelihood $B_{1}$. Histograms are made with $50$ bins and
  normalised to unit integral. The dotted, dot-dashed, and dashed
  lines represent the $68.27\%,95.45\%,99.73\%$ 1D and 2D HPDI. The
  difference between green (darker) and red (lighter) lines gives an idea of the uncertainty on the HPDI due to finite sampling. Numbers for the $68.27\%$ HPDI for the parameters in the two samples are reported above the 1D plots.}
\label{fig:corner_toy_lik_exact_vs_DNN_B1_params_resampling}\end{center}
\end{figure*}

\begin{figure*}[htbp!]
\begin{center}
\includegraphics[width=0.93\textwidth]{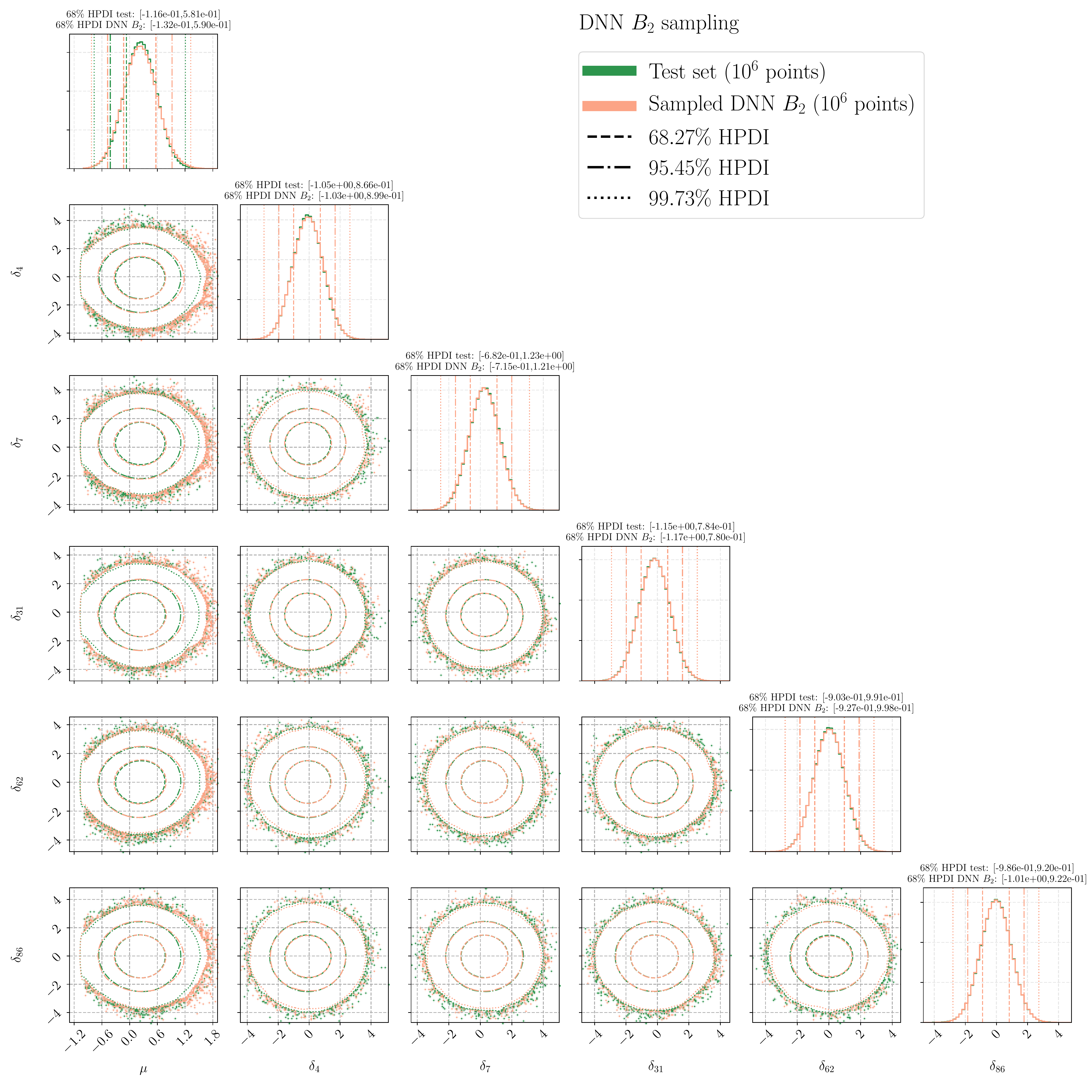}
\caption{Same as Figure \ref{fig:corner_toy_lik_exact_vs_DNN_B1_params_resampling} but for the DNNLikelihood $B_{2}$.}
\label{fig:corner_toy_lik_exact_vs_DNN_B2_params_resampling}\end{center}
\end{figure*}

\begin{figure*}[htbp!]
\begin{center}
\includegraphics[width=0.93\textwidth]{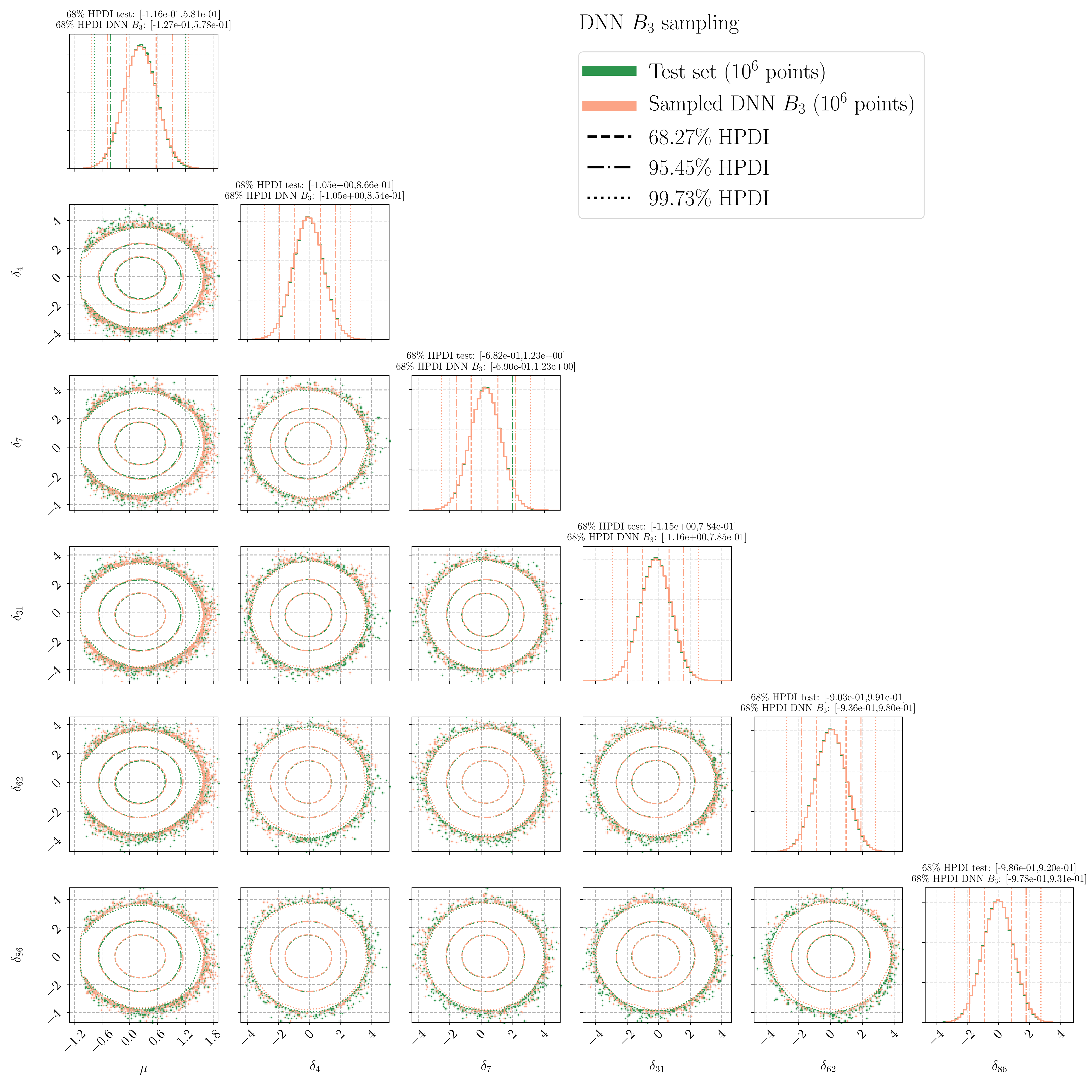}
\caption{Same as Figure \ref{fig:corner_toy_lik_exact_vs_DNN_B1_params_resampling} but for the DNNLikelihood $B_{3}$.}
\label{fig:corner_toy_lik_exact_vs_DNN_B3_params_resampling}\end{center}
\end{figure*}

We have sampled the LF using the DNNLikelihoods $B_{1}-B_{3}$ with the
same procedure used for $S_{1}$.\footnote{Since the effect of
  autocorrelation of walkers is much smaller than the instrinsic error
  of the DNN, to speed up sampling we have used $1024$ walkers with
  $10^{5}$ steps, discarded a burn-in phase of $5\cdot 10^{4}$ steps
  and thinned the remaining $5\cdot 10^{4}$ with a step size of $50$
  to end up with $10^{6}$ samples from each of the DNNLikelihoods.}
Starting from model $B_{1}$ (Figure
\ref{fig:corner_toy_lik_exact_vs_DNN_B1_params_resampling}), Bayesian
inference is well reproduced up to probability intervals of $95.45\%$,
while large deviations start to arise at $99.73\%$. This is
reasonable, since model $B_{1}$ has been trained with only $10^{5}$
points, which are not enough to carefully interpolate in the tails, so
that this region is described by the DNNLikelihood through
extrapolation. Small deviations also arise for smaller probability
intervals, which are again the outcome of a model trained with too few
points. Nevertheless, we want to stress that, considering the very
small training size and the large dimensionality of the LF, model
$B_{1}$ already works surprisingly well. This is a common feature of
the DNNLikelihood, which, as anticipated, works extremely well in
predicting posterior probabilities without the need of a too large
training sample, nor a hard tuning of the DNN hyperparameters. When
going to models $B_{2}$ and $B_{3}$ (Figures
\ref{fig:corner_toy_lik_exact_vs_DNN_B2_params_resampling} and
\ref{fig:corner_toy_lik_exact_vs_DNN_B3_params_resampling})
predictions become more and more reliable, improving as expected with
the number of training points. Notice that the observed deviations are
of the same size, and often smaller, than the ones observed in Figure
\ref{fig:corner_toy_lik_params} between the larger training set and
the smaller test set. Therefore, at least part of the small deviations
observed in the DNNLikelihood prediction have to be attributed to the
finite size of the training set, and are expected to disappear when
further increasing the number of points. Considering the relatively
small training and prediction times shown in Table
\ref{tab:results_DNNLik_Bayes}, it should be possible, once the
desired level of accuracy has been chosen, to enlarge the training and
test sets enough to match that precision. For the purpose of this
work, we consider the results obtained with models $B_{1}-B_{3}$
already satisfactory, and do not go beyond $5 \cdot 10^{5}$ training samples.

In order to allow for a fully quantitative comparison, in Table \ref{tab:HPDI_DNN_B} we summarize the Bayesian 1D HPDI obtained with the DNNLikelihoods $B_{1}-B_{3}$ for the parameter $\mu$ both using reweighting and re-sampling (only upper bounds for the hypothesis $\mu>0$). We find that, taking into account the uncertainty arising from our algorithm to compute HPDI (finite binning) and from statistical fluctuations in the tails of the distributions for large probability intervals, the results of Table \ref{tab:HPDI_DNN_B} are in good agreement with those in Table \ref{tab:HPDI}. This shows that the Bayesian DNNLikelihood is accurate even with a rather small training sample size of $10^{5}$ and its accuracy quickly improves by increasing the training sample size.
\begin{table}[t!]
\begin{center}
\begin{tabular}{c|c|c|c|c}
HPDI 			& $B_{1}$ 	& $B_{2}$ 	& $B_{3}$  \\
\hline\hline
$68.27\%$  	& $0.50$ 	& $0.49$ 	& $0.48$ \\
$95.45\%$  	& $0.92$ 	& $0.91$ 	& $0.88$ \\
$99.73\%$  	& $1.36$ 	& $1.35$ 	& $1.29$ \\
\hline
$68.27\%$  	& $0.49$ 	& $0.49$ 	& $0.49$ \\
$95.45\%$  	& $0.92$ 	& $0.91$ 	& $0.88$ \\
$99.73\%$  	& $1.35$ 	& $1.34$ 	& $1.29$ 
\end{tabular}
\end{center}\caption[]{HPDI obtained from the different DNNLikelihood models $B_{1}-B_{3}$ both by reweighting on the test set of $S_{1}$ (upper block) and by re-sampling (lower block). Results are only shown as upper bound on $\mu$ in the hypothesis $\mu>0$.}\l{tab:HPDI_DNN_B}
\end{table}

\begin{table}[htb!]
\begin{center}
\begin{tabular}{l|c|c|c}
Name													& $F_{1}$				& $F_{2}$				& $F_{3}$				\\
\hline\hline
Sample size ($\times 10^{5}$) 						& $1$					& $2$					& $5$					\\
Epochs													& $183$				& $243$				& $362$					\\
Loss train (MSE) ($\times 10^{-3}$)				& $0.092$				& $0.026$				& $0.030$				\\
Loss val (MSE) ($\times 10^{-3}$)					& $1.18$				& $0.80$				& $0.71$				\\
Loss test (MSE) ($\times 10^{-3}$)				& $1.17$				& $0.80$				& $0.72$				\\
ME train ($\times 10^{-3}$)							& $3.07$				& $0.47$				& $1.1$				\\
ME val	($\times 10^{-3}$)							& $1.78$				& $0.87$				& $0.82$				\\		
ME test ($\times 10^{-3}$)							& $1.50$				& $0.68$				& $0.86$				\\
Median $p$-value of 1D K-S test/pred-train		& $0.53$				& $0.48$				& $0.44$					\\
Median $p$-value of 1D K-S test/pred-val			& $0.15$				& $0.27$				& $0.20$					\\
Median $p$-value of 1D K-S val/pred-test			& $0.13$				& $0.31$				& $0.33$				\\
Mean error on $t_{\mu}$							& $0.11$				& $0.12$				& $0.032$				\\			
Training time (s)										& $1236$				& $2819$				& $7114$					\\
Prediction time ($\mu\text{s}/\text{point}$)		& $11.1$				& $10.8$				& $10.5$				\\
\end{tabular}
\end{center}\caption[]{Results for the best models (Full DNNLikelihood) for different training sample size. All models have been trained for $5$ times to check the stability of the result and the best performing has been quoted. Prediction time is evaluated on a test set with half the size of the training set using the same batch size used in training, and evaluating on a Nvidia Tesla V100 GPU with 32GB of RAM.
All best models have $d_{\text{HL}}=5\cdot 10^{3}$.
}\l{tab:results_DNNLik_Full}
\end{table}

\subsection{Frequentist extension and the full DNNLikelihood}\l{sec:freq_DNNLik}

We have trained the same model architectures considered for the Bayesian DNNLikelihood using the $S_{3}$ sample. 
In Table \ref{tab:results_DNNLik_Full} we show results for the best models we obtained for each training sample size. Results have been obtained by training $5$ identical models and taking the best one. We call these models $F_{1}-F_{3}$ ($F$ may stand for both Full and Frequentist, bearing in mind that these models also allow for Bayesian inference). As we anticipated in the previous Section, the performance gap between models with different number of parameters is very small and often models with less parameters overperform, at least in the hyperparameter space we considered, models with more parameters. This is especially due to our choice of leaving the initial learning rate constant for all architectures, which resulted in a slightly too large learning rate for the bigger models, with a consequently less stable training phase.

\begin{figure*}[htb!]
\begin{center}
\includegraphics[width=0.325\textwidth]{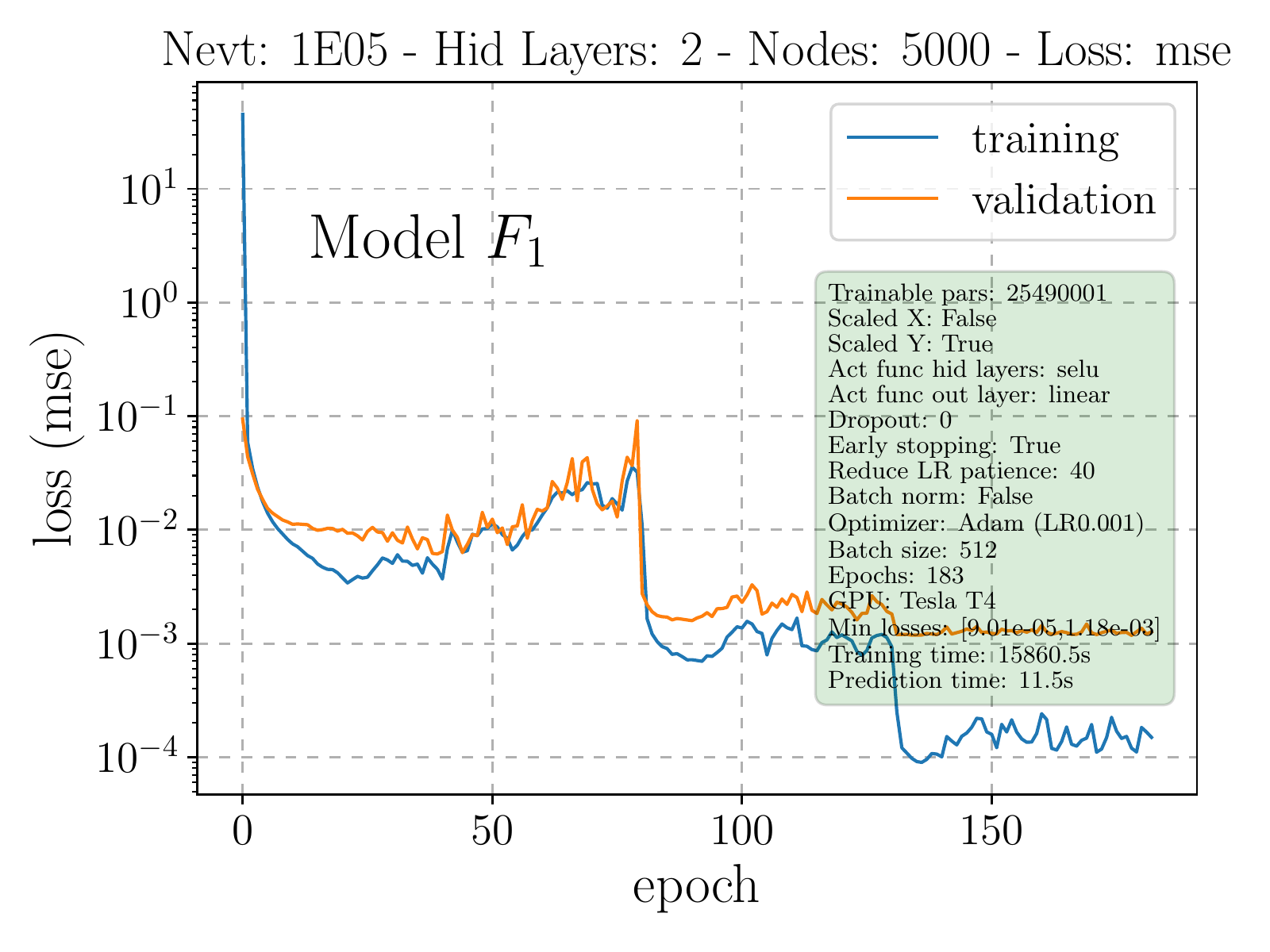}
\includegraphics[width=0.325\textwidth]{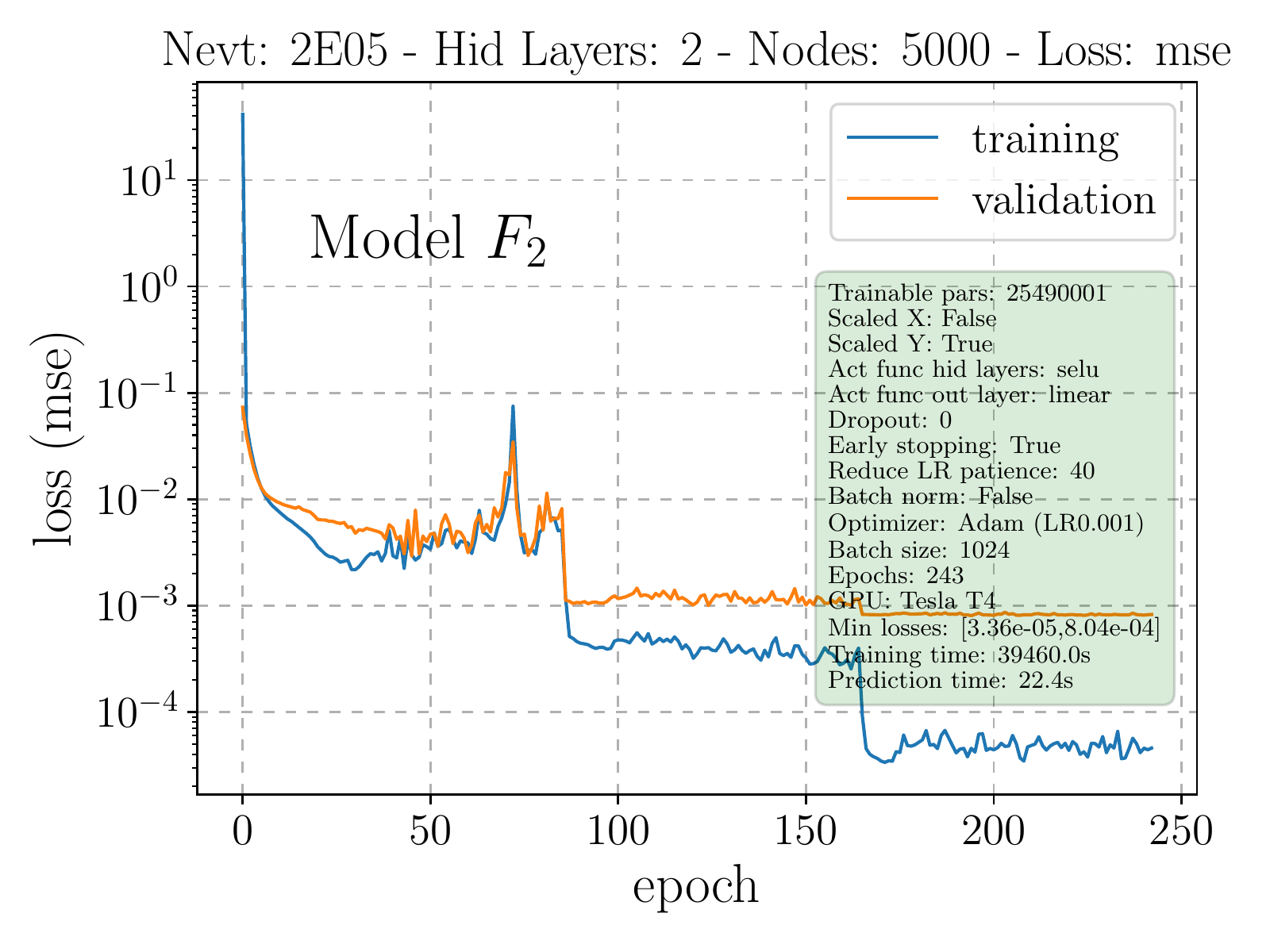}
\includegraphics[width=0.325\textwidth]{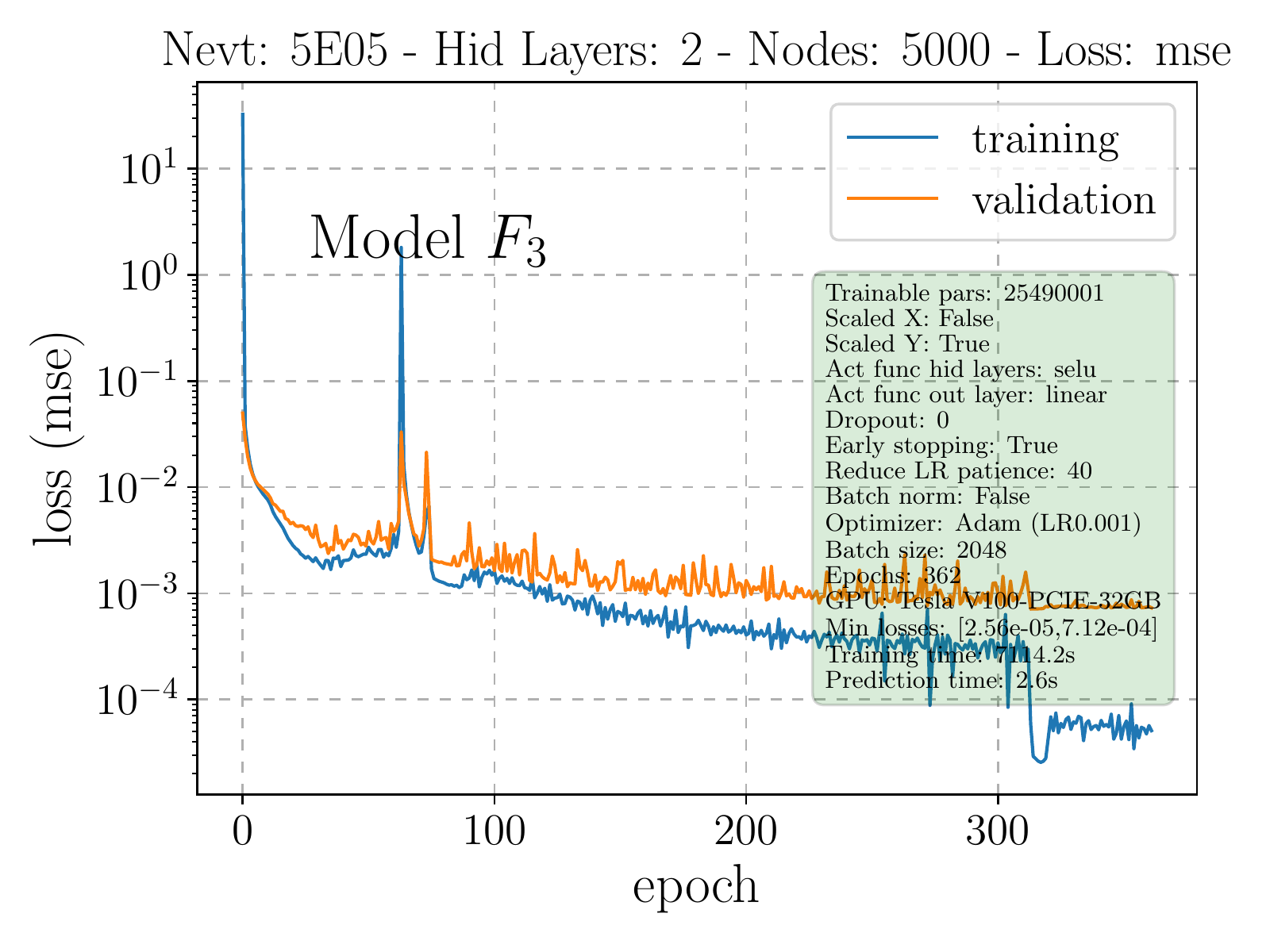}
\caption{Training and validation loss (MSE) vs number of training epochs for models $F_{1}-F_{3}$. The jumps correspond to points of reduction of the \textsc{Adam} optimiser learning rate.}
\label{fig:toy_training_example_Full}\end{center}
\end{figure*}

This can be seen in Figure \ref{fig:toy_training_example_Full}, where
we show the learning curves obtained for the values of the
hyperparameters shown in the legends. The training curves for models
$F_{1}$ and $F_{2}$ are less regular than those of model
$F_{3}$. Nevertheless, as the learning rate gets reduced, they quickly
reach a good validation loss, which promoted them best models for
$10^{5}$ and $2\cdot 10^{5}$ training set sizes. This did not happen
for models trained with $5\cdot 10^{5}$  points, which could have
benefited from reducing further the initial learning rate. However, we
stress that no strong fine-tuning is needed for the DNNLikelihood to
perform extremely well, so that we have chosen the best model $F_{3}$
with less parameters without pushing optimisation further. Final
results are generally very similar to the ones obtained for the
Bayesian DNNLikelihood, with differences arising from the new region
of LF values that is learnt from the DNN. 

\begin{figure*}[htb!]
\begin{center}
\includegraphics[width=0.495\textwidth]{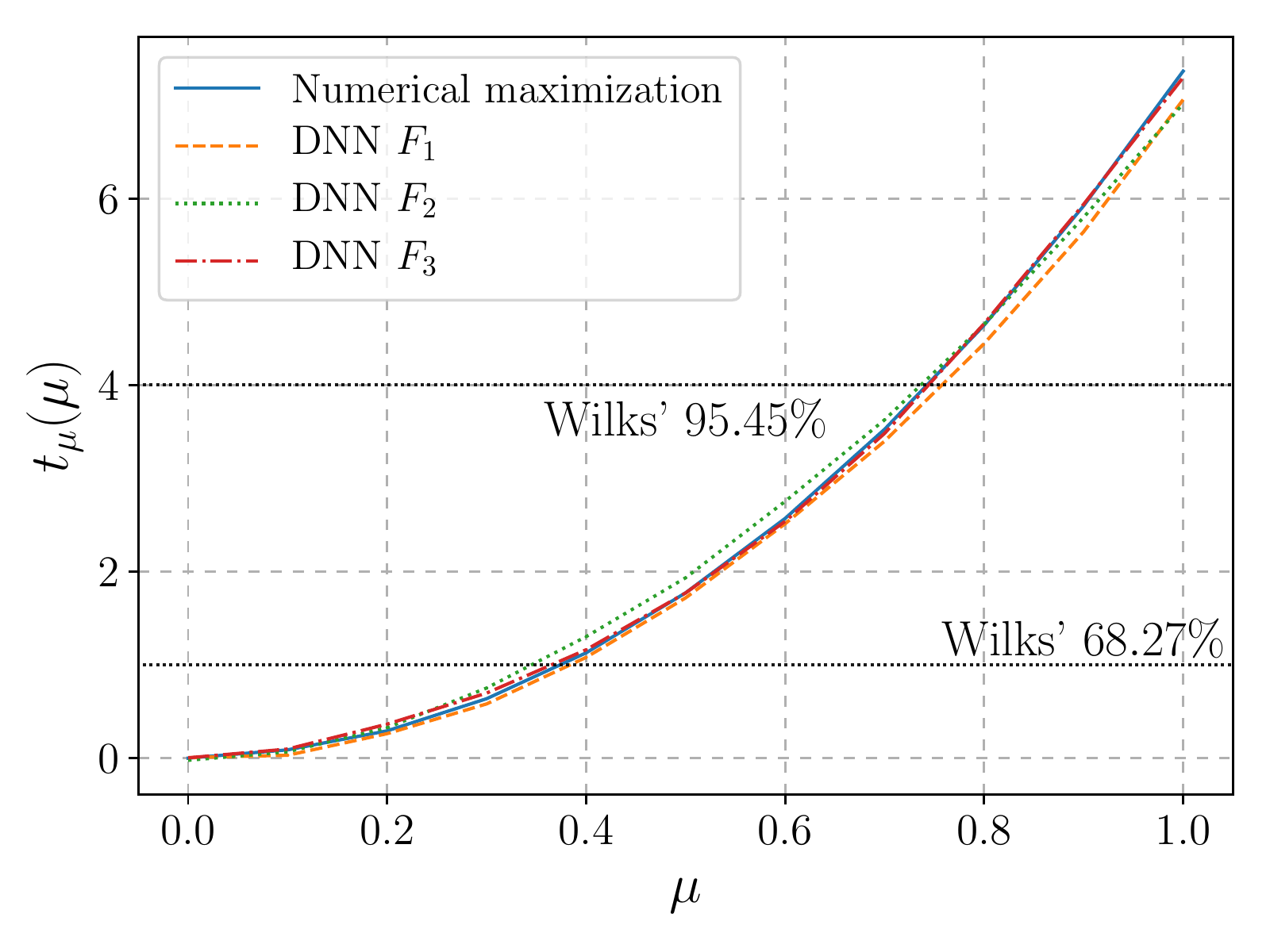}
\includegraphics[width=0.495\textwidth]{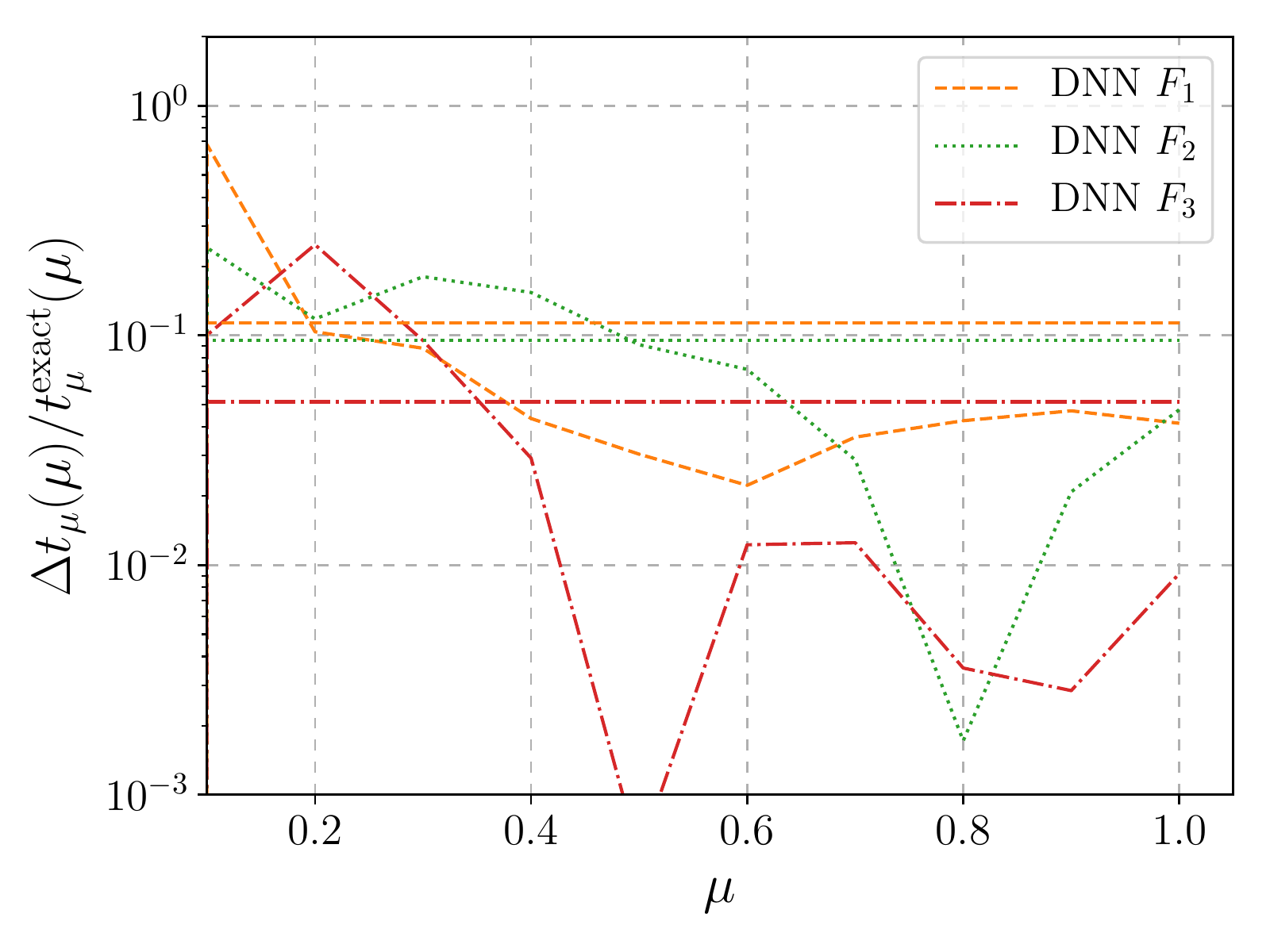}
\caption{Comparison of the $t_{\mu}$ test-statistics computed using
  numerical maximisation of the analytic likelihood and of the
  DNNLikelihoods $F_{1}-F_{3}$. Each of the $t_{\mu}$ prediction
  from the DNNLikelihoods corresponds to the best out of five trained
  models. The left and right plots show respectively the
  $t_{\mu}$ test-statistics and the percentage error computed as
  $|t_{\mu}^{\rm{DNN}}-t_{\mu}^{\rm{exact}}|/t_{\mu}^{\rm{exact}}$. Horizontal
  lines in the right plot represent the mean relative error.}
\label{fig:tmu_DNN_best}\end{center}
\end{figure*}

Before repeating the Bayesian analysis presented for the Bayesian DNNLikelihood, we present results of frequentist inference using the Full DNNLikelihood. We used the models $F_{1}-F_{3}$ to evaluate the test statistics $t_{\mu}$. 
The left panel of Figure \ref{fig:tmu_DNN_best} shows $t_{\mu}$ obtained using the Full DNNLikelihoods, while the right panel of the same Figure shows the relative error with respect to numerical maximisation of the analytic LF. Apart for some visible deviation in the prediction of model $F_{1}$, it is clear that the Full DNNLikelihood is perfectly able to reproduce the test statistics, allowing for robust frequentist inference, already starting with relatively small training sets of few hundred thousand points. Clearly, the larger the training set, the smaller the error, with an average percentage error on $t_{\mu}$ (dashed lines in the right panel of Figure \ref{fig:tmu_DNN_best}) that gets as low as a few percent for our models. We found that ensemble learning can help in reducing differences further (keeping under control statistical fluctuations in the training set and in the DNN weights). However, in the particular case under consideration, results are already satisfactory by taking the best out of five identical models trained with random subsets of the training set. For this reason we do not expand on ensemble learning in this work and just consider it as a tool to improve performance in cases where the LF is extremely complicated or has very high dimensionality.

\begin{figure*}[htbp!]
\begin{center}
\includegraphics[width=0.8\textwidth]{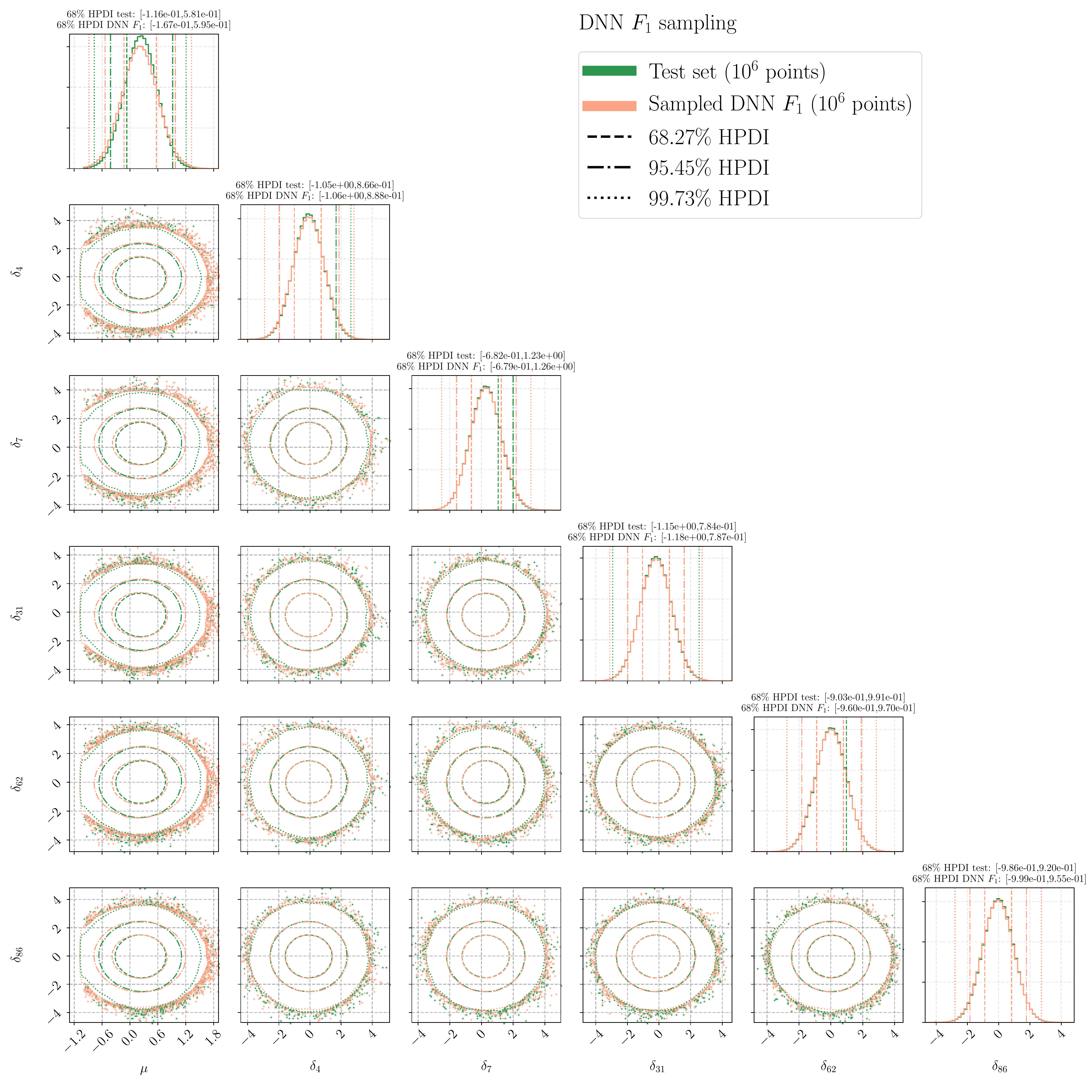}
\caption{1D and 2D posterior marginal probability distributions
  for a subset of parameters from the unbiased $S_{1}$. The green
  (darker) distributions represent the test set of $S_{1}$, while the
  red distributions are obtained by sampling the DNNLikelihood
  $F_{1}$. Histograms are made with $50$ bins and normalised to unit
  integral. The dotted, dot-dashed, and dashed lines represent the
  $68.27\%,95.45\%,99.73\%$ 1D and 2D HPDI. The difference between
  green (darker) and red (lighter) lines gives an idea of the uncertainty on the HPDI due to finite sampling. Numbers for the $68.27\%$ HPDI for the parameters in the two samples are reported above the 1D plots.}
\label{fig:corner_toy_lik_exact_vs_DNN_F1_params_resampling}\end{center}
\end{figure*}

\begin{figure*}[htb!]
\begin{center}
\includegraphics[width=0.8\textwidth]{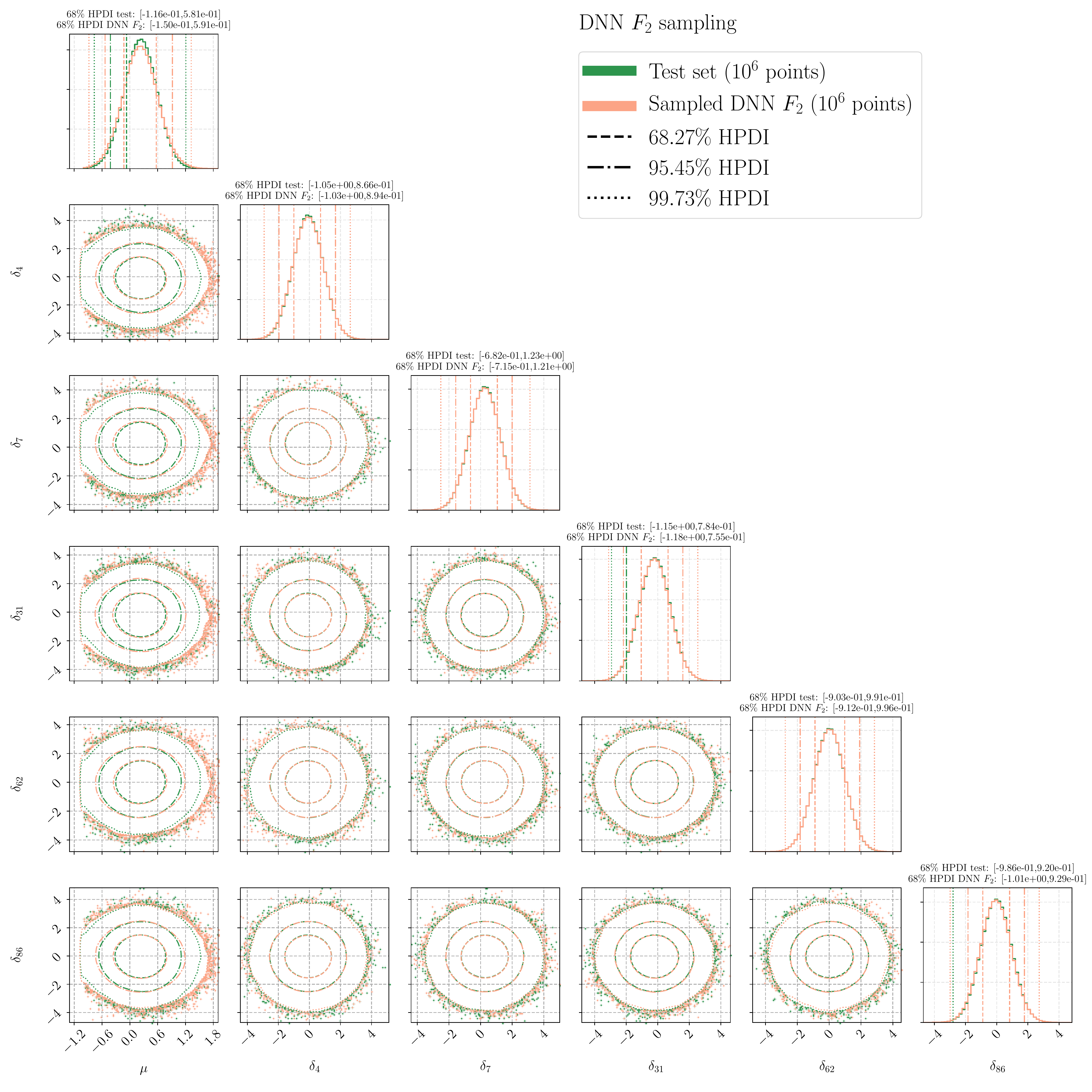}
\caption{Same as Figure \ref{fig:corner_toy_lik_exact_vs_DNN_F1_params_resampling} but for the DNNLikelihood $F_{2}$.}
\label{fig:corner_toy_lik_exact_vs_DNN_F2_params_resampling}\end{center}
\end{figure*}

\begin{figure*}[htb!]
\begin{center}
\includegraphics[width=0.8\textwidth]{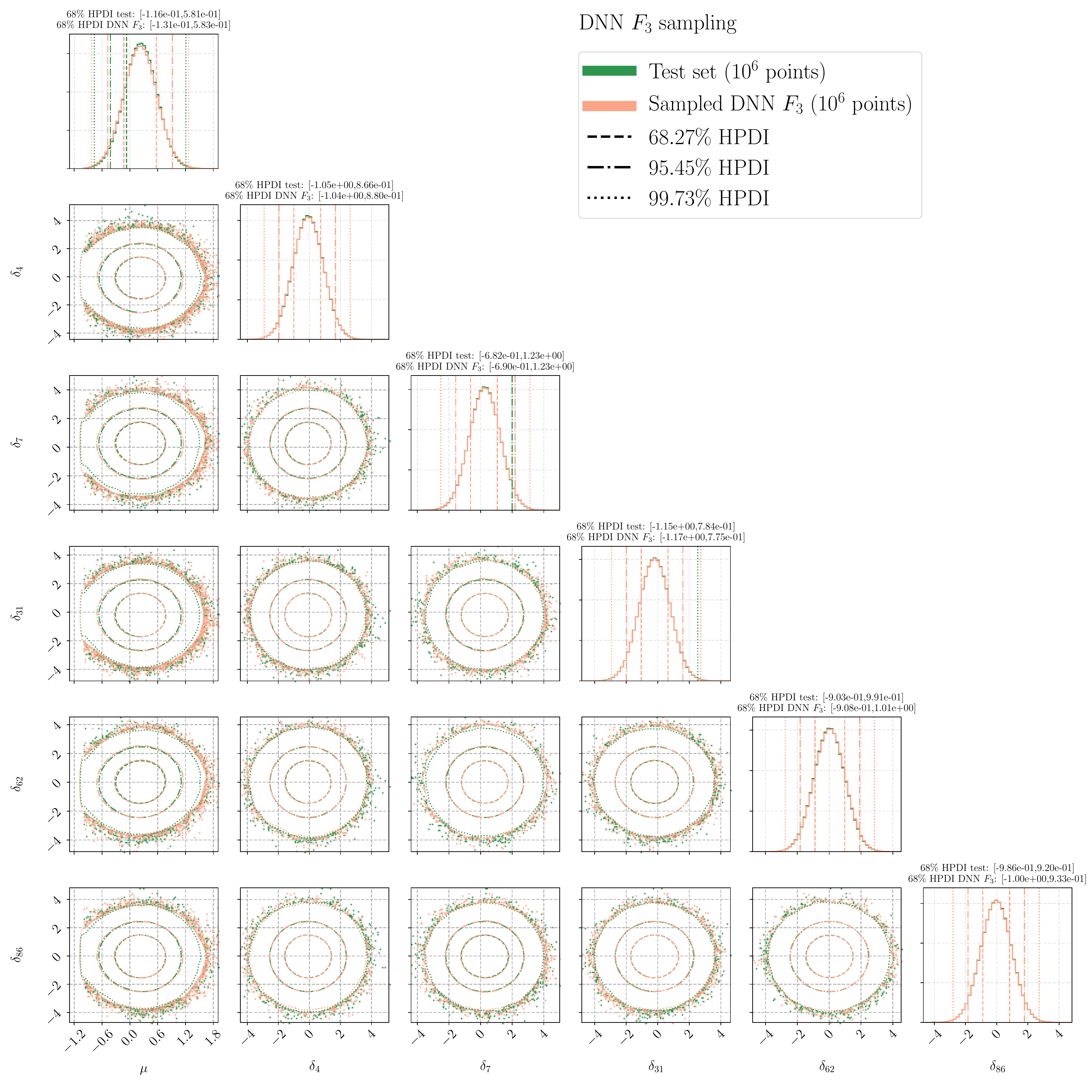}
\caption{Same as Figure \ref{fig:corner_toy_lik_exact_vs_DNN_F1_params_resampling} but for the DNNLikelihood $F_{3}$.}
\label{fig:corner_toy_lik_exact_vs_DNN_F3_params_resampling}\end{center}
\end{figure*}

We now show how the full DNNLikelihood is also able to catch all the
features necessary for Bayesian inference. We do so by repeating the
analysis done for models $B_{1}-B_{3}$ for the full DNNLikelihoods
$F_{1}-F_{3}$. Plots obtained by sampling the DNNLikelihood are shown
in
Figs. \ref{fig:corner_toy_lik_exact_vs_DNN_F1_params_resampling}-\ref{fig:corner_toy_lik_exact_vs_DNN_F3_params_resampling}. Results
are quantitatively unchanged with respect to the Bayesian
DNNLikelihood. For completeness we also summarize in Table
\ref{tab:HPDI_DNN_F} the Bayesian 1D HPDI obtained with the
DNNLikelihoods $F_{1}-F_{3}$ for the physics parameter $\mu$. Results
are compatible with what we found in Table \ref{tab:HPDI_DNN_B} for
the Bayesian DNNLikelihood.

\begin{table}[htb!]
\begin{center}
\begin{tabular}{c|c|c|c}
HPDI 			& $F_{1}$ 	& $F_{2}$ 	& $F_{3}$  \\
\hline\hline
$68.27\%$  	& $0.50$ 		& $0.50$ 		& $0.49$ 	\\
$95.45\%$  	& $0.93$ 		& $0.92$ 		& $0.89$ 	\\
$99.73\%$  	& $1.39$ 		& $1.37$ 		& $1.31$ 	\\
\hline
$68.27\%$  	& $0.50$ 		& $0.50$ 		& $0.48$ 	\\
$95.45\%$  	& $0.93$ 		& $0.93$ 		& $0.88$ 	\\
$99.73\%$  	& $1.35$ 		& $1.37$ 		& $1.28$ 	
\end{tabular}
\end{center}\caption[]{HPDI obtained from the different DNNLikelihood models $F_{1}-F_{3}$ both by reweighting on the test set of $S_{1}$ (upper block) and by re-sampling (lower block). Results are only shown as upper bound on $\mu$ in the hypothesis $\mu>0$.}\l{tab:HPDI_DNN_F}
\end{table}

\section{Conclusion and future work} \label{sec:conclusion}

Publishing and distributing likelihoods in a simple yet general way is
becoming a key issue in many fields, and in particular in the physics
of fundamental interactions, where experimental (and phenomenological)
results typically involve complicated (and often multi-modal or
degenerate) likelihoods which depend on hundreds of parameters. We have
introduced the DNNLikelihood framework, in which the full likelihood,
binned or unbinned, including the dependence on all nuisance
parameters, is published by providing a suitably trained DNN
predictor. Distributing the results of experimental or phenomenological
analyses using the DNNLikelihood framework allows for the combination
of different analyses, for the reinterpretation of the results under
different hypotheses, and for the use of the likelihood in a different
statistical framework, without loss of information.

We have illustrated the power of the DNNLikelihood discussing in
detail the toy experiment presented in ref.~\cite{Buckley:2018vdr}, which mimics a realistic LHC-like NP search. We found that the DNNLikelihood is able to catch the main features of the true LF, allowing for both frequentist and Bayesian inference, already with a limited amount of training data.

A \textsc{Jupyter} notebook, together with \textsc{Python} source files which allow to reproduce all results presented above
are available on GitHub \href{https://github.com/riccardotorre/DNNLikelihood}{\faGithub}. A dedicated \textsc{Python} package allowing to sample LFs and to build, optimize, train, and store the corresponding DNNLikeliihoods is in preparation. This will allow not only allow to construct and distribute DNNLikelihoods,
 but also to use them for inference both in the Bayesian and frequentist frameworks.

We plan to release soon the first examples of the use of the DNNLikelihood for true experimental likelihoods,
including unbinned, multimodal likelihoods, in forthcoming
publications.

As a final remark, we stress the fact that the proposed approach supports and complements the remarkable step taken by the ATLAS collaboration in publishing full experimental likelihoods~\cite{Aad:2019pfy,Cranmer:1456844,ATL-PHYS-PUB-2019-029}. On one hand, the DNNLikelihood framework can be seen as a way to export a HistFactory likelihood to a software environment that doesn't meet the needed dependencies, or as an alternative option for experimental collaborations that don't use HistFactory. On the other, the use of a DNN model doesn't imply specific choices on the likelihood function (e.g.\ it covers equally well unbinned likelihoods and/or likelihoods built from analytical functions and not represented as histograms).

Last but not least, using a DNNLikelihood could be a viable solution to distribute the
outcome of phenomenological studies, such as the ones presented in
refs. \cite{Ciuchini:2000de,Hocker:2001xe,Charles:2004jd,Bona:2005vz,Bona:2007vi,Ciuchini:2013pca,Baak:2014ora,deBlas:2016ojx,Falkowski:2017pss,Ellis:2018gqa,Ciuchini:2019usw,Brivio:2019ius}.

On a long term, a large scale adoption of likelihood publishing towards the DNNLikelihood framework would motivate the
possibility of publishing DNN models on HEPData (which, more generically, would be beneficial for reproducibility issues related to physics analysis using DNNs for selection, etc.), for instance supporting the ONNX format. In this respect, and considering that there could be interest in likelihood publishing in other domains (e.g.\ for the $\Lambda_{CDM}$ likelihood in cosmology), it might be worth considering a less hyerarchical submission procedure for HEPData (e.g.\ allowing individuals outside a structured organization/experimental collaboration to submit a likelihood function)
or the opportunity to create a separate (and not HEP specific) likelihood function repository, e.g.\ based on Zenodo.

\section*{Acknowledgements}
We are indebted to the authors of Ref.~\cite{Buckley:2018vdr} for providing us with details and data concerning the LHC-like experiment considered in this paper. This project has received funding from the European Research Council (ERC) under the European Union's Horizon 2020 research and innovation program (grant agreement n$^o$ 772369) and from the Italian Ministry of Research (MIUR) under grant PRIN 20172LNEEZ. The work of R.~Torre has been partially supported by the INFN Grant MALHEPHYCA.

\appendix

\section{On the multivariate normal distribution} \label{app:multvarnorm}
For the univariate normal distribution (centered in zero and with unit variance)
\be
\mathcal{N}_{1}(x)=\f{1}{\sqrt{2\pi}}e^{-\f{x^{2}}{2}}\,,
\ee
the confidence intervals at $n\sigma$ are defined by the quantiles of the $\chi_k^2$ distribution with $k=1$ degrees of freedom by
\be\l{eq:ci2univariate}
n^2 = 2 Q^{-1}(\f{1}{2},0,1-\alpha)\,,\qquad \alpha = Q(\f{1}{2},0,\f{n^{2}}{2})
\ee
where $1-\alpha$ is the area of the distribution within $\pm n\sigma$ from the mean. 
Generalization to a multivariate normal distribution in $l$ dimensions with zero vector mean and identity matrix variance $\Sigma=I_{l}$
\be
\mathcal{N}_{l}(\bm{x})=\f{1}{(2\pi)^{l/2}}e^{-\f{1}{2}x_{i}\Sigma^{ij} x_{j}}\,,
\ee
is obtained by considering the $\chi_{k}^2$ distribution with $k=l$ degrees of freedom. Equation \eqref{eq:ci2univariate} then becomes
\be\l{eq:ci2multivariate}
n^2 = 2 Q^{-1}(\f{l}{2},0,\alpha)\,,\qquad \alpha = Q(\f{l}{2},0,\f{n^{2}}{2})\,.
\ee
The typical size of $N_{l}$ outside of the confidence interval at $n\sigma$ is given by
\be
\mathcal{N}_{l}\(\bm{x}\sim \|\mu-n\sigma\|\)\approx \f{1}{(2\pi)^{l/2}}e^{-\f{1}{2}2Q^{-1}(\f{l}{2},0,1-\alpha))}\,,
\ee
where we have denoted by $\bm{x}\sim \|\mu-n\sigma\|$ a point that is approximately $n\sigma$ away from the $l$ dimensional mean.
The target set of the distribution within $n\sigma$ from the mean is therefore given by:
\be
\mathcal{N}_{l}\(\bm{x}\lesssim \|\mu-n\sigma\|\) \in (2\pi)^{-l/2}\{\min(\mathcal{N}_{l}(\bm{x}))|_{\bm{x}\lesssim \|\mu-n\sigma\|},1\}=\{e^{-\f{1}{2}2Q^{-1}(\f{l}{2},0,1-\alpha))},1\}\,,
\ee
where $\min(\mathcal{N}_{l}(\bm{x}))|_{\bm{x}\lesssim \|\mu-n\sigma\|}$ is the minimum of $\mathcal{N}_{l}$ for $\bm{x}$ within $n\sigma$ from the mean divided by the normalisation factor $(2\pi)^{-l/2}$. In Table \ref{tab:minloglik} we show the value of $-\log_{10}(\min(\mathcal{N}_{l}(\bm{x}))|_{\bm{x}\lesssim \|\mu-n\sigma\|}) $ for $l=1,10,100,1000$ dimensions and for confidence intervals up to $6\sigma$.
\begin{table}[htb!]
\begin{center}
\begin{tabular}{c|c|c|c|c}
$n\backslash l$ & $1$ & $10$ & $100$ & $1000$ \\
\hline
$1$  & $0$ & $3$   & $23$  & $222$\\
$2$  & $1$ & $4$   & $27$  & $234$\\
$3$  & $2$ & $6$   & $31$  & $245$\\
$4$  & $3$ & $8$   & $36$  & $256$\\
$5$  & $5$ & $10$ & $40$  & $268$\\
$6$  & $8$ & $13$ & $45$  & $279$
\end{tabular}
\end{center}\caption{Value of $-\log_{10}(\min(\mathcal{N}_{l}(\bm{x}))|_{\bm{x}\lesssim \|\mu-n\sigma\|}) $ for $l=1,10,100,1000$ dimensions and for confidence intervals up to $6\sigma$. See the text for details.}\l{tab:minloglik}
\end{table}

\section{Pseudo-experiments and frequentist coverage} \label{app:coverage}

In order to check the validity of the asymptotic approximation given by Wilks' theorem, and to obtain more robust upper bounds from the $t_{\mu}$ test-statistics for the LHC-like search discussed in Section \ref{sec:toy}, we need to generate several pseudo-experiments for each hypothesised ``true'' value of $\mu$, compute the log-likelihood and the test statistics $t_{\mu}$ for each pseudo-experiment, and study the pdf of $t_{\mu}$, denoted by $f(t_{\mu}|\mu)$. The cumulative distribution of $f(t_{\mu}|\mu)$ determines the coverage properties of the test-statistics $t_{\mu}$ through the relation
\be\l{eq:pvaluetmu}
1-p_{\mu}=\int_{0}^{t_{\mu,\text{obs}}}f(t_{\mu}|\mu)dt_{\mu}\,,
\ee
where $t_{\mu,\text{obs}}$ is the $t_{\mu}$ of our actual experiment at the given value of $\mu$. By solving this equation for $\mu$ at specified value of $p_{\mu}=\alpha$ we get the value of $\mu$ excluded at a CL of $1-\alpha$.

\begin{figure*}[t!]
\begin{center}
\includegraphics[width=0.495\textwidth]{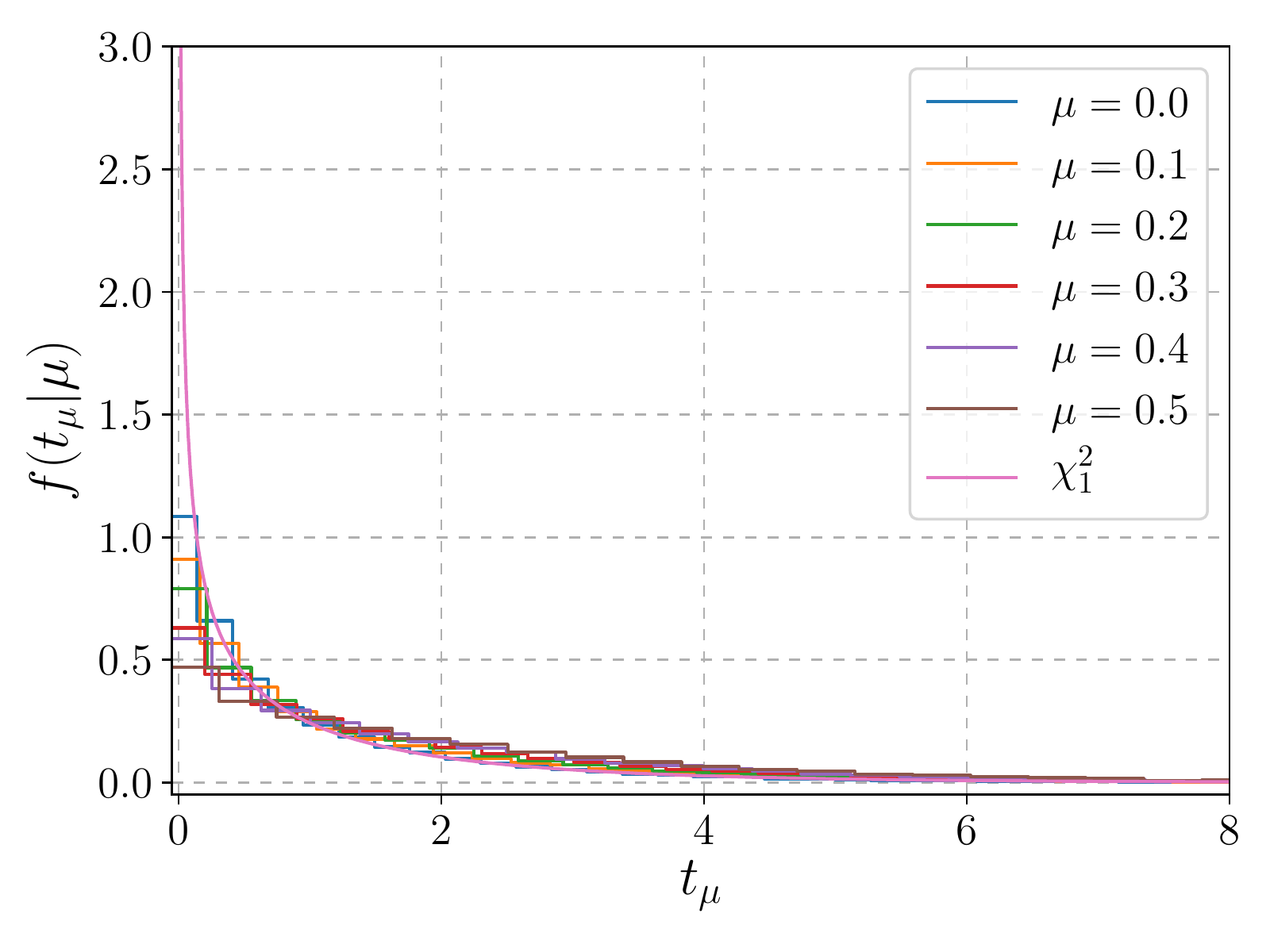}
\includegraphics[width=0.495\textwidth]{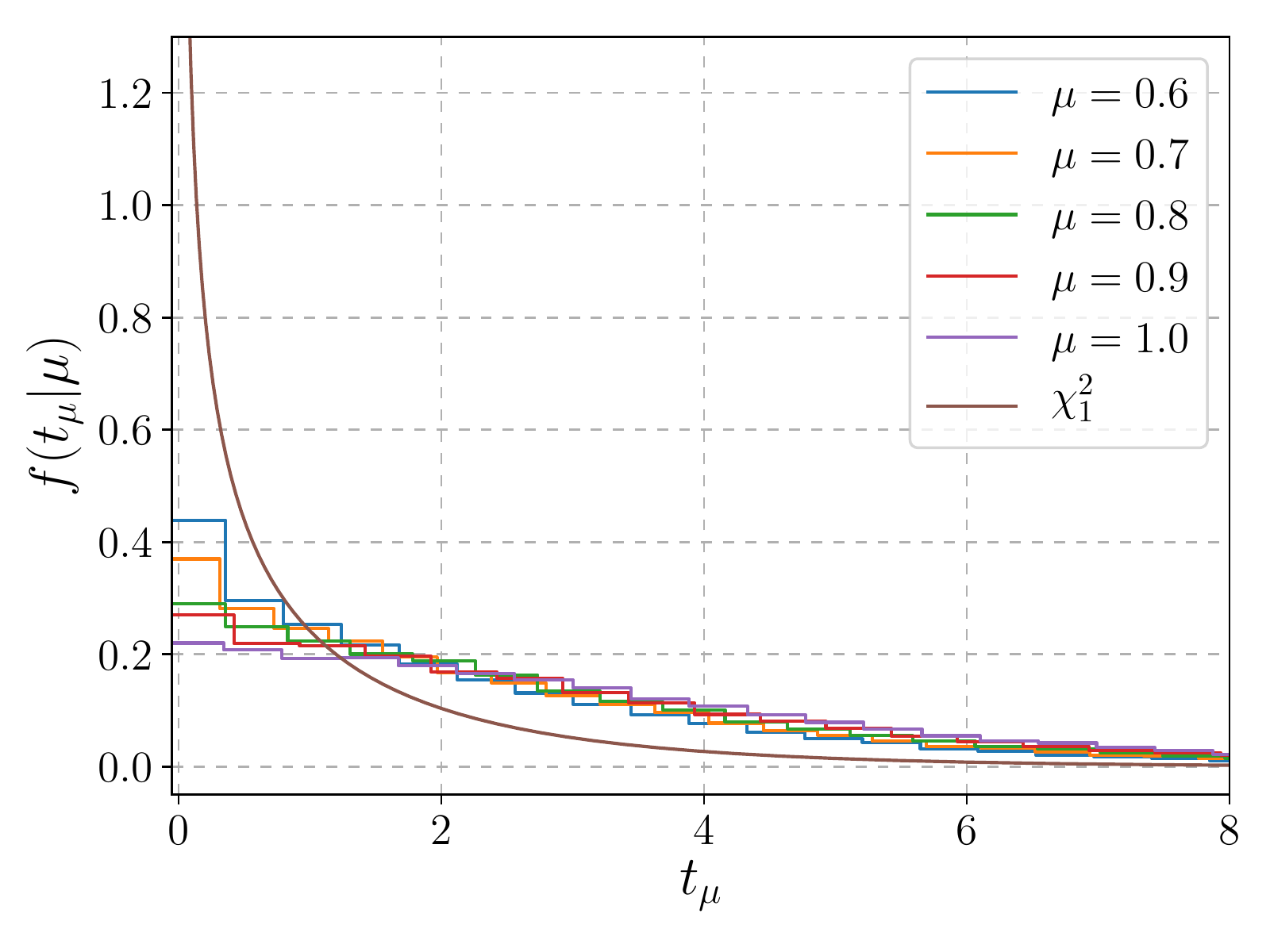}\vspace{-3mm}\\
\includegraphics[width=0.495\textwidth]{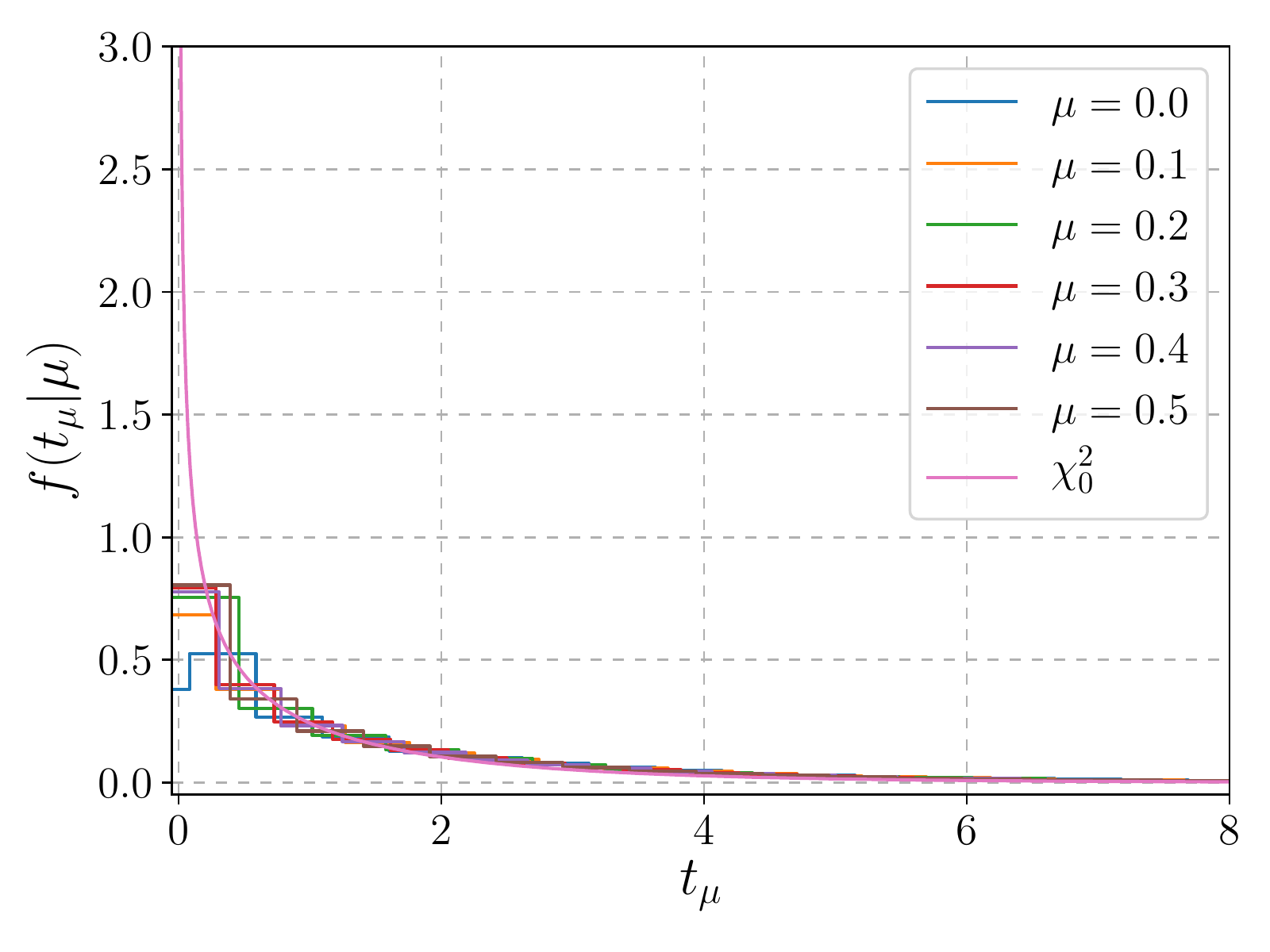}
\includegraphics[width=0.495\textwidth]{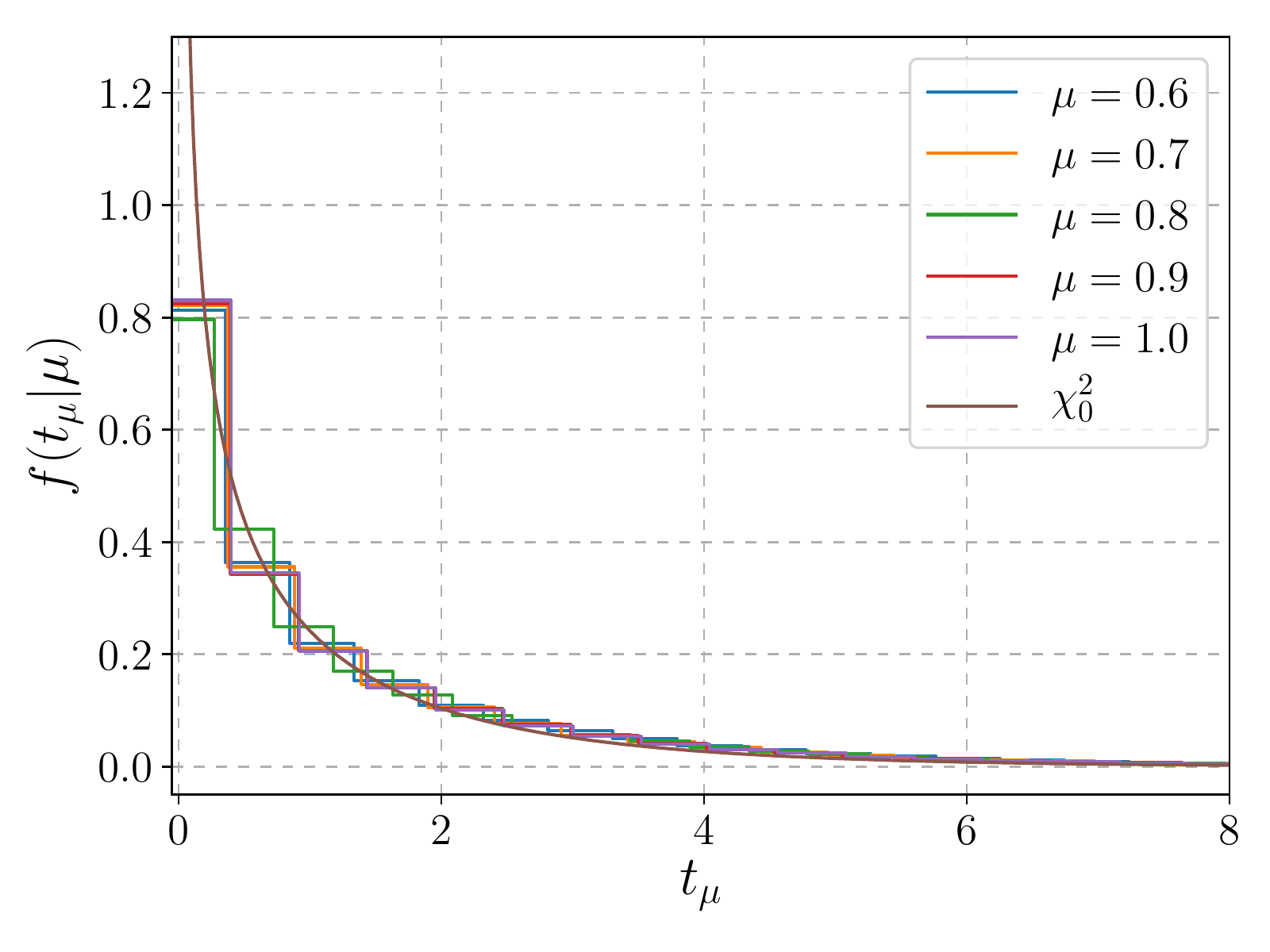}
\caption{Distributions $f(t_{\mu}|\mu)$ for the LHC-like new physics search discussed in Section \ref{sec:toy}, obtained from pseudo-experiments employing the profile construction (upper panel) and the hybrid frequentist-Bayesian tratement of nuisance parameters (lower panel). A $\chi^{2}_{1}$ distribution is shown for comparison.}
\label{fig:coverage}\end{center}
\end{figure*}

We generate pseudo-experiments following two different procedures for the treatment of nuisance parameters and compare the results. 
\begin{itemize}
\item {\bf Profile construction}\\
First we consider a frequentist treatment of nuisance parameters, referred to as profile construction \cite{Cranmer:2005hi,Tanabashi:2018oca}: we determine the values $\bm{\hat{\theta}}_{\mu}$ of the nuisance parameters at the maximum of the LF of our actual experiment for different values of $\mu$, and keep them fixed to generate several pseudo-experiments for each $\mu$ according to eq.~\eqref{eq:expobs}. This approach encodes statistical fluctuations arising in repeated experiments, but does not take into account systematic uncertainties. Although this approach violates the ``anticipation criterion'' \cite{Demortier:2003vc}, it is expected to work well, and have good coverage properties when the uncertainty is statistically dominated.\footnote{This was the approach employed, for instance, in the LHC Higgs boson search combination in 2011 \cite{ATLAS:2011tau}.} Indeed the coverage of this approach grows as the profiled value $\bm{\hat{\theta}}_{\mu}$ approaches the ``true'' value of $\bm{\theta}$. This happens when systematic uncertainties become less and less relevant compared to statistical fluctuations.

Following this procedure we generated $5\cdot 10^{4}$ pseudo-experiments for each value of $\mu$ in the range $[0,1]$ with steps of $0.1$.
\vspace{3mm}
\item {\bf Hybrid frequentist-Bayesian (marginal model)}\\
In order to understand the impact of systematic uncertainties we also consider the hybrid frequentist-Bayesian treatment of nuisance parameters (referred to as marginal model in ref.~\cite{Tanabashi:2018oca}): pseudo-experiments are generated including both variations of the nuisance parameters according to their distribution, and statistical uncertainty through eq.~\eqref{eq:expobs}. This allows to include the effect of systematic uncertainties in the generation of pseudo-experiments. 

In this case we generated, for the same values of $\mu$ considered above, $10^{4}$ expected counts (for all $90$ bins) corresponding to variations of the nuisance parameters over their multivariate distibution, and subsequently used each of these expected counts to generate $10$ pseudo-experiments according to eq.~\eqref{eq:expobs}. This delivers a total of $10^{5}$ pseudo-experiments for each vaue of $\mu$, encoding both statistical and systematic uncertainties.
\end{itemize}
%
\begin{table}[t!]
\begin{tabular}{lrrrrrrrrrrr}
$\mu$ 						& $0.0$ 	& $0.1$ 	& $0.2$ 	& $0.3$ 	& $0.4$ 	& $0.5$ 	& $0.6$ 	& $0.7$ 	& $0.8$ 	& $0.9$ 	& $1.0$ \\
\hline\hline 
$t_{\mu,\text{obs}}$ 		& $0.0$ 	& $0.09$ 	& $0.29$ 	& $0.63$ 	& $1.13$ 	& $1.77$ 	& $2.57$ 	& $3.53$ 	& $4.64$ 	& $5.92$ 	& $7.37$ \\
\hline
coverage (profile) 		& $0.0$ 	& $0.15$ 	& $0.27$ 	& $0.37$ 	& $0.49$ 	& $0.59$ 	& $0.68$ 	& $0.76$ 	& $0.83$ 	& $0.89$ 	& $0.93$ \\
coverage (hybrid) 			& $0.0$ 	& $0.15$ 	& $0.30$ 	& $0.47$ 	& $0.61$ 	& $0.73$ 	& $0.83$ 	& $0.90$ 	& $0.94$ 	& $0.97$ 	& $0.99$ \\
coverage ($\chi^{2}_{1}$) 	& $0.0$ 	& $0.23$ 	& $0.41$ 	& $0.57$ 	& $0.71$ 	& $0.82$ 	& $0.89$ 	& $0.94$ 	& $0.97$ 	& $0.99$ 	& $0.99$ \\
\end{tabular}
\caption{Values of $\mu$ with corresponding $t_{\mu,\text{obs}}$ and CL coverage for the LHC-like new physics search discussed in Section \ref{sec:toy}, obtained from pseudo-experiments employing the profile construction and the hybrid frequentist-Bayesian tratement of nuisance parameters. The last row shows the asymptotic result obtained with a $\chi^{2}_{1}$ distribution.}\l{tab:coverage}
\end{table}
\begin{table}[t!]
\begin{center}
\begin{tabular}{lrrr}
CL		 							& $68.27\%$		& $95.45\%$		& $99.73\%$ \\
\hline\hline
$t_{\mu}$ (profile)				& $2.71$			& $>7.37$		& $>7.37$  \\
$\mu$ (profile)	 				& $0.61$			& $>1.00$		& $>1.00$  \\
\hline
$t_{\mu}$ (hybrid)				& $1.52$			& $5.26$			& $>7.37$  \\
$\mu$ (hybrid) 					& $0.46$			& $0.85$			& $>1.00$  \\
\hline
$t_{\mu}$ ($\chi^{2}_{1}$)	& $1.00$			& $4.00$			& $9.00$  \\
$\mu$ ($\chi^{2}_{1}$) 		& $0.37$			& $0.74$			& $>1.00$  \\
\end{tabular}
\caption{Upper bounds on $t_{\mu}$ and corresponding $\mu$ at different CL for the LHC-like new physics search discussed in Section \ref{sec:toy}, obtained from pseudo-experiment employing a profiled frequentist and hybrid frequentist-Bayesian tratement of nuisance parameters. The last two rows show the asymptotic result obtained with a $\chi^{2}_{1}$ distribution.}\l{tab:coverage_result}
\end{center}
\end{table}
 Figure \ref{fig:coverage} shows the distributions $f(t_{\mu}|\mu)$ for each value of $\mu$, while the probability values covered for each value of $\mu$ and the corresponding $t_{\mu,\text{obs}}$ are given in Table \ref{tab:coverage}. Using the distributions $f(t_{\mu}|\mu)$ we performed the integral in eq.~\eqref{eq:pvaluetmu} for $1-p_{\mu}=0.6827,0.9545,0.9973$, getting the upper bounds on $\mu$ reported in Table \ref{tab:coverage_result}.
All results are shown for both the profile construction and the hybrid frequentist-Bayesian approach.

As expected, due to the small statistics in several bins, the asymptotic result is inaccurate, and in particular tends to undercover the true value. This delivers a too ``aggressive'' upper bound for $\mu$. The hybrid approach gives relatively more conservative upper bounds, very similar (expectedly, given the marginal model used for the nuisance parameters) to the Bayesian result reported in Table \ref{tab:HPDI}. Finally, the profile construction gives the most conservative bound, that is substantially more conservative than the asymptotic one. We do not dare here to discuss which upper bound is to be quoted, since we are only interested in assessing the validity of the asymptotic approximation.



\bibliographystyle{mine}
\bibliography{likelihoods}
\end{document}